\documentclass[]{mn2e}

\usepackage{amsfonts}
\usepackage{amsmath}
\usepackage{graphicx}

\usepackage{appendix}
\usepackage{longtable}
\usepackage{url}

\title[Sersic + Exponential]{Systematic effects on the size-luminosity relations of early and late type galaxies: dependence on model fitting and morphology}
\author[Bernardi et al.]{\parbox{\textwidth}{M. Bernardi$^{1}$\thanks{E-mail: bernardm@sas.upenn.edu},
A. Meert$^{1}$, V. Vikram$^{1}$, M. Huertas-Company$^{2}$, 
S. Mei$^{2}$, F. Shankar$^{2}$  \& R. K. Sheth$^{1,3}$}\vspace{0.4cm}\\
\parbox{\textwidth}{$^{1}$Department of Physics and Astronomy, University of Pennsylvania, 
Philadelphia, PA 19104, USA\\
$^{2}$GEPI, Observatoire de Paris, CNRS, Univ. Paris Diderot;
Place Jules Janssen, 92190 Meudon, France\\
$^{3}$The Abdus Salam International Center for Theoretical Physics, 
      Strada Costiera 11, 34151 Trieste, Italy\\ }}

\begin{document}
 \date{Accepted June 2014}

\maketitle

\label{firstpage}

\begin{abstract}

We quantify the systematics in the size-luminosity relation of galaxies in the SDSS main sample (i.e. at $z\sim 0.1$) which arise from fitting different one- and two-component model profiles to the $r$-band images.  For objects brighter than $L_*$, fitting a single Sersic profile to what is really a two-component SerExp system leads to biases:   the half-light radius is increasingly overestimated as $n$ of the fitted single component increases; it is also overestimated at B/T$\sim 0.6$.  For such objects, the assumption of a single Sersic component is particularly misleading.  However, the net effect on the $R-L$ relation is small, except for the most luminous tail.  We then study how this relation depends on morphology.  Our analysis is one of the first to use Bayesian-classifier derived weights, rather than hard cuts, to define morphology.  For the $R-L$ relation Es, S0s and Sa's are early-types, whereas Sbs and Scds are late, although S0s tend to be 15\% smaller than Es of the same luminosity, and faint Sbs are more than 25\% smaller than faint Scds.  Neither the early- nor the late-type relations are pure power-laws:  both show significant curvature, which we quantify.  This curvature confirms that two mass scales are special for both early- and late-type galaxies: $M_* \sim 3\times 10^{10}M_\odot$ and $2\times 10^{11}M_\odot$.  Also, although the $R_{\rm disk}-L_{\rm disk}$ and $R_{\rm disk}-M_{*\rm disk}$ relations of disks of disk-dominated galaxies run parallel to the corresponding relations for the total light in late-types (i.e., they are significantly curved), $R_{\rm bulge}-L_{\rm bulge}$ and $R_{\rm bulge}-M_{*\rm bulge}$ for bulge dominated systems show almost no curvature (i.e., unlike for the total light of earl-type galaxies).  Finally, the intrinsic scatter in the $R-L$ relation decreases at large $L$ and/or $M_*$ and should provide additional constraints on models of how the most massive galaxies formed.

\end{abstract}


\section{Introduction}
The spatial (and color) distribution of star light in a galaxy is thought to encode information about its formation history, so there has been considerable interest in developing accurate descriptions of the projected surface brightness distribution of galaxies.  

One approach to this problem is to fit the free parameters of a predetermined functional form to the observed surface brightness profile.  These derived free parameters (typically, these are expressed in terms of the scale which contains half the total light, and the surface brightness at this scale) are more useful if the functional form itself actually does provide a good fit to the profile.  A simple version of this approach is to fit many different functional forms to the data, and then select the one which provides the best fit (in some suitably quantified way).  For example, the Sloan Digital Sky Survey (SDSS; Stoughton et al. 2002) reports fits of both exponential ($I(\theta) \propto \exp(-\theta/\theta_1)$) and de~Vaucouleurs ($\propto \exp[-(\theta/\theta_4)^{1/4}]$) profiles to the image, along with an estimate of which fits better.  

The exponential and de~Vaucouleurs (1948) profiles are special cases ($n=1$ and $4$) of the Sersic (1968) profile 
\begin{equation}
 I(\theta) = I_n\, \exp[-(\theta/\theta_n)^{1/n}].  
\end{equation}
With sufficiently good data, it is possible to simply fit a Sersic profile to the data, leaving the fitting procedure to determine $n$ as well as $I_n$ and $\theta_n$.  If galaxies really are intrinsically single Sersics with a wide range of $n$, then the parameters (e.g. half-light radius) returned by forcing $n=1$ or $4$ in the single component fits will generally be biased.  Across the population as a whole, the derived value of $n$ spans a wide range, sometimes being as large as $\sim 8$ or 10 (e.g. Simard et al. 2011 and references therein), suggesting that forcing $n=1$ or $4$ is ill-advised.  

Of course, it is not obvious that the light profile should be fit using a single component.  The stellar kinematics in many galaxies indicate that the stars define more than one dynamical component.  Examples include counter-rotating disks, as well as disk systems with bulges or bars in their centers (e.g. Emsellem et al. 2011).  Evidence for more than one component is often seen in the chemical composition as well (e.g. Johnston et al. 2012).  In such galaxies, it is interesting to see if the light profile also indicates the presence of more than one component.  

This has motivated studies which model the observed profile as the sum of an exponential and a deVaucouleurs profile; what we will call the {\tt deVExp} model.  (Of course, since there are now more free parameters to be fit, better, higher resolution data are required.  In this context, it is worth noting that Sersic's initial motivation was to fit a functional form with fewer free parameters which would allow one to interpolate between two-component systems having varying fractions of an $n=4$ bulge and an $n=1$ disk.)  It is common to report the result of such two-component fits in terms of the fraction of the total light that is in the bulge (de~Vaucouleurs) component:  B/T.  Correlations of these B/T values with other parameters (e.g. luminosity) are then used to constrain formation history scenarios. 

On the other hand, if galaxies really are single component Sersics, and one attempts to fit them with two component {\tt deVExp} profiles, then one will infer an entirely spurious B/T value (the profile was, after all, just a single component).  This spurious B/T will correlate with other parameters if $n$ itself does, complicating the interpretation of such correlations.  Indeed, some have argued that the evidence for two-components in the light profile is sometimes just a consequence of trying to fit what is really a single component Sersic with a linear combination of exponential and deVaucouleurs profiles (e.g. Graham et al. 2003), although this leaves unanswered the question of why dynamically or chemically distinct components do not leave a signature in the light.

In recent years, the correlation between size and luminosity for early-type galaxies has received much attention, because high redshift early-types appear to be more compact than their counterparts at low redshift (e.g. Trujillo et al. 2006; van Dokkum et al. 2008; Cimatti et al. 2008; Bruce et al. 2012).  However, both the size and the luminosity estimates, $R$ and $L$, are derived parameters, obtained by fitting to the observed surface brightness distribution.  As a result, they depend on assumptions about the intrinsic shape of the surface brightness profile.  E.g., if the fit assumes that galaxies are made up of two components or just one, and if two, whether they are modelled as the sum of an exponential and a deVaucouleurs (1948) profile, an exponential and a Sersic (1968), or two Sersics.  

The main goal of this paper is to quantify the systematics on the local $R-L$ relation which are associated with the choice of a particular model.  In practice, `local' means the $5\times 10^5$ galaxies at $z\sim 0.1$ in DR7 of the SDSS Main Galaxy sample (Abazajian et al. 2009).  Because this sample is apparent magnitude limited ($m_r<17.75$), in practice, by $R-L$ relation we always mean $\log_{10}R$ fitted as a function of absolute magnitude (see Sheth \& Bernardi 2012 for a simple description of the bias which would arise from fitting $L$ as a function of $R$).  And $R$ denotes the radius which encloses half the total light $L$.  (For exponential disks, this radius is 1.67 times the scale length of the exponential.  In the case of two components, $R$ is a complicated function of the light in each component and the two scale radii.)  

Our goal implies that we must fit the observed profiles and determine the associated $R-L$ relation using a variety of different models.  Section~\ref{sdss} summarizes the relevant properties of the SDSS DR7 sample we study.  Section~\ref{2components} provides an analysis of the light profiles of SDSS DR7 galaxies which we believe strongly suggests that fitting to a SerExp model returns the least biased answers.  
Section~\ref{lr} compares the $R-L$ relation based on single Sersic, Sersic + exponential, deVaucouleur + exponential, and single deVaucouleur fits, showing that the relations from single Sersic fits (the standard to date) are offset to larger sizes and those from single deVaucouleur fits to smaller sizes, compared to those from the two-component fits.  



There is no particular reason why systems supported by rotational motions should define the same $R-L$ relation as those supported by random motions.  Therefore, it makes little sense to speak of a single $R-L$ relation for the entire galaxy population.  Indeed, the $R-L$ relation has long been known to depend on morphological type (e.g. Shen et al. 2003).  Section~\ref{lrMorph} quantifies this morphological dependence, and then focuses on the differences between the relations defined by early-type bulge dominated systems, and later type disk-dominated systems.  
Disk dominated galaxies have small bulges, and bulge dominated galaxies have extended second components. In Section~\ref{bd} we use our Sersic-exponential fits to study the $R_{\rm bulge}-L_{\rm bulge}$ relation of early-types, $R_{\rm disk}-L_{\rm disk}$ of late types, and the ratio of the bulge size to that of the other component in both early- and late-types. Note that we do not distinguish between bars and bulges in late-type galaxies but none of our science goals depend on this distinction.
A final section summarizes our findings.

\section{The SDSS DR7 sample}\label{sdss}
In this paper, we study the galaxies in the Seventh Data Release (DR7) of the SDSS Main Galaxy sample (Abazajian et al. 2009).  This sample contains $\sim 5\times 10^5$ galaxies at $z\sim 0.1$, and is apparent magnitude limited to $m_r<17.75$, where $m_r$ is a Petrosian magnitude. This limit is sufficiently bright that surface brightness related selection effects are negligible.

The DR7 database provides crude estimates of the (galactic extinction corrected) apparent brightness and angular size of each galaxy in the catalog.  We will use the SDSS-based {\tt cmodel} magnitudes and sizes (a weighted combination of separate fits to exponential and deVaucouleurs profiles) defined in Bernardi et al. (2010).  When converting these to physical sizes and luminosities, we assume a flat $\Lambda$CDM model with $\Omega_m=0.3$ and a Hubble constant whose present value is $H_0 = 70$~km~s$^{-1}$Mpc$^{-1}$.  The luminosities are $k$-corrected following Bernardi et al. (2003a).  For colors we use SDSS DR7 {\tt model} magnitudes (corrected for galactic extinction).  The database also provides estimates of the stellar velocity dispersion of each galaxy.  We follow custom and correct these values for aperture effects following Bernardi et al. (2003a).  


In the first half of this paper, we describe a number of other estimates of the total light associated with each $r$-band image (analysis of the images in other bands is ongoing).  However, recent work has focussed on stellar masses $M_*$ rather than luminosity $L$.  Our estimated stellar masses come from combining our estimates of $L_r$ with $(M_*/L_r)$ estimated following Bernardi et al. (2010) assuming a Chabrier IMF.

\subsection{Estimates of photometric parameters}
One of our primary goals is to quantify how the $R-L$ relation depends on how $R$ and $L$ were estimated.  To this end, we perform fits to the SDSS DR7 Main Galaxy Sample images using the {\tt PyMorph} package, which can fit seeing convolved two-component models to observed surface brightness profiles (Vikram et al. 2010; see their Figs.4--6 for a discussion of the steps involved, and choices made regarding e.g., the pixels to be masked, centering and alignment of the two components, etc.).  The algorithm is described and tested in Meert et al. (2013) who show that when the fitted functional form is the same as the one used to generate the image, then {\tt PyMorph} returns accurate values of the free-parameters (e.g., background sky-level, total light, half-light radius, Sersic index, axis-ratio, bulge-total ratio).  

We use {\tt PyMorph} to fit single component deVaucouleurs and Sersic profiles, and two component exponential + deVaucouleur (deVExp) and exponential + Sersic (SerExp) profiles to each image.  It is conventional to speak of the two components as being `bulge' and `disk' components; while this is accurate for disk-dominated systems (typically later-type galaxies), it may be better to think of the `disk' component in bulge-dominated systems (typically early-type galaxies) as simply being a second component that is not necessarily a (thin, inclined) disk (e.g. Oemler 1976; Schombert 1986; Gonzalez et al. 2005).  

Before moving on, we note that there is an analytic expression for the light enclosed within a given distance of the center of a single circular Sersic profile (e.g. Ciotti \& Bertin 1999).  From this, the half light radius can be obtained easily.  However, if the object has axis ratio $b/a\ne 1$, where $b$ and $a$ are the half-light radii along the principal axes of the image, then the corresponding expression must be integrated numerically.  Since this can be time-consuming, it is usual to approximate this case by using the expression for a circle, but with a suitably chosen effective circular radius.  The most common choice is $\sqrt{ba} = a\, \sqrt{b/a}$, but Saglia et al. (2010) have recently shown that $(b+a)/2$ is more accurate:  for bulge dominated systems the difference matters little, but it does matter for disks.  Therefore, we use $(b+a)/2$ except in Section~\ref{lrSersic} where, to fairly compare with previous work, we use $\sqrt{ba}$.  

This raises the question of what we should do when we have two components?  
A natural choice would be to circularize each component using its own $(b+a)/2$, and to then determine the half light radius of the sum of the circularized components, where each is weighted by the fraction of the total light that it contains (e.g. equation~\ref{rhBT} in the Appendix).  We have found that this approximation is quite accurate, so we use it throughout.  

\subsection{Morphologies}
A secondary goal of this paper is to quantify the role of galaxy type or morphology on the $R-L$ relation.  In practice this is difficult, because unambiguous determinations of the morphological type are not straightforward, although the task is slightly easier for bulge dominated systems.  Previous work has used crude proxies for morphological type:  these include isophotal shape and central concentration (Strateva et al. 2001), the Sersic index $n$ (Shen et al. 2003), the color, spectral features, and some combination of the above (Bernardi et al. 2003a; Baldry et al. 2004).  

In what follows, we use the Bayesian Automated Classifications (hereafter BAC) of Huertas-Company et al. (2011) which are available for our full DR7 sample.  The BAC classifications are particularly interesting, because they are expressed as probabilistic weights (determined from an object's $k$-corrected $g-r$ and $r-i$ colors, and its isophotal shape and light concentration in the $i$-band) 
-- something we expect will become increasingly common in the next generation of large datasets.  We explore the use of hard cuts based on these weights as indicators of morphology, as well as simply weighting each galaxy by the BAC-probability that it is one type or another.  



For instance, we will study an `early-type' sample defined on the basis of hard conservative cuts on two parameters which are available for each galaxy:  the value of $n$ returned by {\tt PyMorph} when fitting a single Sersic profile to the image, and the BAC probability $p$(E+S0) that the object is an early-type.  We require 
 $$n>3 \qquad {\rm and}\qquad p({\rm E+S0}) > 0.85.$$
These cuts by no means select all early-type galaxies; they are simply designed to select a population which is very unlikely to be contaminated by later-types.  Since our goal is to select objects of a single type, we are willing to sacrifice completeness for purity.  

To assess how these BAC-based hard-cuts perform, we use the eye-ball classifications of Fukugita et al. (2007; hereafter F07) and of Nair et al. (2010; hereafter N10).
These are based on analysis of a much smaller patch of the SDSS sky, and a brighter magnitude limit (e.g. F07 has only $\sim 7000$ objects restricted to $m_r<16$; N10 has about twice as many), but for our purposes, the important point is that they are both magnitude limited.  

Whereas BAC classifies galaxies into 4 (E,S0,Sab,Scd) morphological types, F07 use $0<$T$<7$ in steps of 0.5.  To convert, we assign E (T = 0 and 0.5), S0 (T = 1), Sa (T = 1.5 and 2), Sb (T = 2.5 and 3), and Scd (T = 3.5, 4, 4.5, 5, and 5.5).  Similarly, N10 use the T-Type classification ($-5<$T$<7$) from the modified RC3 classifiers; we assign E (T$=-5$ and T$=-4$), S0 (T$= -3$, T$=-2$ and T$=-1$), Sa (T$=0$, T$=1$ and T$=2$), Sb (T$=3$ and T$=4$), and Scd (T$= 5$, T$=6$ and T$=7$).

\begin{table}
\begin{tabular}{lccccc}
   \hline
   Selection &  E & S0 & Sa & Sb & Scd \\
   \hline
   \hline
   EARLY-TYPES & & & & & \\
   \hline
   Selected & & & & & \\
   \hline
P(E+S0)$ > 0.85$ AND $n > 3$ &   0.70 &  0.21 &  0.08 &  0.01 &  0 \\
$n > 2.5$ &   0.44 &  0.18 &  0.20 &  0.13 & 0.05\\
   \hline
   Missed & & & & & \\
   \hline
P(E+S0)$ < 0.85$ OR $n < 3$  &   0.10 & 0.43 & &\\
$n < 2.5$  &   0.02 & 0.12 & &\\
\hline
\hline
   LATE-TYPES & & & & & \\
\hline
   Selected & & & & & \\
   \hline
P(E+S0)$ < 0.15$ AND $n < 3$ &   0 & 0.01 &  0.08 &  0.36 &  0.51\\
$n < 2.5$ &   0.01 &  0.04 &  0.11 &  0.33 & 0.45\\
   \hline
   Missed & & & & & \\
   \hline
P(E+S0)$ > 0.15$ OR $n > 3$  & & & 0.86 & 0.47 & 0.21\\
$n > 2.5$  & & &  0.75 & 0.41 & 0.16\\
   \hline
   \hline
  \end{tabular}
\caption{Eyeball morphological classifications from Fukugita et al.  
(2007). We set E (T = 0 and 0.5), S0 (T = 1), Sa (T = 1.5 and 2), Sb  
(T = 2.5 and 3), and Scd (T = 3.5, 4, 4.5, 5, and 5.5).}
 \label{F07}
\end{table}

Table~\ref{F07} shows the mix of F07 morphological types in samples which are defined by the hard cuts (on $n$ and BAC $p$(type)) given above.  Table~\ref{N10} reports a similar analysis which is based on the eye-ball classifications of N10 instead of F07.  These Tables show that 91\% and 86\% of the resulting sample are indeed either E and S0.  In contrast, requiring only $n>2.5$ (as done in the past) yields a sample in which the E$+$S0 fraction is just 62\% and 56\% respectively.  (A small fraction of the objects are Irregulars, which is why the numbers do not always add up to 100\%.)  Clearly our selection is much purer.  

As a measure of its incompleteness, we also indicate the fraction of objects classified as Es and S0s which do not make the cut.  These fractions are 10\% and 43\% for the F07 classifications, and 14\% and 43\% for N10.  This `missed' fraction is much smaller if we only require $n<2.5$, but we believe the price to pay in purity is unacceptable.

The bottom halves of the two tables show a similar analysis of BAC and $n$ cuts which are designed to produce a pure sample of later types.  In this case requiring 
 $$n<3 \qquad {\rm and}\qquad p({\rm E+S0}) < 0.15$$
yields a sample in which Sa + Sb + Scd account for 95\% of the objects.  If we only require $n<2.5$, and do not use the BAC-probability at all, then  Sa + Sb + Scd account for 89\% of the objects so, for later types, the use of the BAC analysis does not make such a dramatic difference.  

Whereas the extremes of the morphological mix are relatively easy to define, Tables~\ref{F07} and~\ref{N10} indicate that the intermediate regime, the Sa/Sb class, will be difficult to define cleanly.  Indeed, we have found that more than a third of the objects with BAC $p$(S0)~$>0.6$ are Sa's, and  about a fifth of the objects with $p$(Scd)~$>0.6$ are Sb's.  Conversely, of the objects which have $p$(Sab)~$>0.6$, about one third are Scd's.  For this reason, when we provide these BAC weight-derived relations in Tables~\ref{LRfits} and~\ref{MsRfits}, we refer to them as being for E, S0/Sa, Sa/Sb/Scd and Scd samples.  

We will use these hard cuts, as well as the BAC-weights themselves, in Section~4.  Note that all the results which follow are based on the full SDSS DR7 sample:  we use the smaller F07 and N10 samples again only to perform sanity checks (as we did here) in Section~\ref{lrMorph}.

\begin{table}
\begin{tabular}{lccccc}
   \hline
   Selection &  E & S0 & Sa & Sb & Scd \\
   \hline
   \hline
   EARLY-TYPES & & & & & \\
   \hline
   Selected & & & & & \\
   \hline
P(E+S0)$ > 0.85$ AND $n > 3$ &   0.57  &   0.29 &    0.14 &  0 & 0 \\
$n > 2.5$ &   0.32 &  0.23 &  0.29 &   0.12 &  0.03\\
   \hline
   Missed & & & & & \\
   \hline
P(E+S0)$ < 0.85$ OR $n < 3$  &   0.14 & 0.43 & &\\
$n < 2.5$  &   0.02 & 0.07 & &\\
\hline
\hline
   LATE-TYPES & & & & & \\
\hline
   Selected & & & & & \\
   \hline
P(E+S0)$ < 0.15$ AND $n < 3$ &   0 &  0 &  0.15 &  0.41 & 0.39\\
$n < 2.5$ &   0.01 &  0.03 &  0.17 &   0.35 &  0.36\\
   \hline
   Missed & & & & & \\
   \hline
P(E+S0)$ > 0.15$ OR $n > 3$  & & &  0.79 & 0.35 & 0.17\\
$n > 2.5$  & & &  0.71 & 0.33 & 0.10\\
   \hline
   \hline
  \end{tabular}
\caption{Eyeball morphological classifications from Nair et al. (2010)  
who used T-Type classification using the modified RC3 classifiers. We  
set E (T$=-5$ and T$=-4$), S0 (T$= -3$, T$=-2$ and T$=-1$), Sa (T$=  
0$, T$=1$ and T$=2$), Sb (T$=3$ and T$=4$), and Scd (T$= 5$, T$=6$ and  
T$=7$).}
 \label{N10}
\end{table}

\section{Sersic index and B/T ratio in SDSS galaxies: Evidence for two components in the surface brightness profile}\label{2components}
In this Section, we provide an analysis of the light profiles of SDSS DR7 galaxies which we believe strongly suggests that fitting to a SerExp model returns the least biased answers.  

\subsection{How many components?}

\begin{figure*}
\centering
\includegraphics[scale = 0.5]{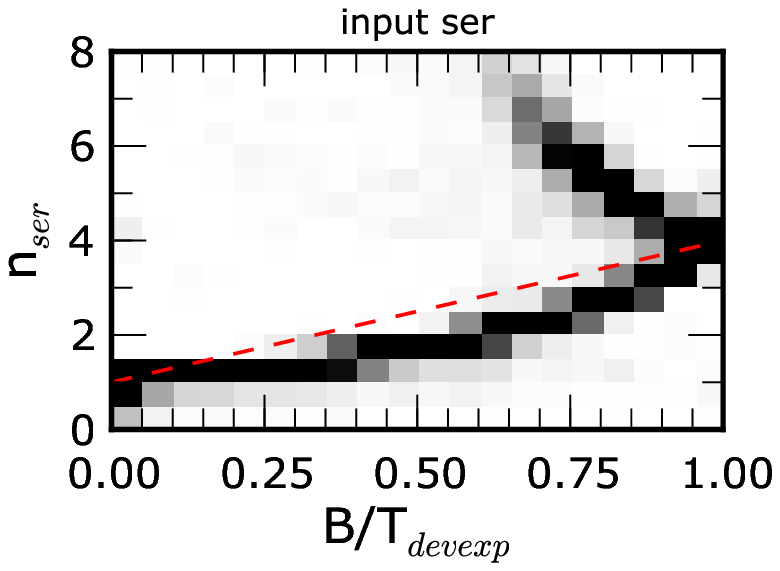}
\includegraphics[scale = 0.5]{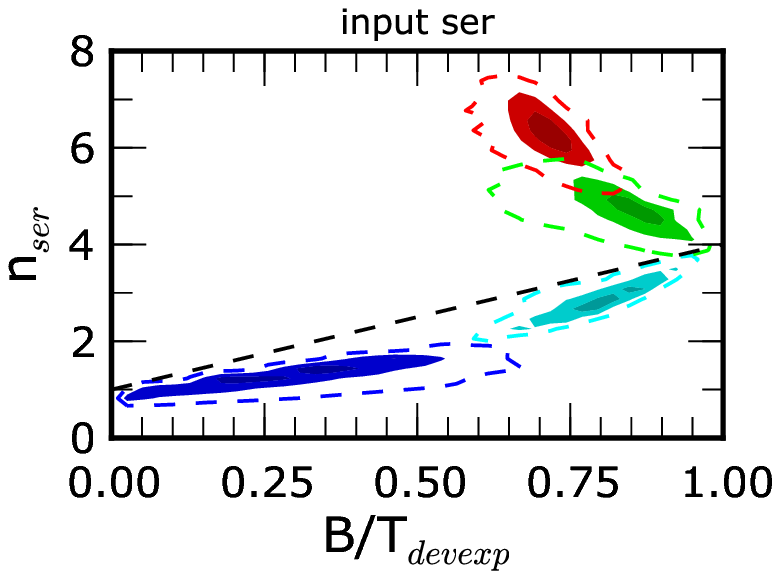}
\includegraphics[scale = .5]{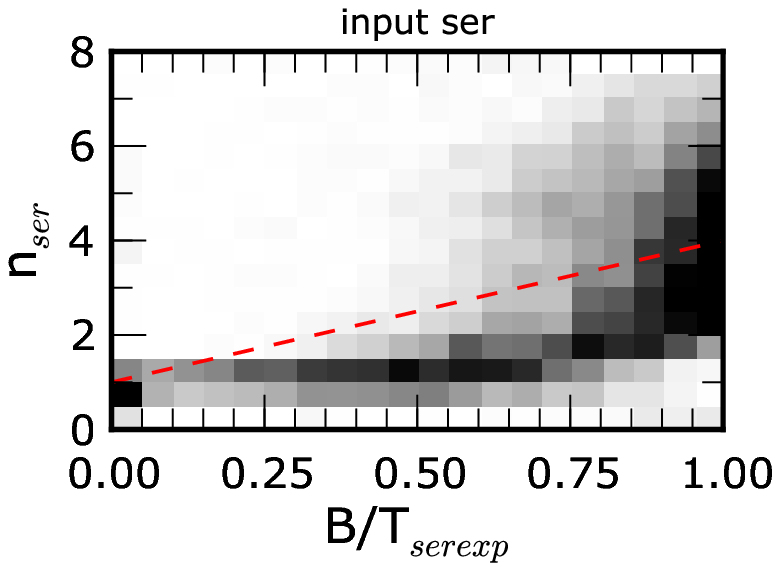}
\includegraphics[scale = .5]{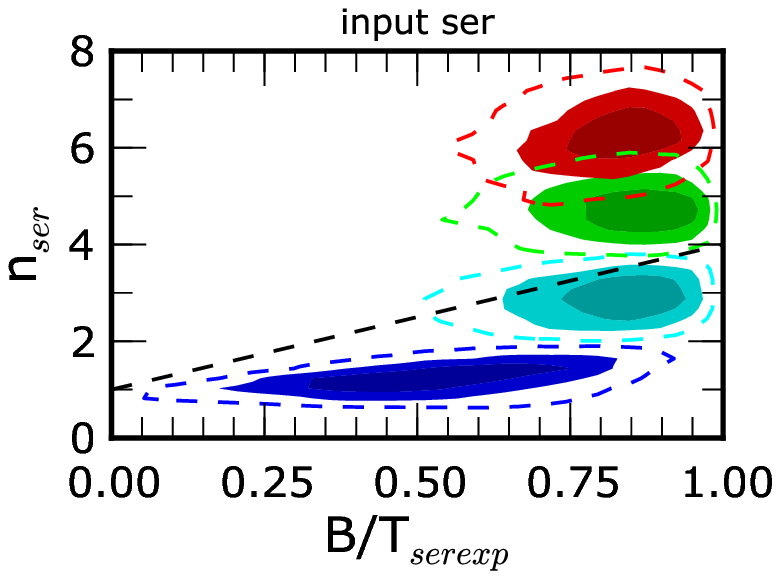}
\includegraphics[scale = 0.5]{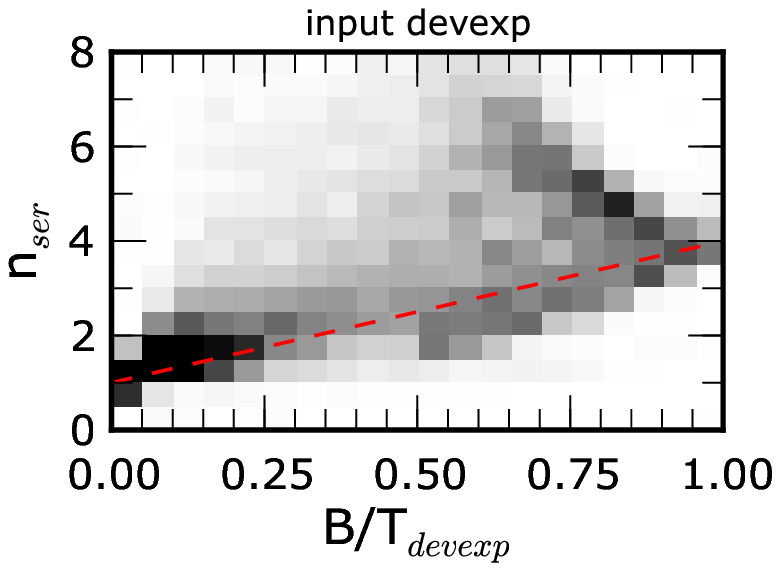}
\includegraphics[scale = 0.5]{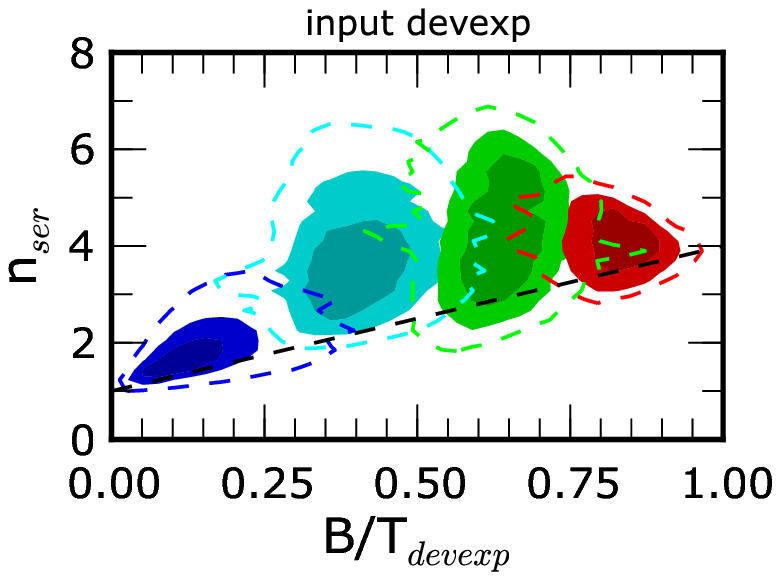}
\includegraphics[scale = .5]{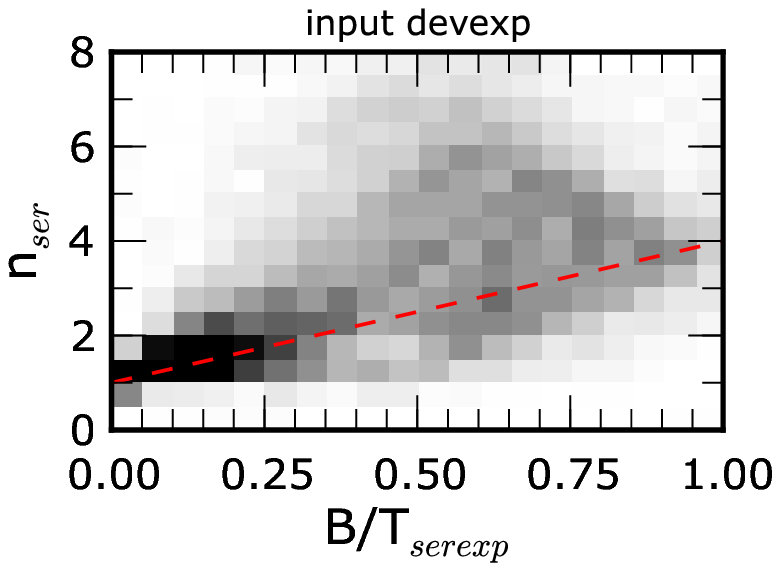}
\includegraphics[scale = .5]{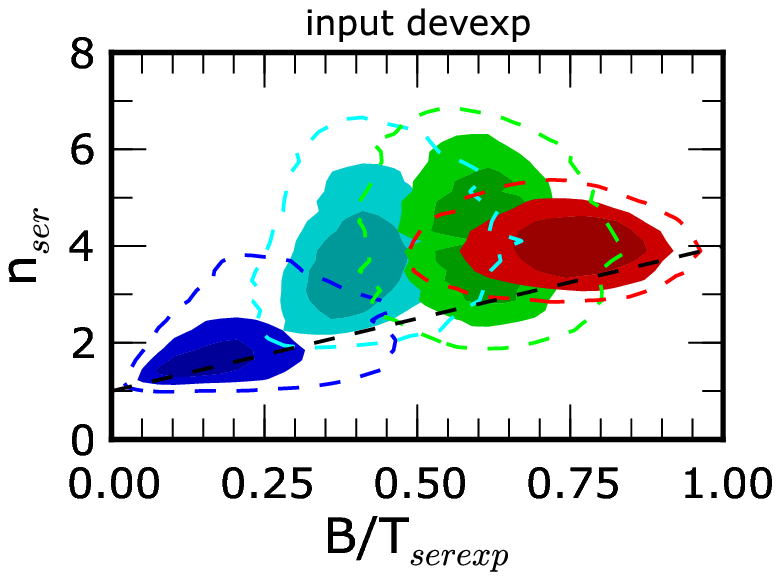}
\includegraphics[scale = 0.5]{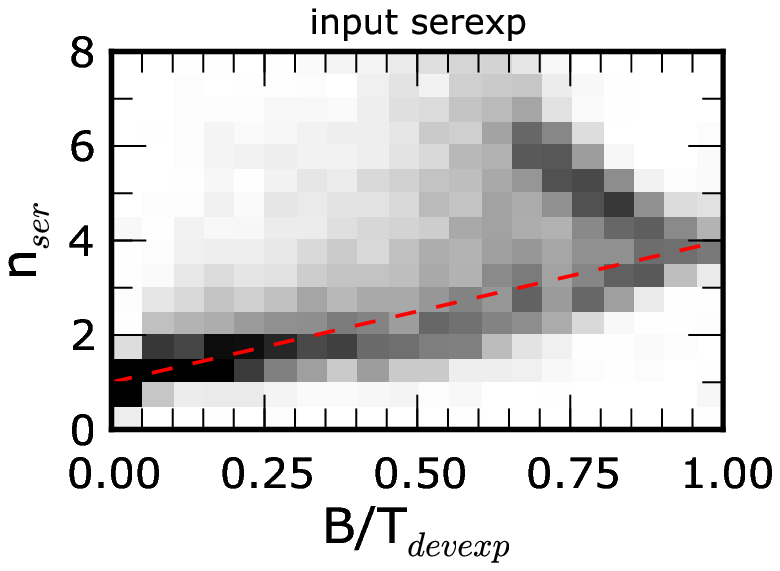}
\includegraphics[scale = 0.5]{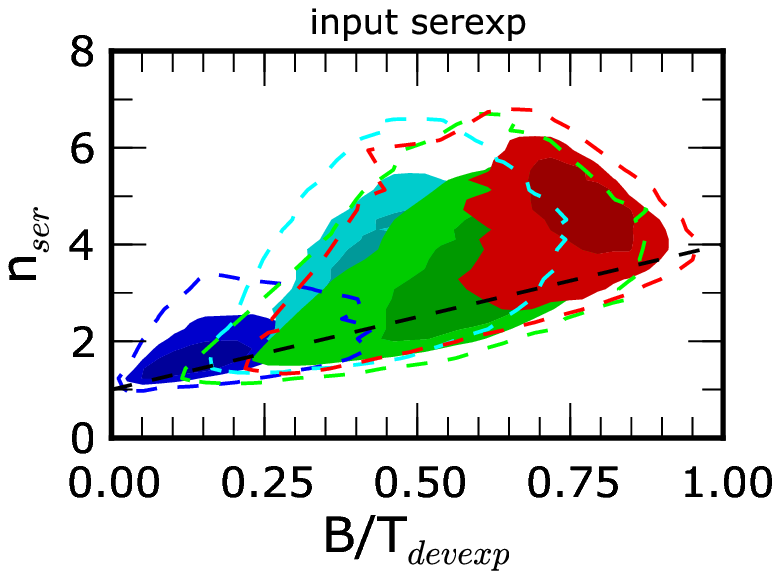}
\includegraphics[scale = .5]{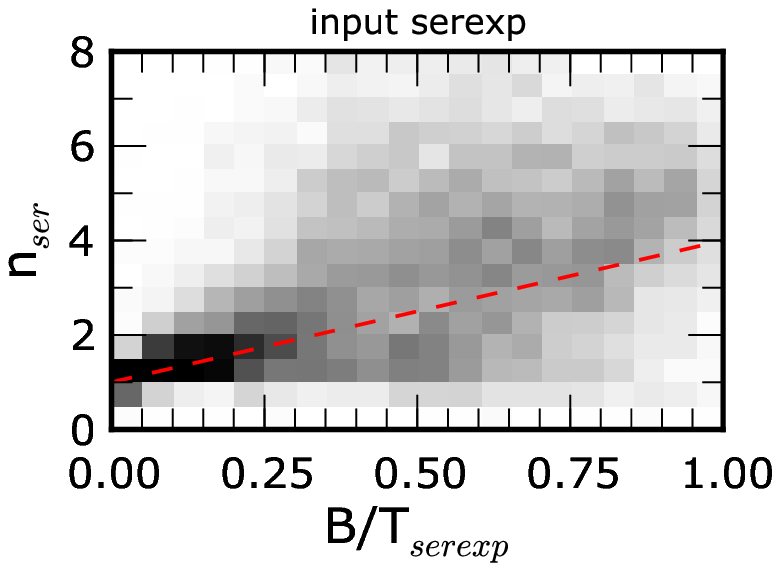}
\includegraphics[scale = .5]{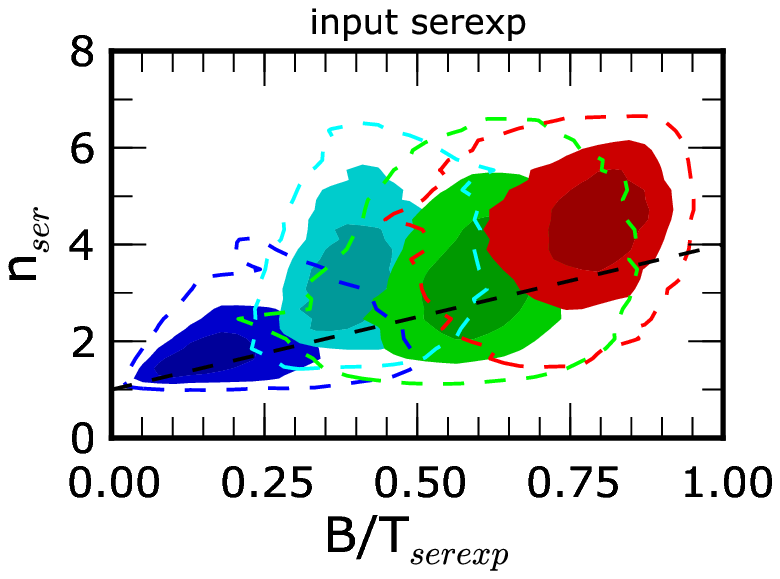}
\caption{Fitted n$_{\rm ser}$ vs fitted B/T for simulated images which were 
generated using a single component Sersic profile (top), 
or two-component {\tt deVExp} (middle) or {\tt SerExp} profiles (bottom). 
The two left columns show $n_{\rm ser}$, returned by fitting a single 
Sersic profile to the image, versus B/T, returned from fitting a 
{\tt deVExp} profile; 
the two right columns show the same $n_{\rm ser}$, but now B/T comes from 
a {\tt SerExp} fit. 
For each pair of columns the left column shows the density across the full 
sample, while the right column shows the density for four bins, colored by 
input $n_{\rm ser}$ (top; the bins have width $\Delta n_{\rm ser}=2$ and run 
from $0-8$, with red showing the largest $n_{\rm ser}$) and 
input B/T (middle and bottom; bins have width $\Delta$B/T=$0.25$ and 
run from 0 to 1 with red showing the largest B/T).  For each color, 
the darkest region contains 25\% of the sample, the lighter region 
surrounding it contains 50\%, and the outermost dashed line contains 
75\% of the sample.  The straight dashed line, same in each panels, 
serves mainly to guide the eye, and to facilitate comparison between 
panels.  Clearly, the distribution of $n_{\rm ser}$ vs B/T depends on 
the input model. }
 \label{sersic}
\end{figure*}

\begin{figure*}
 \centering
 \includegraphics[scale = .5]{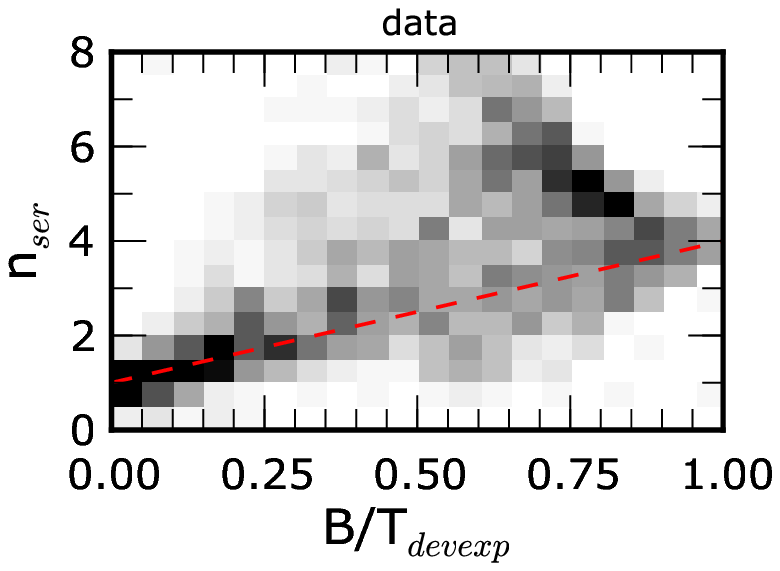}
 \includegraphics[scale = .5]{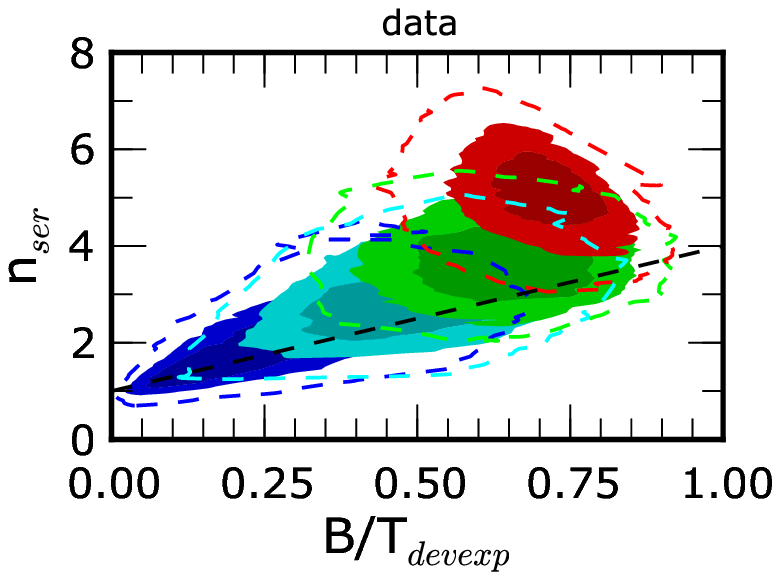}
 \includegraphics[scale = .5]{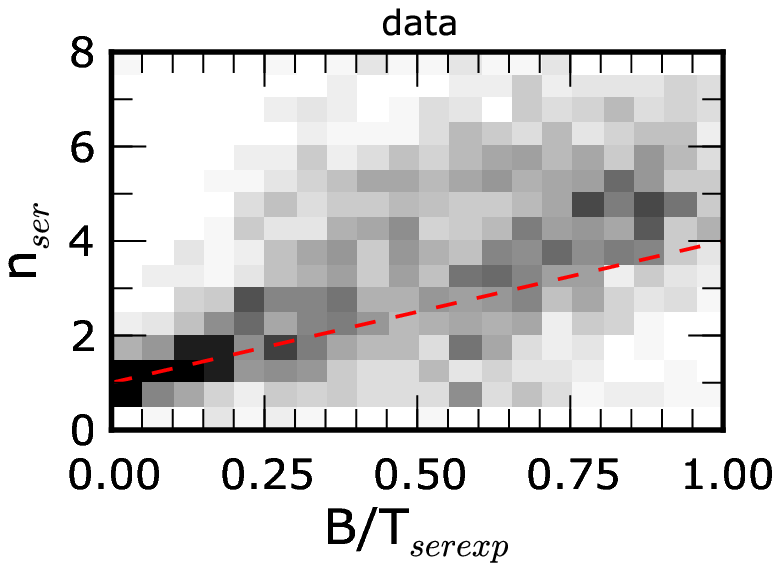}
 \includegraphics[scale = .5]{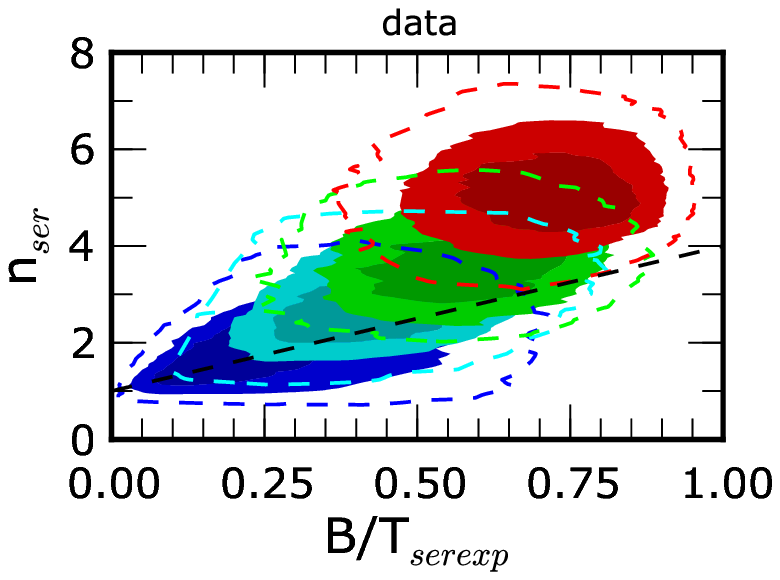}
 \caption{Similar to Fig. \ref{sersic} but for real galaxies.
          Fitted (single component) n$_{ser}$ vs fitted B/T using the two 
          component {\tt deVExp} fit (two left columns) and the two 
          component {\tt SerExp} fit (two right columns).  
          Colors represent the probability that the galaxy is an early-type 
          (four adjacent bins in $p$(E+S0) bins, each of width 0.25, with 
           red showing the highest probability bin).  
          Comparison with the previous figure 
          shows greatest agreement with the row of bottom panels in which 
          the input model was a two component {\tt SerExp}.}
 \label{sdss}
\end{figure*}

As we noted in the Introduction, it is not at priori obvious whether one or multiple components are physically reasonable or necessary.  One way to address this question is to fit the image with the sum of two Sersic profiles, each with its own value of $n$, and then see if allowing for the second component does indeed provide a statistically significant improvement in the accuracy of the fit (once one accounts for the increase in the number of fitted parameters).  In what follows we will perform a slightly simpler version of this:  we force one of the components to have $n=1$, while leaving the other to be determined by the fitting procedure.  We then provide a novel argument which indicates that this {\tt SerExp} model is indeed a better approximation to the surface brightness profiles of real galaxies than is either a single Sersic, or the {\tt deVExp} model.  

We are not the first to have come to this conclusion; e.g. Allen et al. (2006) argued that at least half of the $\sim 10^4$ galaxies at $z\sim 0.1$ in the Millenium Galaxy Catalog are two component {\tt SerExp} systems, and Simard et al. (2011) have recently performed a similar analysis of $\sim 10^6$ SDSS galaxies.  But our argument for why we believe two components are needed is new.  

To gain intuition, Section~\ref{mocks} shows the result of fitting a variety of synthetic images (generated using either a single or two-component models) with single Sersic, {\tt deVExp} and {\tt SerExp} profiles.  Section~\ref{sdss2cs} presents a similar analysis of SDSS DR7 galaxies.  Section~\ref{biases} discusses some biases which arise from fitting the image with a single Sersic.

\subsection{Fits to synthetic images}\label{mocks}
In this section we show the result of using {\tt PyMorph} (Meert et al. 2013)
to fit a variety of synthetic (mock) and real (SDSS) galaxies.   We contrast what happens when {\tt PyMorph} is forced to fit an image using only a single Sersic component, to when it is allowed to use two Sersic components, one with $n=1$ and the other free:  the {\tt SerExp} model.  
For the two-component fits, we first show results when $n$ of the Sersic component is set to $4$, since this corresponds to the traditional `deVaucouleurs bulge + exponential disk' {\tt deVExp} fits, and then when $n$ is allowed to be a free parameter, determined by the fit.   

In all the results which follow, the parent distribution is essentially 
a random subset of the SDSS DR7 main galaxy sample, which is magnitude 
limited to $m_r<17.7$.  We fit each object in this sample using three 
different models:  a single Sersic, a {\tt deVExp} and a {\tt SerExp}.  
We then use the best-fit parameters from these different fits to 
generate three synthetic images for each object.  In this way, we have, 
in effect, three different mock SDSS catalogs (see Meert et al. 2013 
for detailed tests).  
If galaxies were, in reality, e.g. two-component {\tt deVExp} models, then 
only our {\tt deVExp} mock catalog would be realistic -- performing 
profile fits (e.g., using the other two models) to this catalog should 
return results which are similar to those when fitting to the SDSS data.  
Moreover, although all three catalogs will contain correlations between 
$n$, total luminosity, half-light radius, etc., these correlations are 
only guaranteed to be like those in the SDSS data for this (in this case, 
{\tt deVExp}) mock catalog.

\subsubsection{Fitting to a profile which is truly a single Sersic}
We begin with the case in which {\tt PyMorph} is asked to fit what is 
in reality a single Sersic profile of index $n$ (i.e. we use the mock 
galaxies generated using a single Sersic profile) with a single component 
Sersic, and with {\tt deVExp} and {\tt SerExp} profiles.  
The distribution of input $n$ values used to simulate the mock galaxies 
is that which one obtains from fitting single Sersics to the parent 
(magnitude limited) sample. 
Rather than showing the fits themselves, we present our results in 
the parameter space of the best-fit $n$ versus best-fit B/T.  
In all cases, darker shading indicates regions in the parameter space 
that are more heavily populated.  

The top row in Figure~\ref{sersic} shows results for input single Sersic
mock galaxies.
The two panels on the left show B/T values determined from the {\tt deVExp} 
fits, and the two on the right are from {\tt SerExp} fits.  We describe 
the {\tt deVExp} results first.  
The top left panel of  Figure~\ref{sersic} shows the distribution of 
the sample in best fit $n-$B/T space, and the next panel to the right 
shows the result of restricting the analysis to narrow ranges of input $n$.  
The different colors show the distribution in fitted $n$ and B/T for input 
$n$ in the range $0-2$, $2-4$, $4-6$ and $6-8$ (we show the regions which 
enclose 25\%, 50\% and 75\% of the points).  Comparison with the values 
along the y-axis shows that {\tt PyMorph} correctly returns the input $n$ 
values.  

The distribution in the $n-$B/T plane is clearly non-trivial.  
For $n<4$ there is a tight correlation between the value of $n$ 
returned by the single component and B/T from the 
deVaucouleurs-exponential fit:  B/T$\to 1$ as $n\to 4$.  
But as $n$ increases beyond $4$, B/T begins to decrease again.  
I.e., B/T is not a monotonic function of $n$.  
Since the {\tt deVExp} profile only has $n=1$ or $n=4$ components, 
to fit $n>4$ profiles {\tt PyMorph} requires more and more of an 
exponential-like component, i.e. B/T decreases.  (The figure does 
not show this, but the fit returns bulge half-light radii which 
are ever smaller fractions of the half-light radius of that of the 
input Sersic profile.)  
As a result, for $1/2 <$B/T$<1$, the distribution of $n$ at fixed B/T 
appears bimodal.  This shows that, unless one is certain that large 
values of $n$ do not occur in nature, then, especially around 
B/T$\sim 0.7$, B/T values may be misleading, if not meaningless.  

\begin{figure}
 \centering
 \includegraphics[scale = .45]{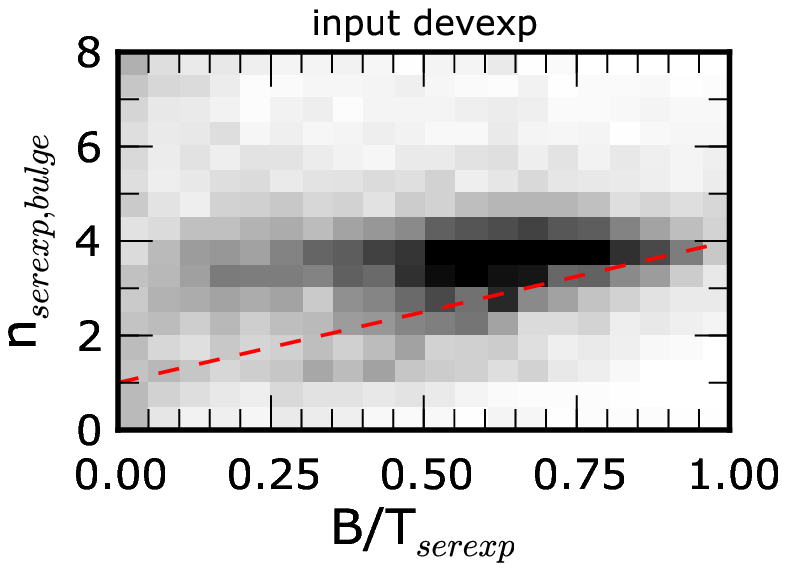}
 \includegraphics[scale = .45]{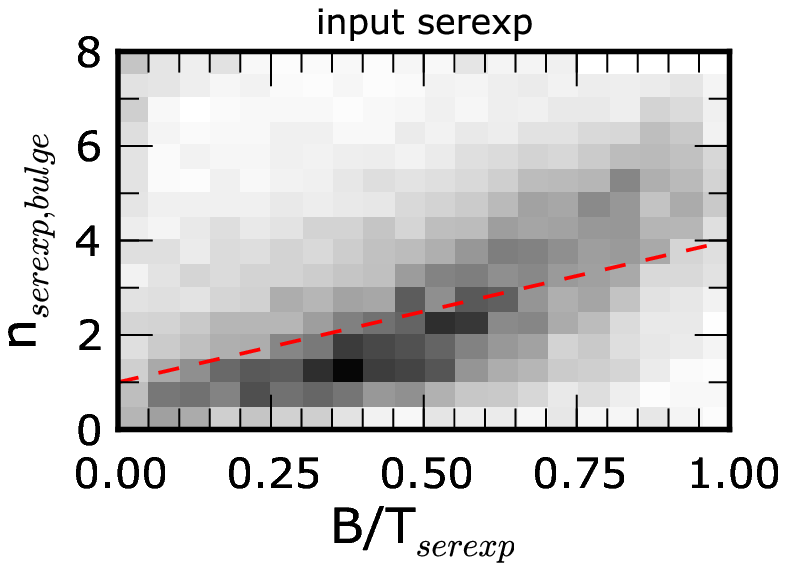}
 \caption{Parameters $n_{\rm ser}$ of the bulge and B/T obtained from 
          fitting the two-component {\tt SerExp} model to mock 
          galaxies generated using input {\tt deVExp} (left) and 
          {\tt SerExp} models (right).  In the panel on the left, 
          the fits correctly return values of $n_{\rm ser}\sim 4$; 
          in the panel on the right, the distribution resembles the 
          input one:  notice that this one indicates that bulges do 
          not necessarily have $n=4$.  The red dashed line, same in 
          both panels, serves only to guide the eye.  }
 \label{simLumB}
\end{figure}

The two panels on the right show the corresponding distribution 
for {\tt SerExp}; they are clearly different from those for 
{\tt deVExp}.  This is primarily because {\tt PyMorph} correctly 
assigns the entire profile to the bulge (Sersic) component, except 
when the input $n\sim 1$, since then which of the two $n=1$ 
components should be called the bulge is ambiguous.  (We have 
checked that, when $n\sim 1$ and B/T $< 1$, then the half-light 
radius of the `bulge' component is indeed the same as that of the 
total:  i.e., the two components differ only by the value of B/T.)  
The fact that B/T is not exactly equal to unity is a measure of the 
error in B/T which comes from the extra degree of freedom associated 
with having a second component with which to fit the profile.  

\subsubsection{Fitting to a profile which is truly a {\tt deVExp}}
The second row shows results when the input profile used to simulate the 
mock galaxies is a two component {\tt deVExp} model (the distribution of 
input B/T values is obtained from fitting {\tt deVExp} models to the 
SDSS parent magnitude limited sample).  This two-component profile is 
then fit with a single Sersic to get $n$; B/T comes from fitting a 
{\tt deVExp} model (two panels on left) or a {\tt SerExp} model (two 
panels on right).  The overall (grey-scale) distributions are rather 
different than in the corresponding panels in the top row.  This is the 
first hint that the distribution of fitted $n$-B/T can be used as a 
diagnostic of the true profile shape.  Different colors show results 
for narrow bins in input B/T; these indicate that {\tt PyMorph} indeed 
returns the correct values when it fits the right model.  The 
additional freedom when fitting a {\tt SerExp} profile to what is 
really a {\tt deVExp} means that, in the panel on the far right, the 
distribution of fitted B/T at fixed input B/T is slightly broader than 
when fitting a {\tt deVExp}.

\begin{figure}
 \centering
 \includegraphics[scale = .5]{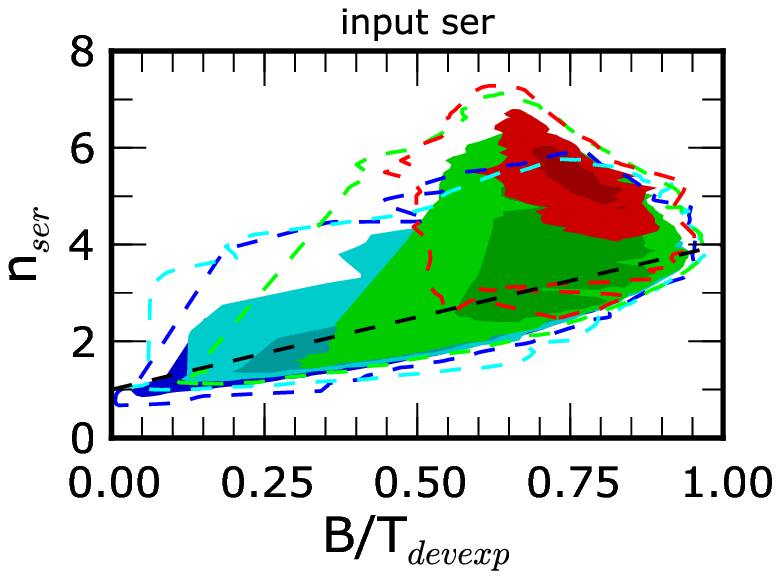}
 \includegraphics[scale = .5]{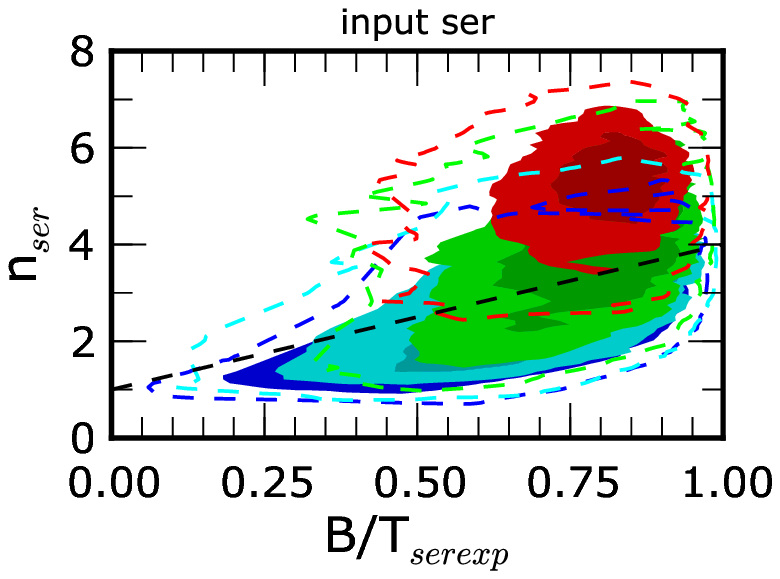}
 \includegraphics[scale = .5]{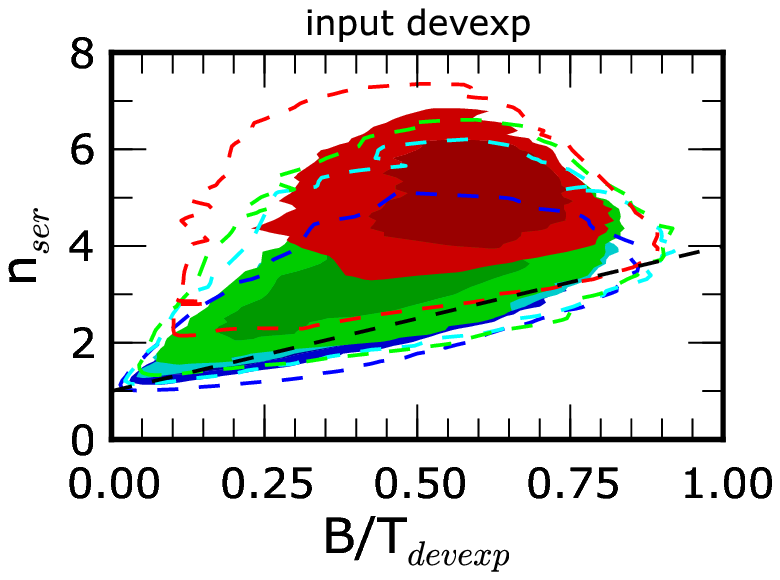}
 \includegraphics[scale = .5]{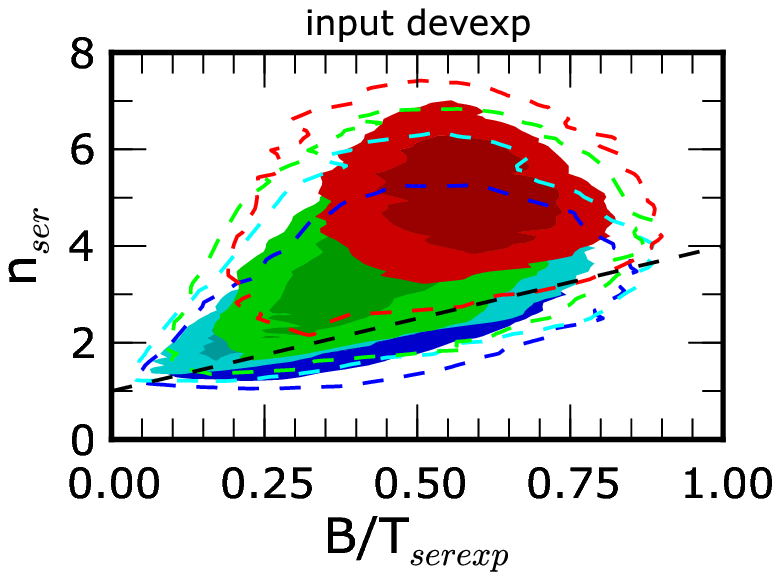}
 \includegraphics[scale = .5]{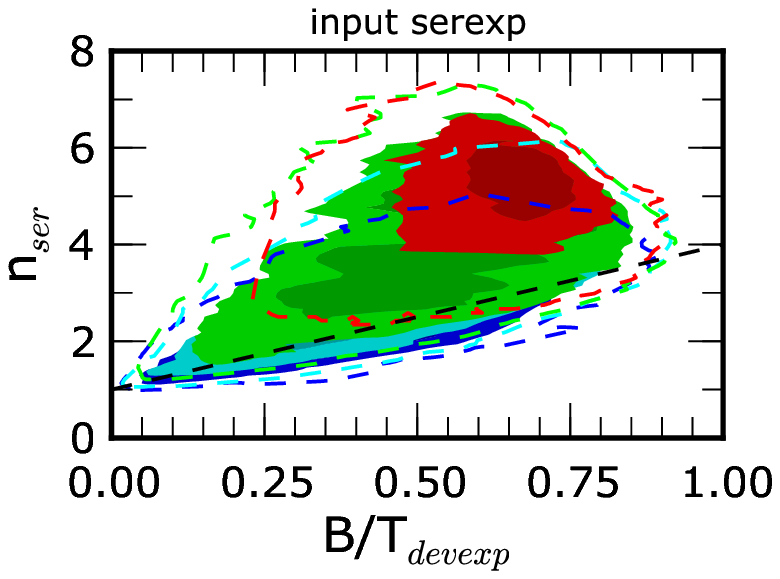}
 \includegraphics[scale = .5]{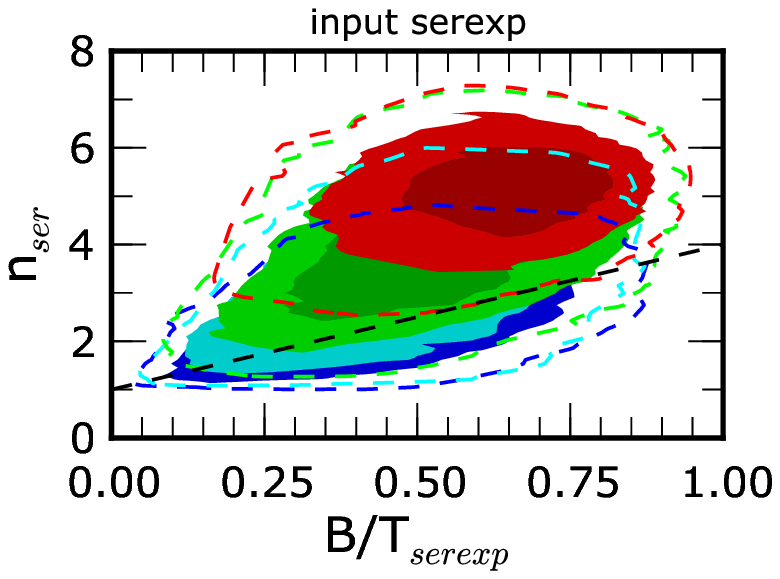}
 \caption{Fitted n$_{ser}$ vs fitted B/T for simulated galaxies that are 
  assumed to be single Sersic profiles (top), 
  two-component {\tt deVExp} profiles (middle) 
  and two-component {\tt SerExp} profiles (bottom). 
  In all cases, the y-axis shows $n_{\rm ser}$ returned by fitting a 
  single Sersic profile to the image.  In the left column, B/T is 
  obtained from fitting a two-component {\tt deVExp} model; 
  the right column, B/T is determined from fitting a {\tt SerExp} 
  model.  
  The density is shown in four bins colored by output absolute magnitude: 
   $-24 < M_r < -23$ (red), $-23 < M_r < -22$ (green), 
   $-22 < M_r < -21$ (cyan), $-21 < M_r < -20$ (blue).
  The top panel is clearly rather different from the other two.}
 \label{simLum}
\end{figure}

\subsubsection{Fitting to a profile which is truly a {\tt SerExp}}
Finally, the bottom row shows results when the input model used to simulate 
the mock galaxies was a {\tt SerExp} (with $n$ and B/T values chosen from 
fitting the SDSS parent 
sample to a {\tt SerExp} model).  The results here differ from those 
in the row above in subtle ways, perhaps most appreciably in the upper 
right corner (large fitted $n$ and B/T) of the bottom right plots.  

In this case, we also show (Figure~\ref{simLumB}) the $n_{bulge}-$B/T plane, 
where both $n_{bulge}$ and B/T come from fitting a {\tt SerExp} model to mock
images generated using input {\tt deVExp} (left) and {\tt SerExp} (right) 
profiles.  The panel on 
the left shows that {\tt PyMorph} correctly returns $n\sim 4$ when it 
should; we have checked that the distribution in the panel on the right 
is similar to the input one, again suggesting that {\tt PyMorph} is working 
well (Meert et al. 2013).

\begin{figure}
 \centering
 \includegraphics[scale = .5]{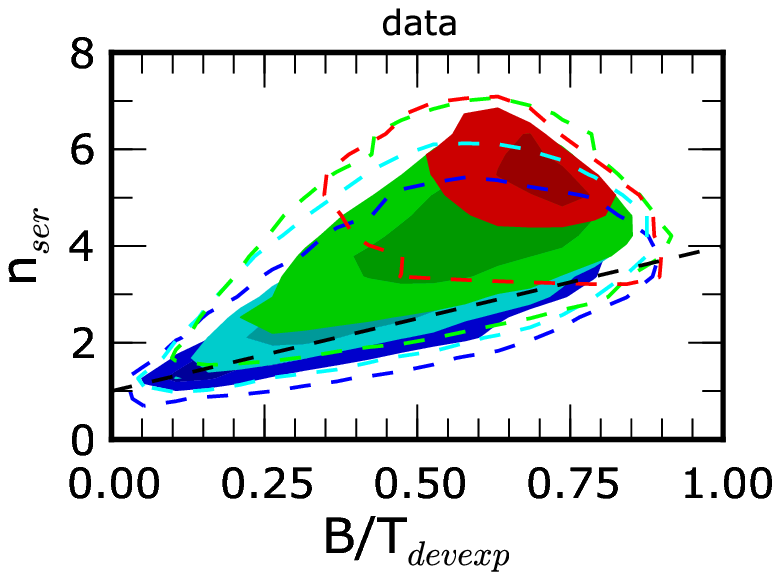}
 \includegraphics[scale = .5]{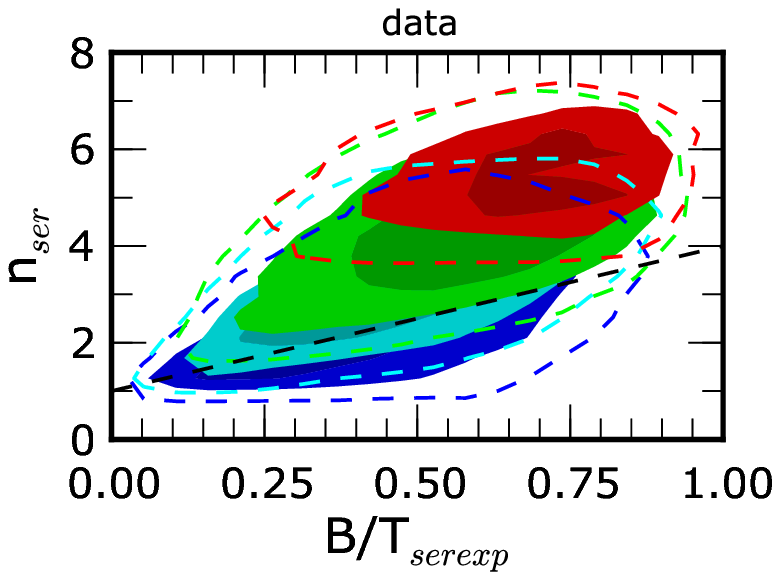}
 \caption{Similar to Figure~\ref{simLum} but for real SDSS galaxies.
          These data are most similar to the bottom panel of 
          Figure~\ref{simLum}, for which the input model was a 
          two-component {\tt SerExp} profile.}
 \label{dataLum}
\end{figure}

\subsection{Fitting to SDSS images}\label{sdss2cs}
Figure~\ref{sdss} shows a similar analyis of SDSS images.  In the two 
panels on the left, $n$ comes from fitting a single component Sersic, 
and B/T from fitting a two-component {\tt deVExp}.  In the panels on 
the right, B/T comes from fitting a two-component {\tt SerExp}.  
Notice that the gray scale plots are very unlike those in the top 
row of Figure~\ref{sersic}, and most like those in the bottom row.  
This suggests that SDSS galaxies are almost certainly not single-component 
systems.  

In addition, of the two-component models, the {\tt SerExp} model 
appears to be more like the data than is the {\tt deVExp}.  
This is because, when B/T comes from fitting a {\tt SerExp}, then 
the SDSS data (third panel from left) populate the large $n-$B/T 
corner which input {\tt SerExp} models also fill, but input {\tt deVExp} 
models do not (c.f. Figure~\ref{sersic}).  There is a more subtle 
difference when B/T comes from the {\tt deVExp} fit (left most 
panels) in Figures~\ref{sersic} and~\ref{sdss}:  the SDSS shows a 
rather well-defined ridge at the boundary of the large $n-$B/T corner, 
which appears to be more separated from the peak at small $n$.  This 
separation is more apparent for the input {\tt SerExp} models than 
for input {\tt deVExp}.  

Since we cannot classify the objects by the true value of $n$ or B/T, the colors (contours in Fig.~\ref{sdss}) show the result of restricting the analysis to objects which are most likely to be early-types (red) to least likely (blue), as determined by Huertas-Company et al. (2011).  This shows that the early-types do indeed have large values of $n$, and spirals the lowest, as expected.  

To provide a slightly more straightforward comparison between simulations and data, we have considered the $n-B/T$ distribution for objects in narrow bins in (output) luminosity.  Figures~\ref{simLum} and~\ref{dataLum} show results in simulations (the same fits used for Figure~\ref{sersic}) and in the SDSS (cf. Figure~\ref{sdss}), respectively.  
These too indicate that the two-component models are more like the data, with the {\tt SerExp} marginally favoured (the two panels in Figure~\ref{dataLum} look more like the bottom than the middle panels of Figure~\ref{simLum}).

\begin{figure}
 \centering
\includegraphics[scale = .45]{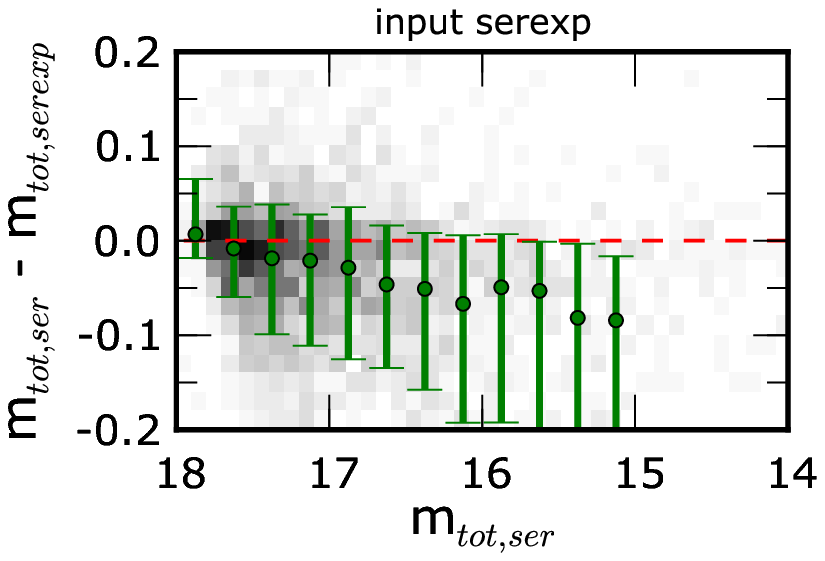}
 \includegraphics[scale = .45]{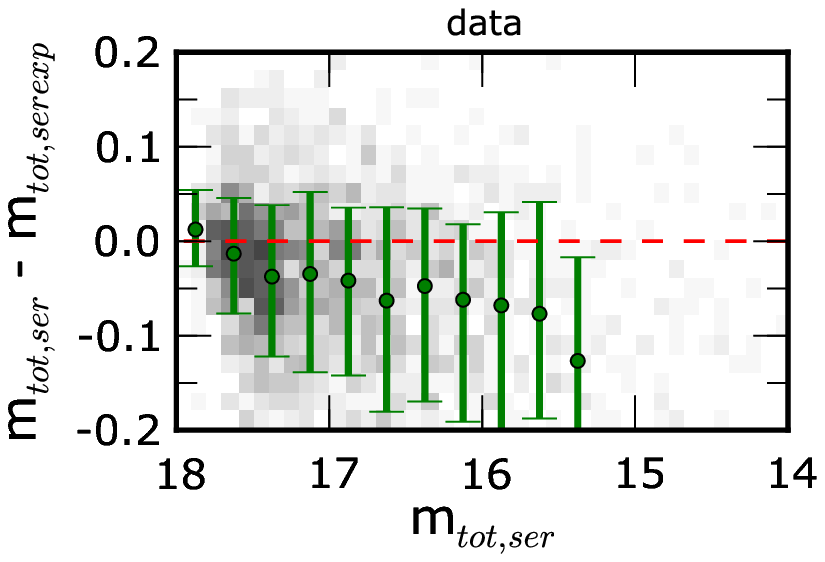}
 \includegraphics[scale = .45]{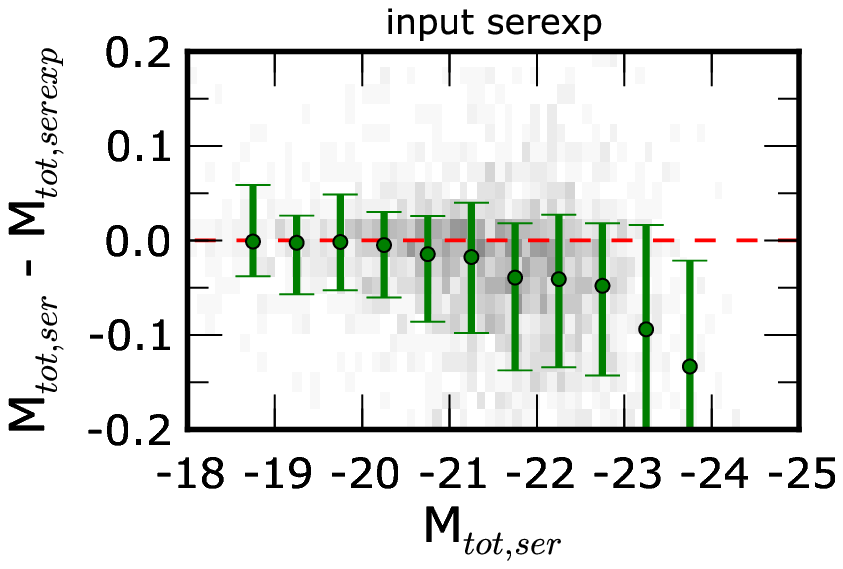}
 \includegraphics[scale = .45]{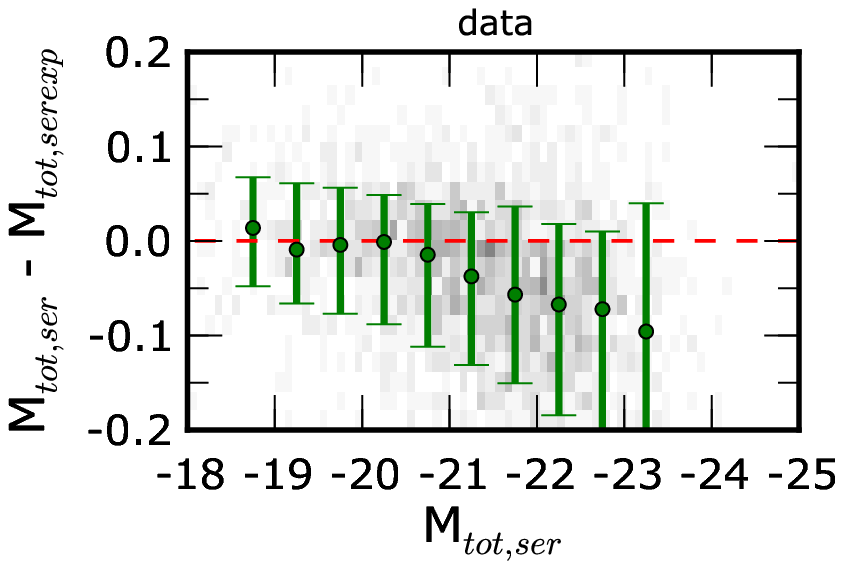}
 \caption{Comparison of total apparent magnitude (top) and luminosity (bottom) returned from single Sersic and {\tt SerExp} fits to simulated {\tt SerExp} (left) and real SDSS (right) galaxies. The error bars show the $1 \sigma$ rms scatter around the median.  The similarity between the panels on the left with their counterparts on the right indicates that the input {\tt SerExp} model is realistic.}
 \label{dataM}
\end{figure}

\begin{figure}
 \centering
 \includegraphics[scale = .475]{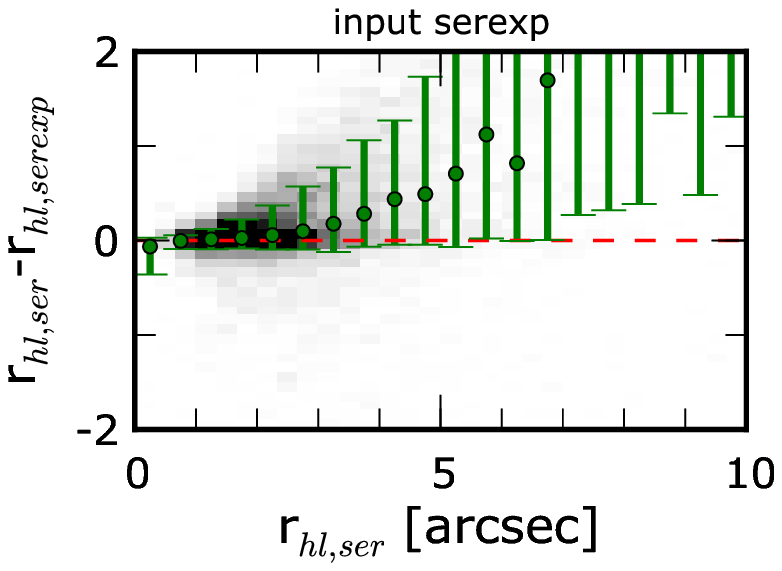}
 \includegraphics[scale = .475]{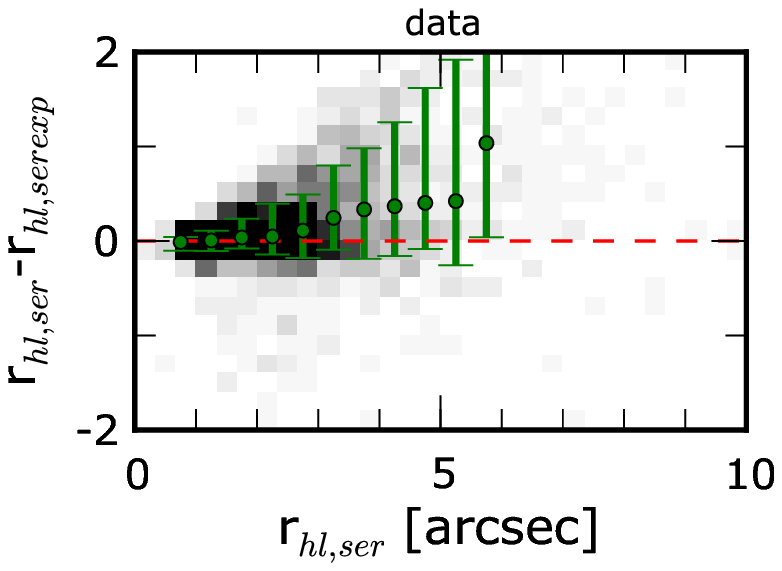}
 \includegraphics[scale = .475]{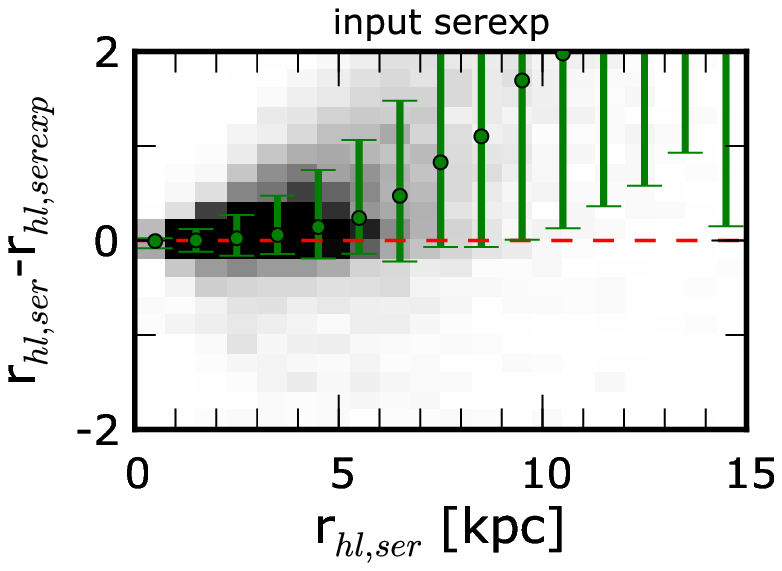}
 \includegraphics[scale = .475]{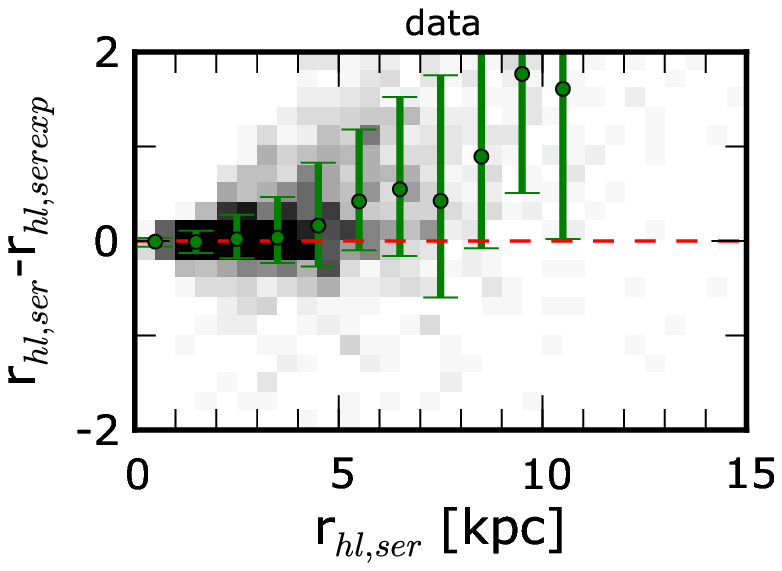}
 \caption{Comparison of angular (top) and physical (bottom) half-light radii returned from single Sersic and {\tt SerExp} fits to simulated {SerExp} (left) and real SDSS (right) galaxies. The error bars show the $1 \sigma$ rms scatter around the median. As for the previous figure, the similarity between the panels on the left with their counterparts on the right indicates that the input {\tt SerExp} model is realistic.}
 \label{dataR}
\end{figure}

\subsection{Biases from fitting single Sersic profiles}\label{biases}

The analysis above shows that a single component Sersic profile is not as good a description of SDSS galaxies as one with two-components.  Since such single component fits are much simpler to perform, and are commonly used, it is interesting to ask if they lead to significant biases in commonly used parameters.  E.g., one might expect the total light to be a reasonably robust quantity, so different models for the shape of the profile may still return consistent values of $L_{tot}$.  

The top left panel of Figure~\ref{dataM} shows that the apparent magnitudes returned by single Sersic fits to the objects in our {\tt SerExp} mock catalog are quite accurate, with a tendency for the Sersic fits to return an overestimate by about ten percent at the bright end.  
The top right panel shows that a similar comparison for the objects in the SDSS produces similar results. The bottom panels show the impact of these small biases on the inferred luminosities.  

\begin{figure}
 \centering
 \includegraphics[scale = .47]{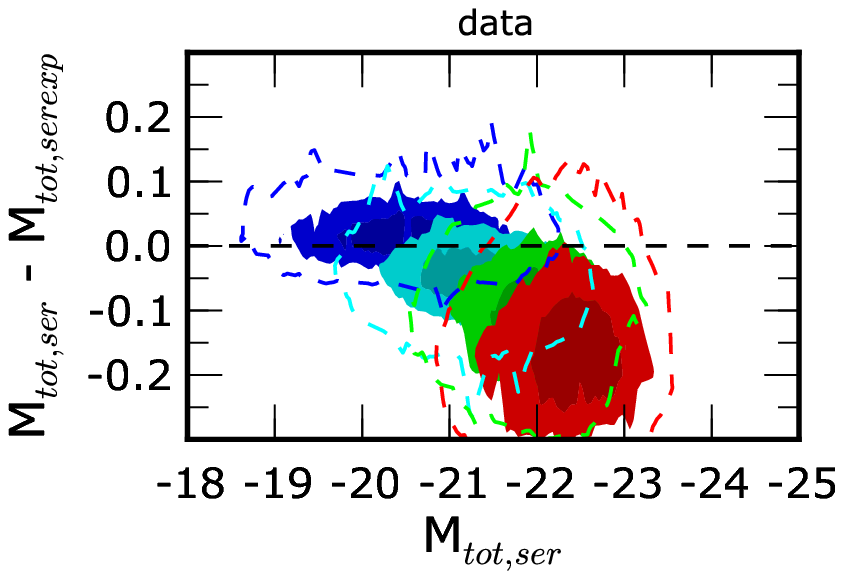}
 \includegraphics[scale = .47]{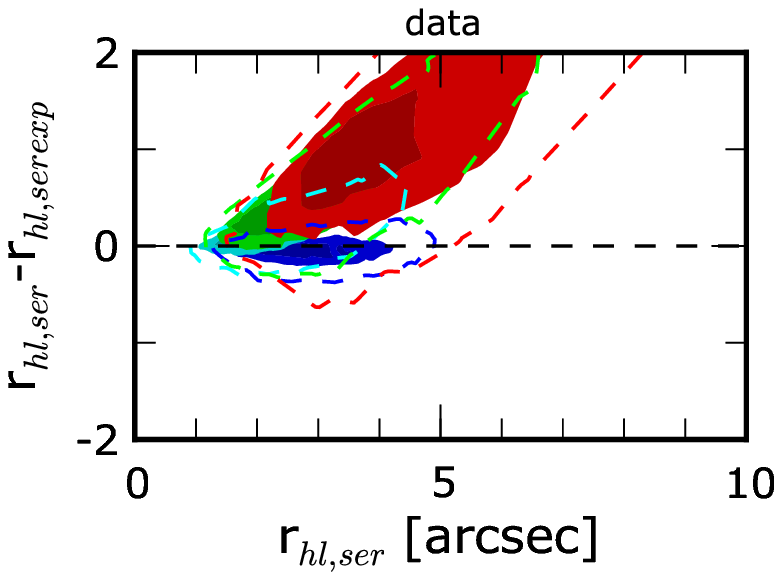}
 \includegraphics[scale = .47]{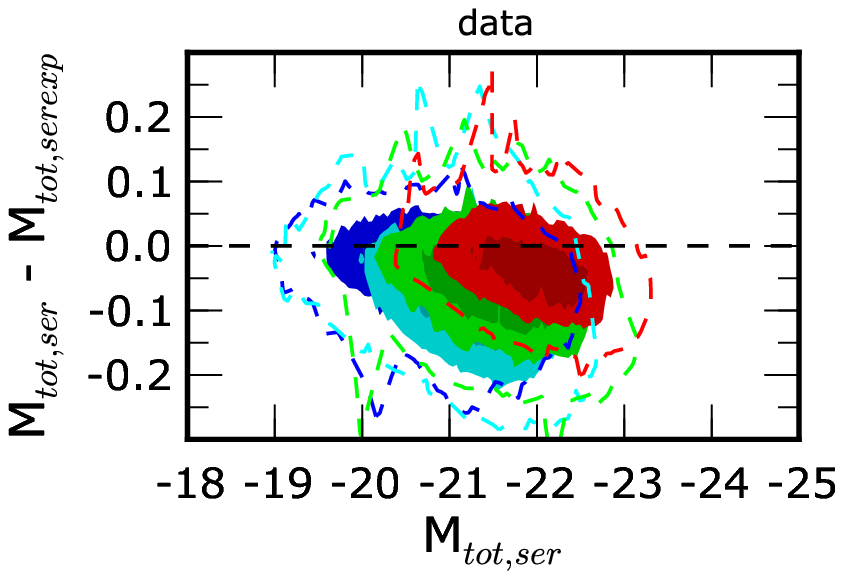}
 \includegraphics[scale = .47]{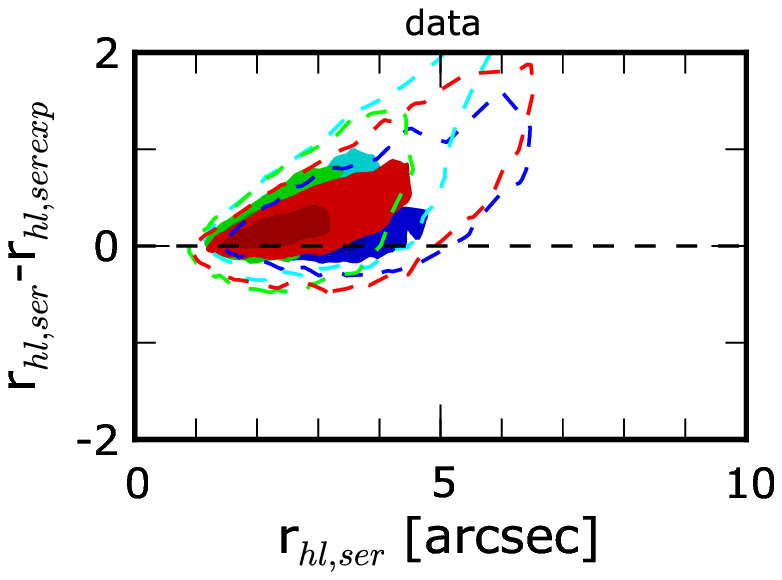}
 \includegraphics[scale = .47]{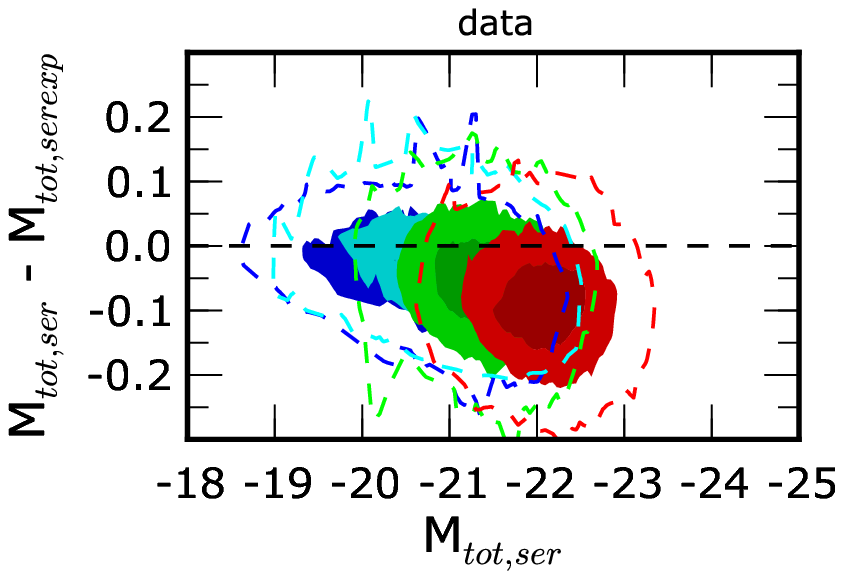}
 \includegraphics[scale = .47]{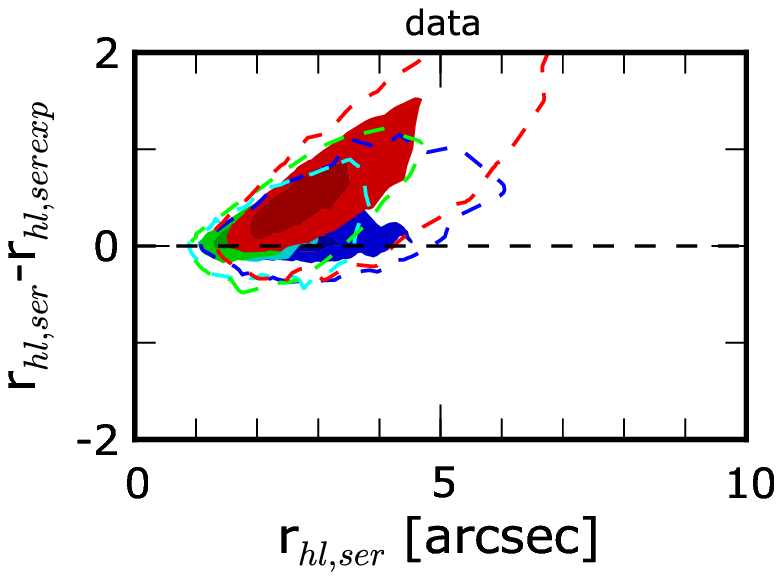}
 \caption{Comparison of luminosities (left) and angular half light radii (right) returned from single Sersic and {\tt SerExp} fits to real SDSS galaxies, color coded by best-fit $n$ (top), best-fit B/T (middle), and $p({\rm E+S0})$ (bottom).  The single Sersic-based luminosities are sytematically brighter at the bright end; this bias is most pronounced when the best-fit $n$ is large. The single Sersic sizes are larger if $n>4$, B/T$>0.5$ or $p({\rm E+S0})>0.75$.}
 \label{dataMnpbt}
\end{figure}

Figure~\ref{dataR} shows a similar analysis of the half-light radii:  the single Sersic fit tends to overestimate the sizes by about ten/fifteen percent, particularly for the largest objects.  The largest and/or most luminous galaxies tend to have large $n$ and/or intermediate to large B/T.  Therefore, this bias is worst for objects that are likely to be ellipticals.  
We show this explicitly in Figure~\ref{dataMnpbt}.

\begin{figure*}
 \centering
 \includegraphics[scale = .8]{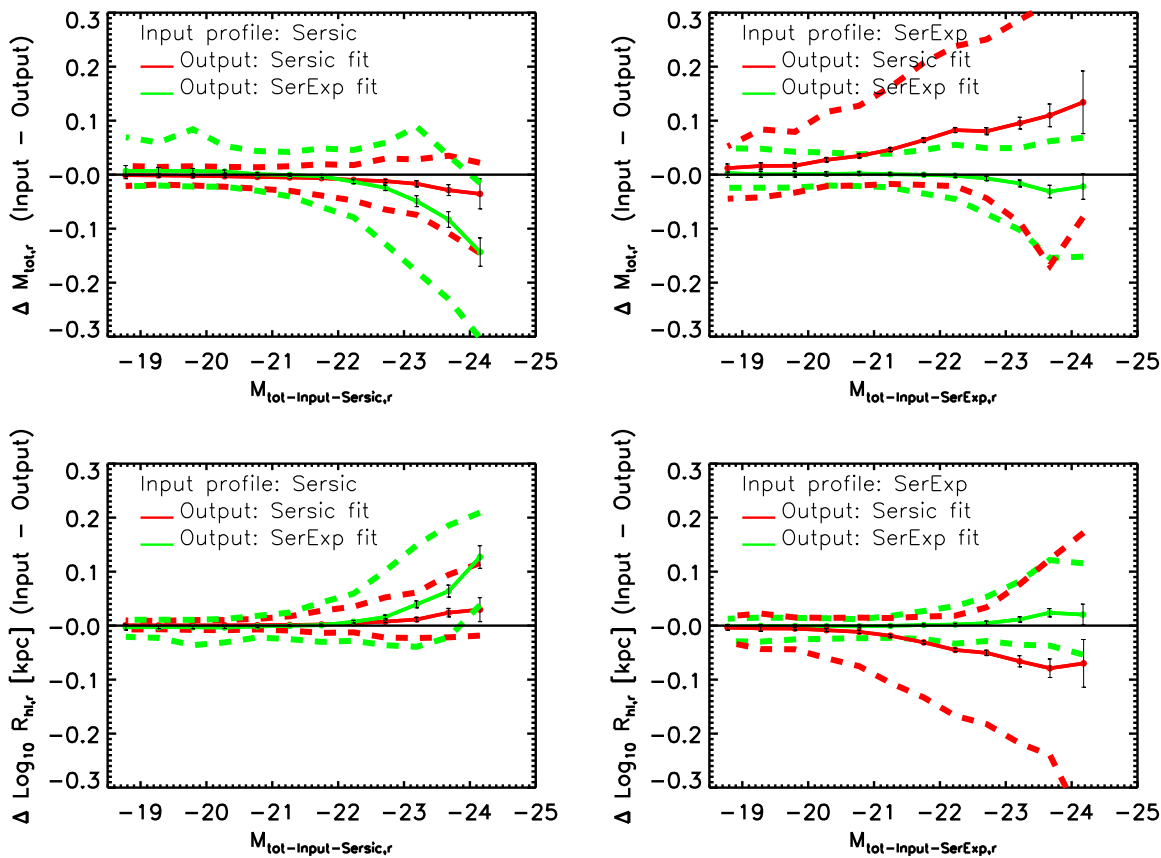}
 \caption{Biases in the estimated luminosities and sizes which come 
          from fitting single Sersic and two-component SerExp profiles 
          to images which are really pure Sersics (left) and two-component 
          {\tt SerExp}s (right).  The latter shows that fitting a single 
          Sersic to what is really a {\tt SerExp} results in an overestimate 
          of the total luminosity and size. }
 \label{SimulationsLR}
\end{figure*}

Figure~\ref{SimulationsLR} illustrates the magnitude of such biases which come from forcing {\tt PyMorph} to fit single Sersic and two-component SerExp models to what is truly a single Sersic (left) or SerExp (right) profile.  These show that when {\tt PyMorph} fits the right model, it tends to return unbiased estimates.  However, biases do arise when fitting an incorrect model, especially at high luminosities.  

This raises the question of whether we should pay more attention to the panels on the left or the right?  Since the SerExp model is a better description of the population as a whole than is a single Sersic (Section~\ref{sdss2cs}), we are most interested in the panels on the right of Figure~\ref{SimulationsLR}.  These show that fitting a single Sersic to what is really a {\tt SerExp} results in overestimates of the total luminosity and size, especially at the bright end.  
These biases have small systematic effects on the size-luminosity correlation of objects that are likely to be early-types, and are presented in the next Section.  In contrast, late-type objects are less likely to be biased.  



\section{The $R-L$ relation at $z\sim 0.1$}\label{lr}

We now study how the $R-L$ relation depends on the functional form for the surface brightness profile that was assumed when estimating $R$ and $L$.  We would especially like to compare the effects of fitting one versus two-component models to the images, and we do this separately for the early and late type samples defined by the hard cuts on $n$ and BAC-$p$(type) described in Section 2.2.


\subsection{Single Sersic fits and comparison with previous work}\label{lrSersic}
To connect with previous work, we begin with a comparison of the {\tt PyMorph} derived $R-L$ relation based on single Sersic-derived parameters, with analogous fits from the literature.  
Using the objects in an earlier SDSS data release, Shen et al. (2003) reported fits to the $R-L$ relation for objects which had $n>2.5$ and $n<2.5$, where $R$, $L$ and $n$ were determined from a single Sersic fit to the light profile. Note that the Sersic parameters used by Shen et al. were estimated from a 1-dimensional radial surface brightness profile ({\tt profMean}), measured in $\sim 5-10$ azimuthally averaged annuli (Blanton et al. 2003). Thus, it is expected to be significantly less accurate than a 2-dimensional fit to the whole galaxy image.  

\begin{figure*}
 \centering
 \vspace{-2.cm}
 \includegraphics[width = 0.45\hsize]{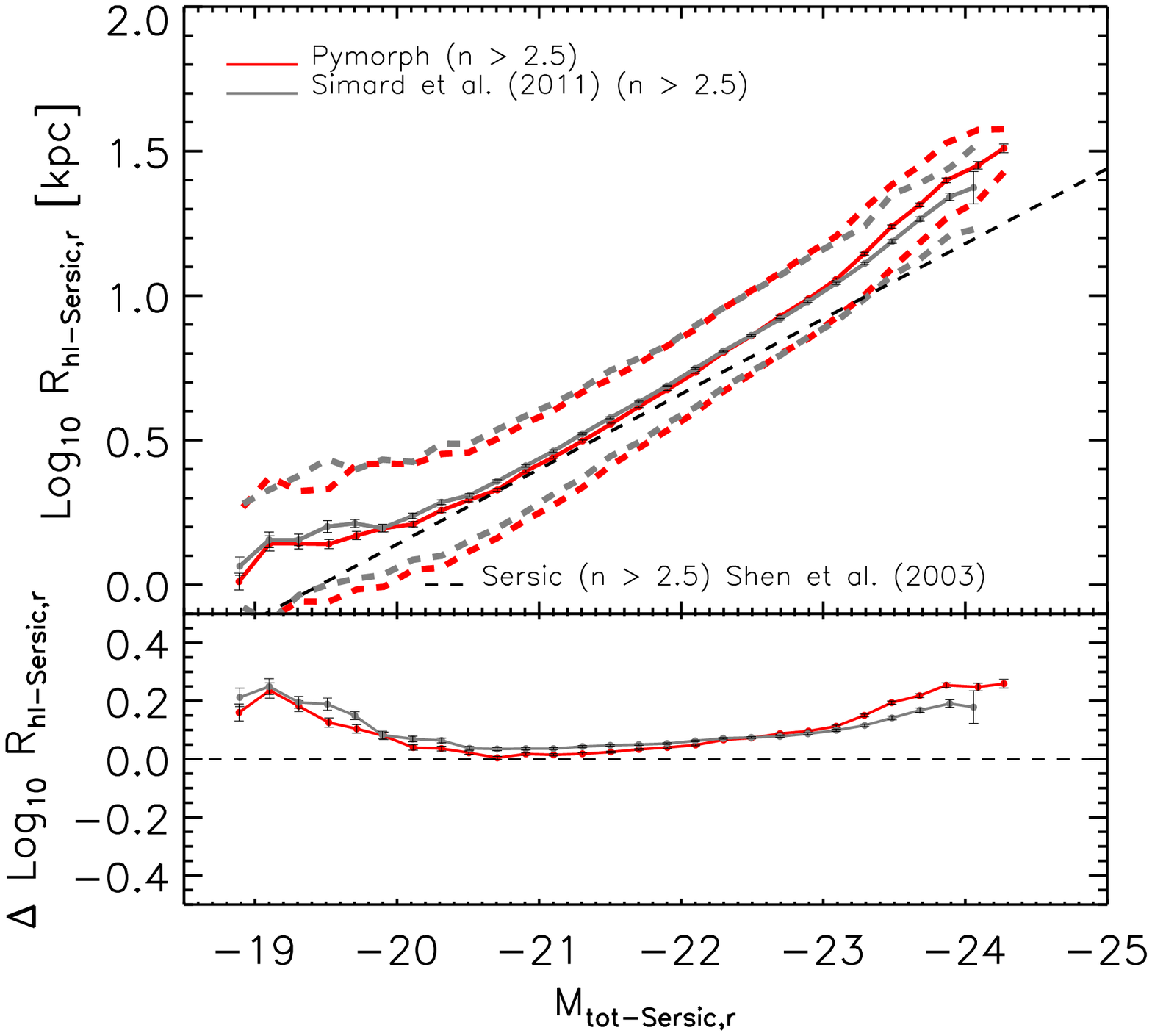}
 \includegraphics[width = 0.45\hsize]{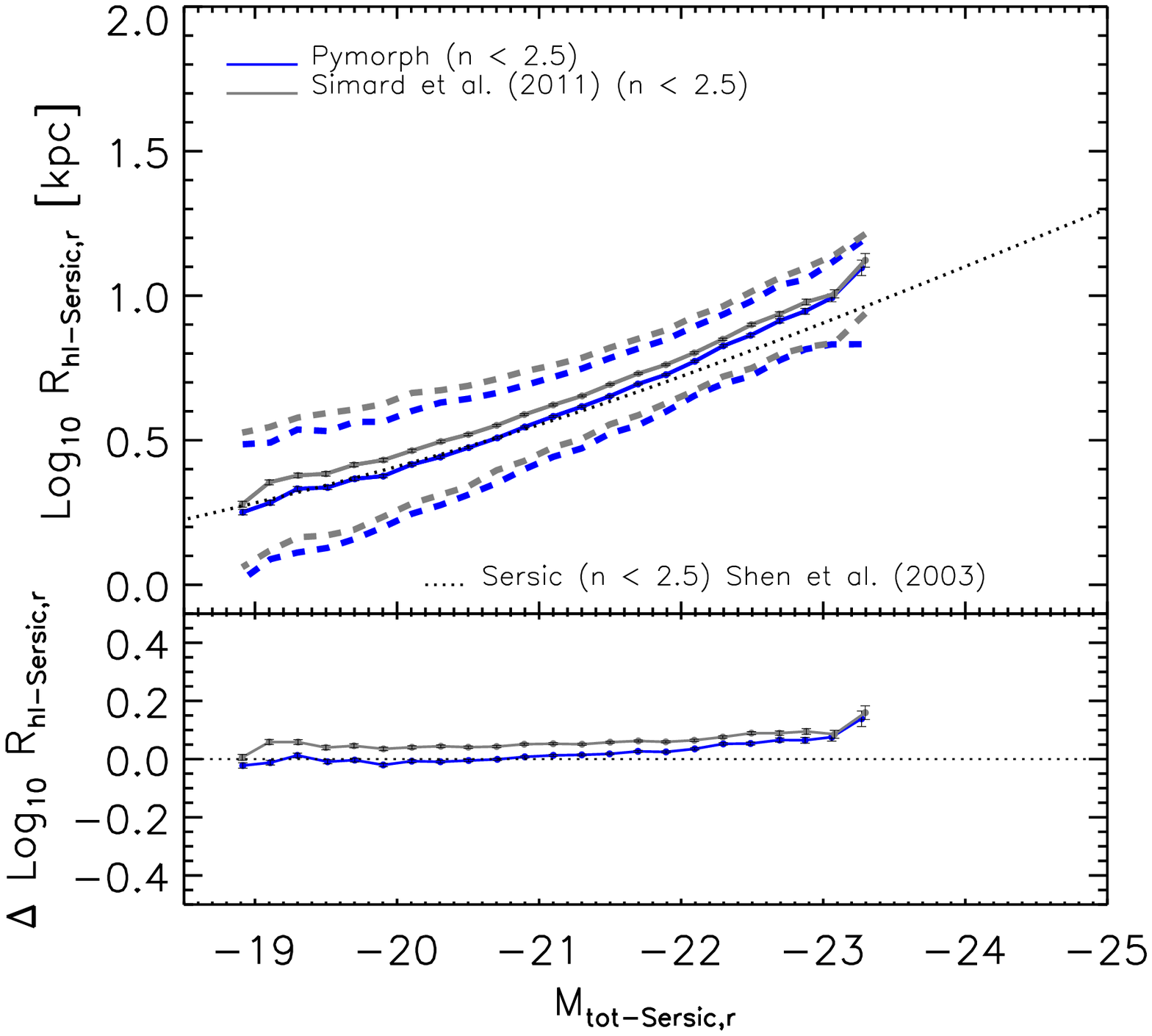}
 \caption{The r-band single-Sersic based half-light radius (R$_{hl}$) 
          versus total absolute magnitude (M$_{{\rm tot}}$) relation for 
          objects with $n>2.5$ (left) and $n<2.5$ (right).  In the panel 
          on the left, our {\tt PyMorph} determination is in good 
          agreement with that based on single-Sersic parameters from 
          Simard et al. (2011), but lies about 0.1~dex above, and is 
          more curved than the fit reported by Shen et al. (2003).
          Symbols with  error bars (joined by a solid curve for clarity) 
          show the median half-light radius in bins of absolute magnitude. 
          Dashed lines show the $16$th and $84$th percentile.  
          In the panel on the right (objects with $n<2.5$), except for 
          the brightest objects, the {\tt Pymorph} relation lies 
          systematically 0.05~dex below that of Simard et al.}
 \label{LRps}
\end{figure*}

The Shen et al. relations for $n>2.5$ and $n<2.5$ are shown as the dashed and dotted lines in the left and right hand panels of Figure~\ref{LRps}, respectively.  The red and blue symbols with error bars show our determination of the single-Sersic based relation, where now $R$, $L$ and $n$ are from our {\tt PyMorph} reductions, and the grey symbols and error bars show the $R-L$ relation which follows from single-Sersic fits performed by Simard et al. (2011, hereafter S11).

For objects with $n<2.5$, the S11-derived relation runs parallel to that from Shen et al., but is offset to larger sizes by 0.05~dex, whereas the {\tt PyMorph}-derived relation transitions from Shen et al. at low luminosities to S11 at high luminosities.  For objects with $n>2.5$ the {\tt PyMorph}-derived relation lies about 0.1~dex above, and is more curved than the fit reported by Shen et al.  The {\tt PyMorph} and S11 based relations depart significantly from Shen et al. at the low and high luminousity ends, where they curve upwards to larger sizes.  For this reason, we are inclined to conclude that, at least at the bright end, Shen et al. is slightly biased. At the low end the curvature could be due to contamination by later-type galaxies.

However, at the highest luminosities, and for objects with $n>2.5$, 
the {\tt PyMorph} and S11 relations are also slightly but significantly 
different from one another.  Appendix~\ref{noSimard} shows that, in 
fact, at high luminosities, the derived magnitudes and sizes can be 
quite different:  the correlated nature of these differences means 
that the $R-L$ relation is only moderately affected.  
Appendix~\ref{noSimard} goes on to show that the S11 reductions appear 
to require rather dramatic evolution in $n$ and $R$:  both are larger 
at $z=0.2$ than at $z=0.05$.  Since we believe this is unphysical, we 
conclude that the {\tt PyMorph} reductions, which show no such 
systematic trend with $z$, are less biased, so we will use them in the 
remainder of this paper.




\subsection{Dependence on model fitting}\label{lrAll}
Figure~\ref{LRps} showed the $R-L$ relation derived from single Sersic fits to the two-dimensional surface brightness profile.  We now compare these to relations based on 
 SDSS fits to a single deVaucouleurs profile;  
 SDSS-based {\tt cmodel} sizes defined by Bernardi et al. (2010), 
 {\tt PyMorph} fits to a two-component deVExp model; 
 {\tt PyMorph} fits to a two-component SerExp model; and
 {\tt PyMorph} fits to a single Sersic profile.  
As a result of the analysis in Section~\ref{2components}, we expect the SerExp reductions to return the least biased estimates of $R$ and $L$, and hence of the $R-L$ relation.  We also show the corresponding $R-M_*$ relations (with $M_*$ estimated as described in Section 2).  

All of these relations show curvature which we quantify by fitting to 
\begin{equation}
 \Bigl\langle \log_{10}\frac{R}{{\rm kpc}}\Big|O\Bigr\rangle
 = p_0 + p_1 O + p_2 O^2;
 \label{RLcurved}
\end{equation}
the coefficients of these fits for $O=M_r$ and $O=\log_{10}(M_*/M_\odot)$ are reported in Tables~\ref{LRfits} and~\ref{MsRfits}.  
Although $p_1$ is the coefficient of the linear part of the relation, the slope on scale $O$ is $p_1 + 2p_2\,O$; this is the value (at some characteristic $O$) which should be compared with the slope of a linear fit to the relation.

\begin{figure*}
 \centering
 \vspace{-2cm}
 \includegraphics[width = .48\hsize]{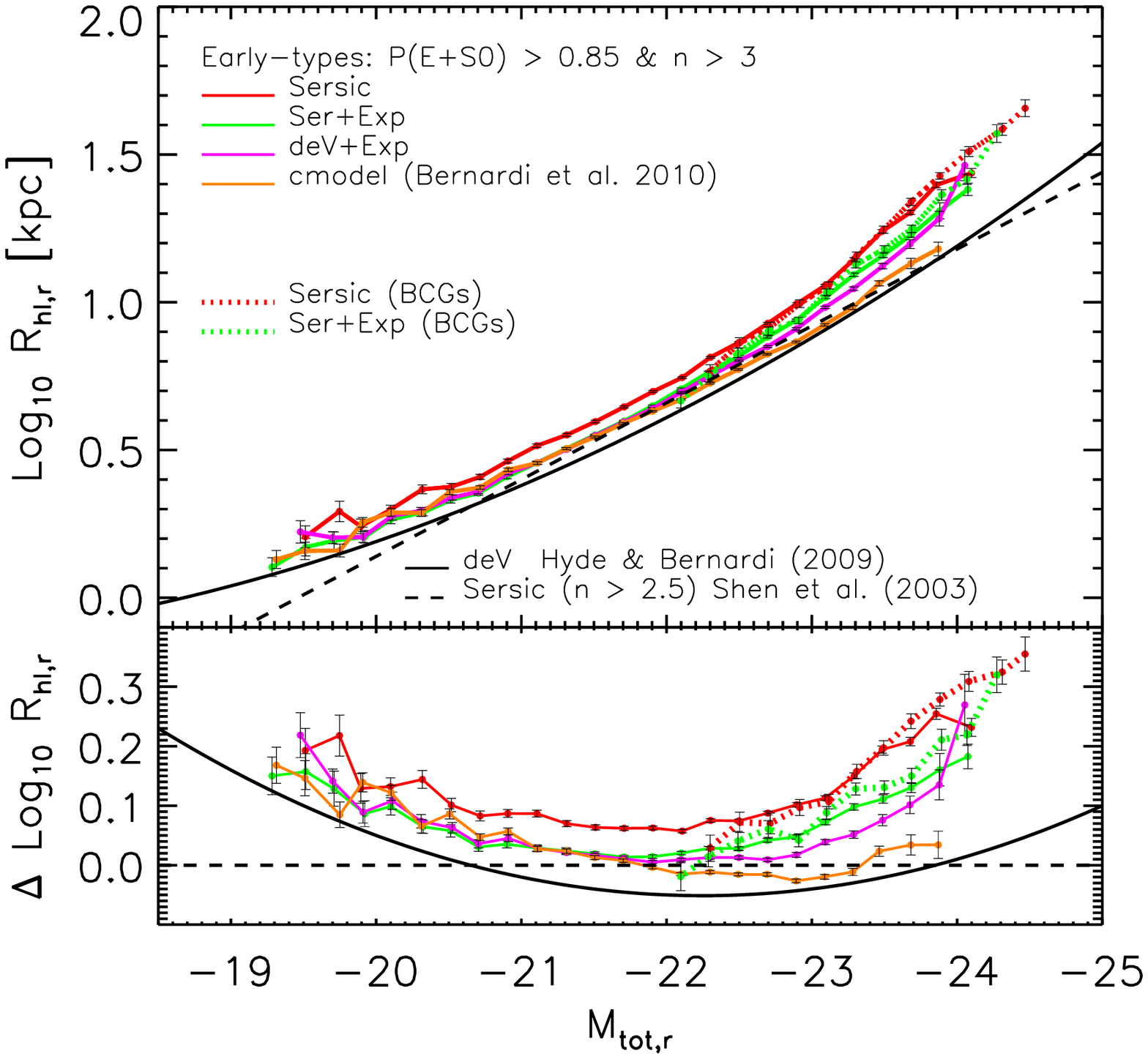}
 \includegraphics[width = .48\hsize]{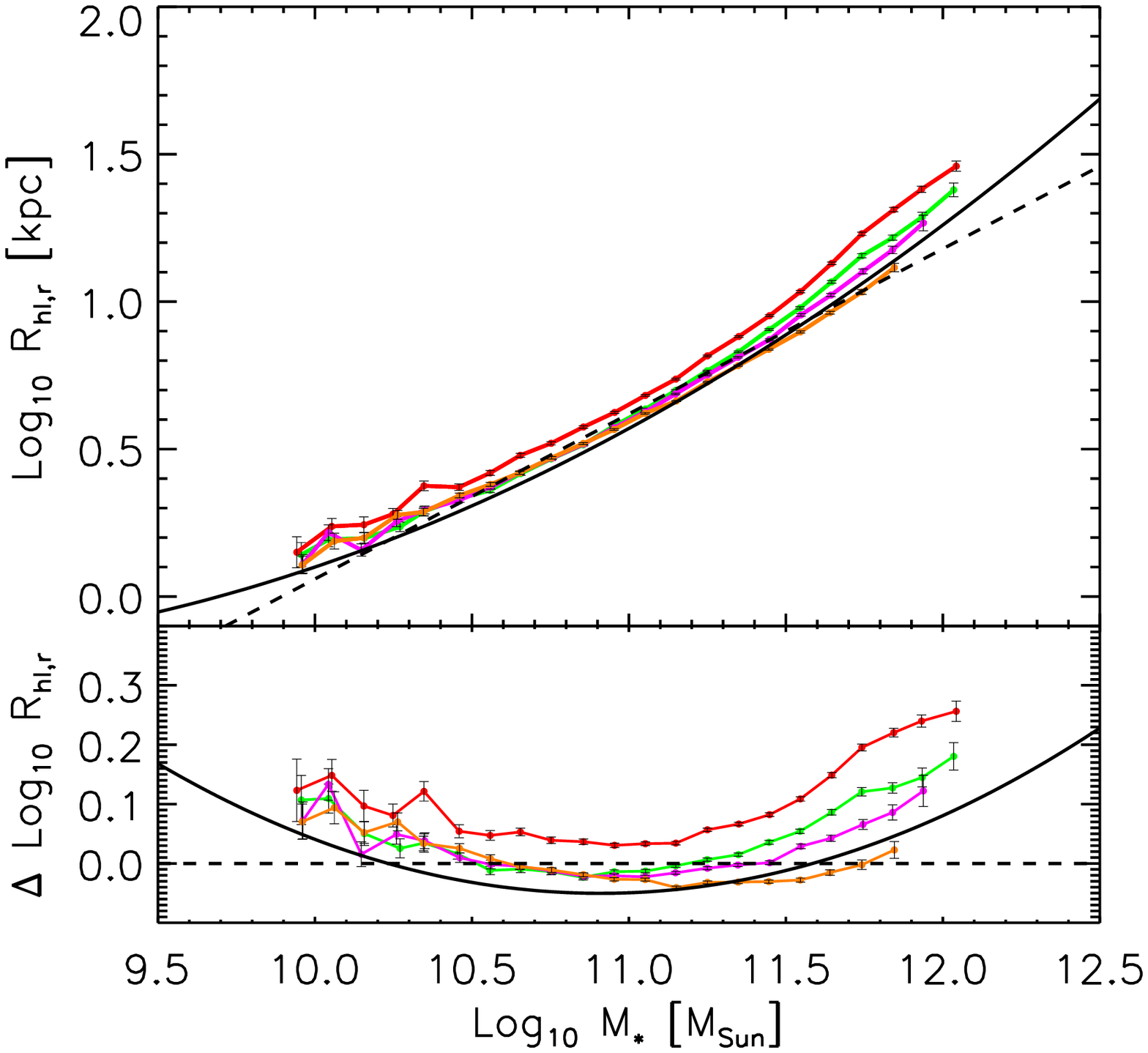}
 \caption{Dependence of derived size-luminosity (left panels)
          and size-stellar mass (right panels) correlations for 
          early-type galaxies on the assumed surface brightness 
          profile. Symbols with error bars (joined by a solid curve for 
          clarity) show the median half-light radius in bins of 
          absolute magnitude (left) and stellar mass (right).
          The SDSS fits to a single deVaucouleurs profile 
          return a relation with the smallest sizes; our PyMorph fits 
          to a single Sersic profile return the largest sizes.  
          Of the relations which lie in between these two extremes, 
          and which are almost indistinguishable at $M<-21.5$, 
          the SDSS based cmodel sizes (defined by Bernardi et al. 2010) 
          are the smallest; those based on our PyMorph fits to a 
          two-component deVExp model are slightly larger; and those 
          based on PyMorph fits to a SerExp model are largest.  
          The curvature at the bright end appears to be due to 
          an increasing incidence of BCGs, which define steeper 
          relations (dotted lines) than the bulk of 
          the early-type population.  To highlight this curvature, 
          bottom panel shows the ratio of all quantities in the top 
          panel to the dashed line.  }
\label{LR}
\end{figure*}

\subsubsection{Early-types}
The panel on the left of Figure~\ref{LR} shows the $R-L$ relation obtained for the early-type sample (i.e. $n>3$ and $p$(E+S0)$>0.85$) 
based on  
a number of single and two-component fits to the images.  
There are clear systematic differences between these relations, with the single Sersic and deVaucouleurs models returning the relations with the largest and smallest sizes, respectively.  The various two-component based relations are in good agreement except at the highest luminosities ($M_r<-22$), where the sample becomes increasingly contaminated by BCGs which are known to define steeper relations than the bulk of the population (e.g. Bernardi et al. 2007; Bernardi 2009).


The panel on the right shows a similar analysis of the $R-M_*$ relation. 
Note that both $R-L$ and $R-M_*$ are significantly curved, consistent with previous work (e.g., Bingelli et al. 1984, and the recent review by Graham 2013).  Of course, if the stellar population models used to estimate $M_*/L$ are incorrect, or if the IMF is mass-dependent, then this will modify the curvature in $R-M_*$.

\begin{figure}
 \centering
 \includegraphics[scale = .4]{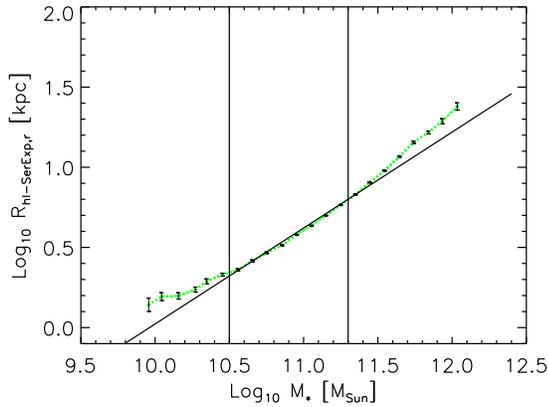}
 \caption{The $R-M_*$ relation for SerExp fits, compared to a 
          power-law showing $R\propto M_*^{0.6}$. 
          Vertical lines show the two mass scales identified by 
          Bernardi et al. (2011) on the basis of a variety of other 
          early-type galaxy scaling relations; the $R-M_*$ relation 
          clearly curves away from the power law at these scales.}
\label{MsRserexp}
\end{figure}

While not the main focus of the present work, we note that 
Figure~\ref{LR} is consistent with recent work showing that a variety 
of other early-type galaxy scaling relations change slope at 
$\sim 3\times 10^{10}M_\odot$ and again at $\sim 2\times 10^{11} M_\odot$ 
(Bernardi et al. 2011).  Crude estimates of these two mass scales are 
given by the values of $M_*$ at which a linear fit intersects the 
parabola given by the quadratic fit.  For the {\tt SerExp} fits to 
the Early-types shown above the best linear fit has 
 $\langle R|M\rangle = - 5.9155 + 0.598\log_{10}(M_*/M_\odot)$, 
so it crosses the associated parabola (parameters from 
Table~\ref{MsRfits}) at $\log_{10}(M_*/M_\odot) = 10.4$ and $11.6$.  
We could assign uncertainties to these scales by propagating the 
uncertainties on the fitted coefficients $p_i$, but these are smaller 
than the systematics associated with this particular choice of defining 
the mass scales.  E.g., another estimate comes from adjusting the 
amplitude of the straight line fit so that it is tangent to the 
parabola, and then identifying the scales on which the parabola lies 
sufficiently far from it.  
Figure~\ref{MsRserexp} shows such an analysis for the SerExp $R-M_*$ 
relation; the vertical lines show $3\times 10^{10}M_\odot$ and 
$2\times 10^{11} M_\odot$.  This demonstrates our main point -- that 
the existence of these two scales is not an artifact of the model 
used to estimate $R$ and $L$.

\begin{figure}
 \centering
 \includegraphics[scale = .4]{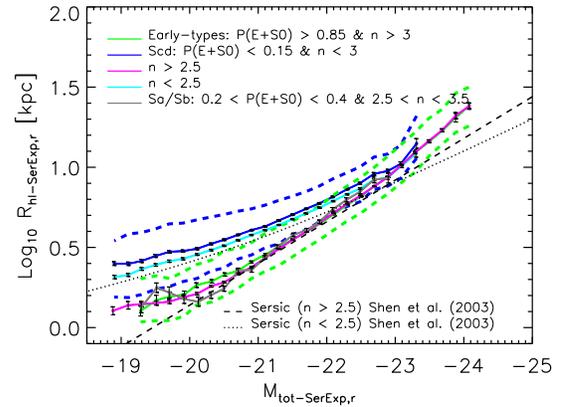}
 \caption{Similar to previous figure, but now objects are selected using different hard cuts which define early-, late- or intermediate-type (i.e. Sa/Sb) samples. Symbols with  error bars (joined by a solid curve for clarity) show the median half-light radius in bins of absolute magnitude. Dashed lines show the $16$th and $84$th percentile. Note that this definition of intermediate's (i.e. Sa/Sb) yields an $R-L$ relation which is essentially the same as for the population with $n>2.5$.  }
\label{LRes}
\end{figure}

\begin{table}
\begin{tabular}{lccc}
  \hline
  Sample/fit & $p_0$ & $p_1$ & $p_2$  \\
  \hline
  Sersic (Early-types)       &  12.8145  &  1.3788  &  0.0377 \\
  Sersic (Late-types) &   8.4847  &   0.9092  &   0.0254\\
  Sersic ($n > 2.5$) &  8.1624  &   0.9821  &   0.0292 \\
  Sersic ($n < 2.5$) &  4.7207  &   0.5601 &   0.0173 \\
  \hline
  Sersic (E)       & 7.0946   &   0.8650  &   0.0262 \\
  Sersic (S0/Sa)     & 10.9232   &    1.2218  &   0.0344 \\
  Sersic (Sa/Sb/Scd) & 13.9656   &    1.4694  &   0.0395 \\
  Sersic (Scd)       & 12.6494   &    1.3128  &   0.0352 \\
  \hline
  SersExp (Early-types)  &  8.6032  &  0.9979  &  0.0290 \\
  SersExp (Late-types) &  7.3204  &  0.7929 &  0.0226 \\
  SersExp ($n > 2.5$) &  6.0716  &  0.7770  &  0.0242 \\
  SersExp ($n < 2.5$) &  4.2848 &  0.5151 & 0.01615\\
  \hline
  SersExp (E)       &  7.4437  &    0.8922 &   0.0266 \\
  SersExp (S0/Sa)     &  9.6010  &    1.0903  &    0.0311 \\
  SersExp (Sa/Sb/Scd) &  9.3135   &   1.0182  &   0.0286 \\
  SersExp (Scd)       &  7.8056  &    0.8396 &    0.0237 \\
  \hline
  SersExp (Early-type-Bulges) &    -2.0733  &    0.0956 &   0.0098 \\
  SersExp (Late-type-Disks) &     6.4982  &    0.6934 &   0.0199 \\
  \hline
 \end{tabular}
\caption{Luminosity-size relation. Early-types: $p$(E+S0)$ > 0.85$ and $n > 3$. Late-types: $p$(E+S0) $< 0.15$ and $n < 3$. Early-type-bulges: the bulge half-light radius versus the bulge luminosity for galaxies with $p$(E+S0) $> 0.85$ and $n > 3$. Late-type-disks: the disk half-light radius versus the disk luminosity for galaxies with $p$(E+S0) $< 0.15$ and $n < 3$.}
 \label{LRfits}
\end{table}

\begin{table}
\begin{tabular}{lccc}
  \hline
  Sample/fit & $p_0$ & $p_1$ & $p_2$  \\
  \hline
  Sersic (Early-types)    &   19.0933  &    -3.9536  &    0.2070 \\
  Sersic (Late-types)   &   13.0054  &    -2.6438  &    0.1393 \\
  Sersic ($n > 2.5$) &   14.4995   &  -3.1767  &   0.1742 \\
  Sersic ($n < 2.5$) &   8.6098   &  -1.8301  &  0.1014 \\
  \hline
  Sersic (E)       &  13.6593  &   -2.9799  &   0.1635 \\
  Sersic (S0/Sa)     &  20.1092  &   -4.1549  &   0.2166 \\
  Sersic (Sa/Sb/Scd) &  22.3082  &   -4.4655  &   0.2275 \\
  Sersic (Scd)       &  17.9815  &   -3.6102  &   0.1862 \\
  \hline
  SerExp (Early-types)  &   13.4131   &  -2.9324 &    0.1607 \\
  SerExp (Late-types) &   11.2699  &  -2.3026 &    0.1227 \\
  SerExp ($n > 2.5$) &  12.5026  &   -2.7875  & 0.1551 \\    
  SerExp ($n < 2.5$) &   9.5210 &   -1.9963  & 0.1090 \\
  \hline
  SerExp (E)       &  12.8394  &   -2.8246  &   0.1557 \\
  SerExp (S0/Sa)     &  19.2830  &   -3.9866  &   0.2079 \\
  SerExp (Sa/Sb/Scd) &  18.6150  &   -3.7425  &   0.1922 \\
  SerExp (Scd)       &  11.7537  &   -2.3957  &   0.1271 \\
  \hline
  SerExp (Early-type-bulges) &    4.0853 &    -1.4159 &  0.0992 \\
  SerExp (Late-type-disks)  &    17.9763  &    -3.5683  &  0.1831 \\
  \hline
 \end{tabular}
\caption{Stellar mass-size relation. Early-types: $p$(E+S0)$ > 0.85$ and $n > 3$. Late-types: $p$(E+S0) $< 0.15$ and $n < 3$. Early-type-bulges: the bulge half-light radius versus the bulge stellar mass for galaxies with $p$(E+S0) $> 0.85$ and $n > 3$. Late-type-disks: the disk half-light radius versus the disk stellar mass for galaxies with $p$(E+S0) $< 0.15$ and $n < 3$.}
 \label{MsRfits}
\end{table}

\subsubsection{Comparison of early- and late-types}
We have repeated this analysis for the late-type sample (defined by requiring $n<3$ and $p$(E+S0)~$<0.15$).  Although we do not show the corresponding plots here, we again see curvature (coefficients of fits to equation~\ref{RLcurved} are given in Tables~\ref{LRfits} and~\ref{MsRfits}).  Rather, we illustrate this in Figure~\ref{LRes}, which compares the SerExp-based $R-L$ relation for our way of selecting early- and late-type samples, with the more traditional cuts on $n$ (larger or smaller than 2.5).  The two ways of selecting the samples lead to very similar results, with the low luminosity early-types having smaller sizes, but defining a steeper relation, so they would cross the $R-L$ relation of late-types at about $M_r<-23$ (beyond which there are few late-types anyway).  

We have also selected an intermediate-type population (i.e. Sa/Sb) by 
requiring $2.5 < n < 3.5$ and $0.2< p{\rm (E+S0)} < 0.4$.  
Notice that this sample defines the same $R-L$ relation as when 
we require our early-type selection 
(i.e., $n > 3$ and $p{\rm (E+S0)} > 0.85$), 
as well as that when we only require $n>2.5$; 
we return to this in Section~\ref{lrMorph}.

\begin{figure}
 \centering
 \includegraphics[width = \hsize]{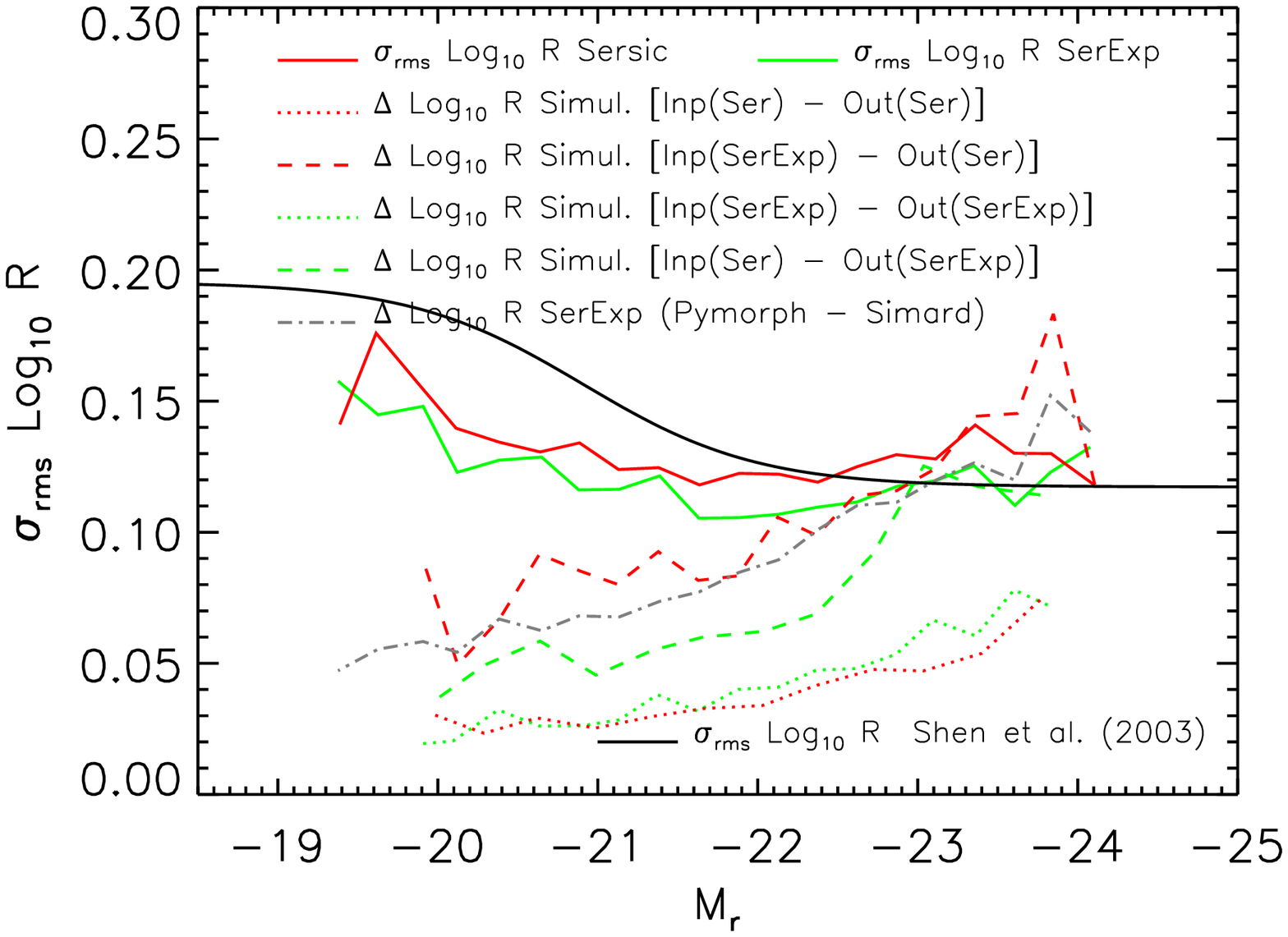}
 \includegraphics[width = \hsize]{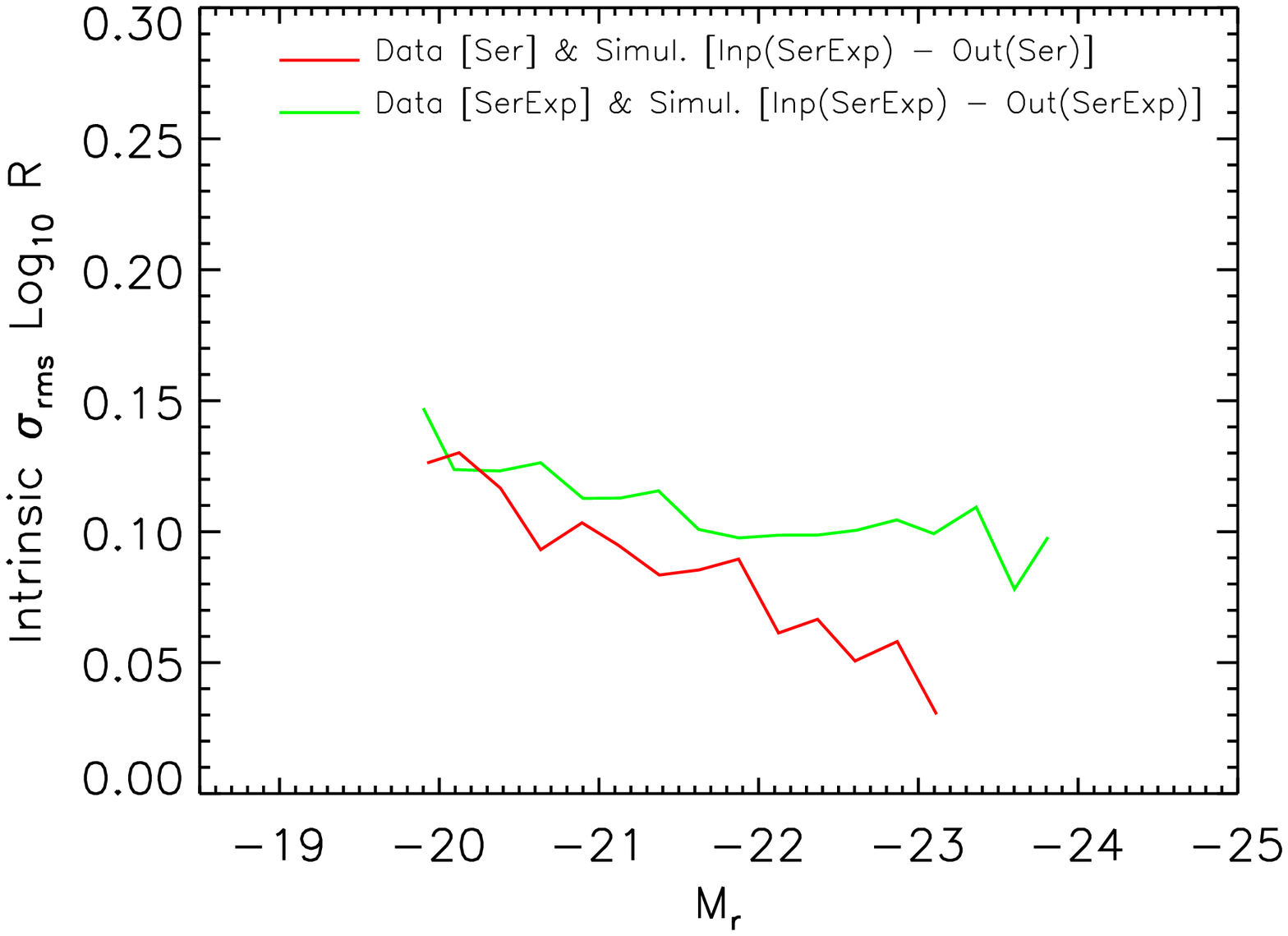}
 \caption{Top:  Observed scatter around the mean $\langle R|L\rangle$ relations for early-types based on fitting Sersic (solid red) and SerExp models (solid green) to the images.  Black solid curve shows the corresponding (error-broadened) measurement from Shen et al. (2003).  Dashed and dotted curves show a number of estimates from simulations of the measurement errors (see text for details).  Grey dot-dashed line in top panel shows the rms difference between {\tt PyMorph} and S11 sizes (both based on fitting a two-component SerExp).  
Bottom:  Estimate of the intrinsic scatter around the Sersic (lower, red curve) and SerExp (upper, green) derived relations for early-types, obtained by subtracting in quadrature the red-dashed and green dotted curves from the corresponding red and green solid curves shown in the top panel.}
 \label{LRscatter}
\end{figure}

\subsection{Scatter in log(size) around the mean relation for early-types}\label{scatter}
In addition to the mean $R-L$ relation, the scatter around the mean relation is expected to constrain galaxy formation models (Shen et al. 2003; Shankar \& Bernardi 2009; Shankar et al. 2012).  As we show below, our analysis of the mean $R-L$ relation allows us to make two interesting statements about the scatter around it for early-type galaxies:  
  $(i)$ there is intrinsic scatter and
 $(ii)$ it is smaller at the high mass end.  

The top two jagged solid curves in the top panel of Figure~\ref{LRscatter} 
show the measured scatter around the mean $R-L$ relation for SDSS 
early-types, when $R$ and $L$ are determined by fits to a single 
Sersic (larger scatter) and to a SerExp model (lower scatter).  
This scatter is broader than the intrinsic one, because it includes 
a contribution from the measurement errors.  For comparison, the 
smooth black curve shows the corresponding (error broadened) scatter 
reported by Shen et al. (2003).  It is in reasonably good agreement 
with ours, except at the faint end, where we believe the enhanced 
scatter is due to increased contamination by spirals, for which the 
scatter is larger (as we show later).  

To estimate the intrinsic scatter, we must account for the broadening 
due to measurement errors.  We estimate the errors on the sizes from 
fitting to the objects in the mock catalogs used in 
Section~\ref{2components}, where we know the input values.  (See 
Meert et al. 2013 for details of how the mocks were generated.)  
The dotted and dashed lines show these simulation-based estimates 
of the measurement error on the sizes for an early-type sample.  
The lowest dotted line shows the rms scatter in $\log_{10}R$ 
around the input value if the input profile is a single Sersic, 
and we fit it with a Sersic.  In this, and all the cases which 
follow, we show this scatter as a function of the fitted (as opposed 
to the input) absolute magnitude.  The other dotted line, which 
lies only slightly above the previous one, shows what happens if 
we fit a SerExp with a SerExp.  These curves certainly underestimate 
the full measurement error, since they are based on fits to smooth 
images, whereas real images may be lumpy, have spiral arms, etc.  

\begin{figure}
 \vspace{-2cm}
 \centering
 \includegraphics[width = 1\hsize]{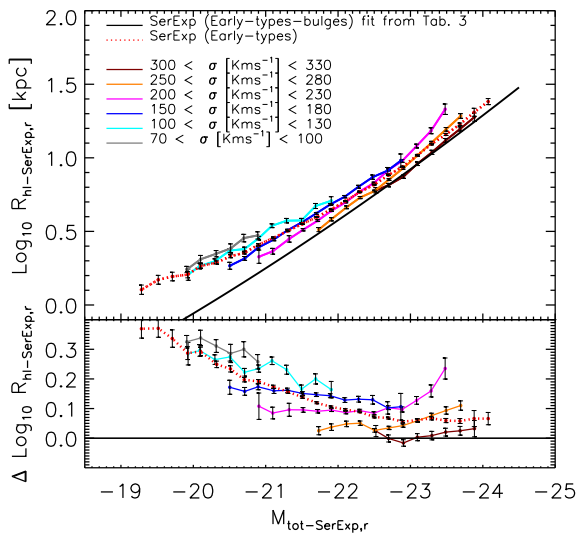}
 \caption{At fixed velocity dispersion $\sigma$, the $R-L$ relation 
          is almost a pure power law whose slope is almost the same 
          for all $\sigma$ (the black solid line shown is very close to 
          $R\propto L^{0.83}$), but whose zero-point increases as 
          $\sigma$ decreases.  
          In contrast, the relation upon averaging over all $\sigma$ 
          (red dotted curve) is much more curved.  Bottom panel shows 
          the result of dividing the measurements in the top panel 
          by the black solid line.  
          }
 \label{LRfixS}
\end{figure}

To get an idea of the magnitude of such effects, the dot-dashed curve 
in the top panel of Figure~\ref{LRscatter} shows the rms difference 
between {\tt PyMorph} and S11 sizes returned by two-component SerExp 
fits to SDSS images, plotted as a function of the {\tt PyMorph} SerExp 
absolute magnitude for the early-type sample.  
This is almost certainly an overestimate of the measurement error on 
the sizes, since it includes systematic effects which arise from the 
differences between {\tt PyMorph} and S11; we have included it just to 
get a sense of the overall magnitude with which systematic rather than 
random errors might affect the scatter in the $R-L$ relation.  

The two dashed curves show results from using {\tt PyMorph} to fit a 
Sersic with a SerExp (lower) and a SerExp with a Sersic (upper).  
The differences between these and the dotted curves give an idea of 
the effect on the scatter of fitting an incorrect model to the data.  
The upper dashed curve is particularly interesting, in view of the fact 
that the SerExp model is more realistic (see Section~\ref{2components}), 
whereas the single Sersic model is most often fit.  Clearly, subtracting 
it in quadrature from the upper solid curve will lead to negative values 
at large luminosities.  This is shown by the lower of the two curves in 
the bottom panel: at $M_r<-23$ or so, the intrinsic scatter is 
consistent with zero.  This, of course, does {\em not} mean that 
the $R-L$ relation is intrinsically a line with negligible scatter.  
Rather, it is entirely a consequence of fitting an incorrect model.  

\begin{figure*}
 \centering
 \includegraphics[scale = .425]{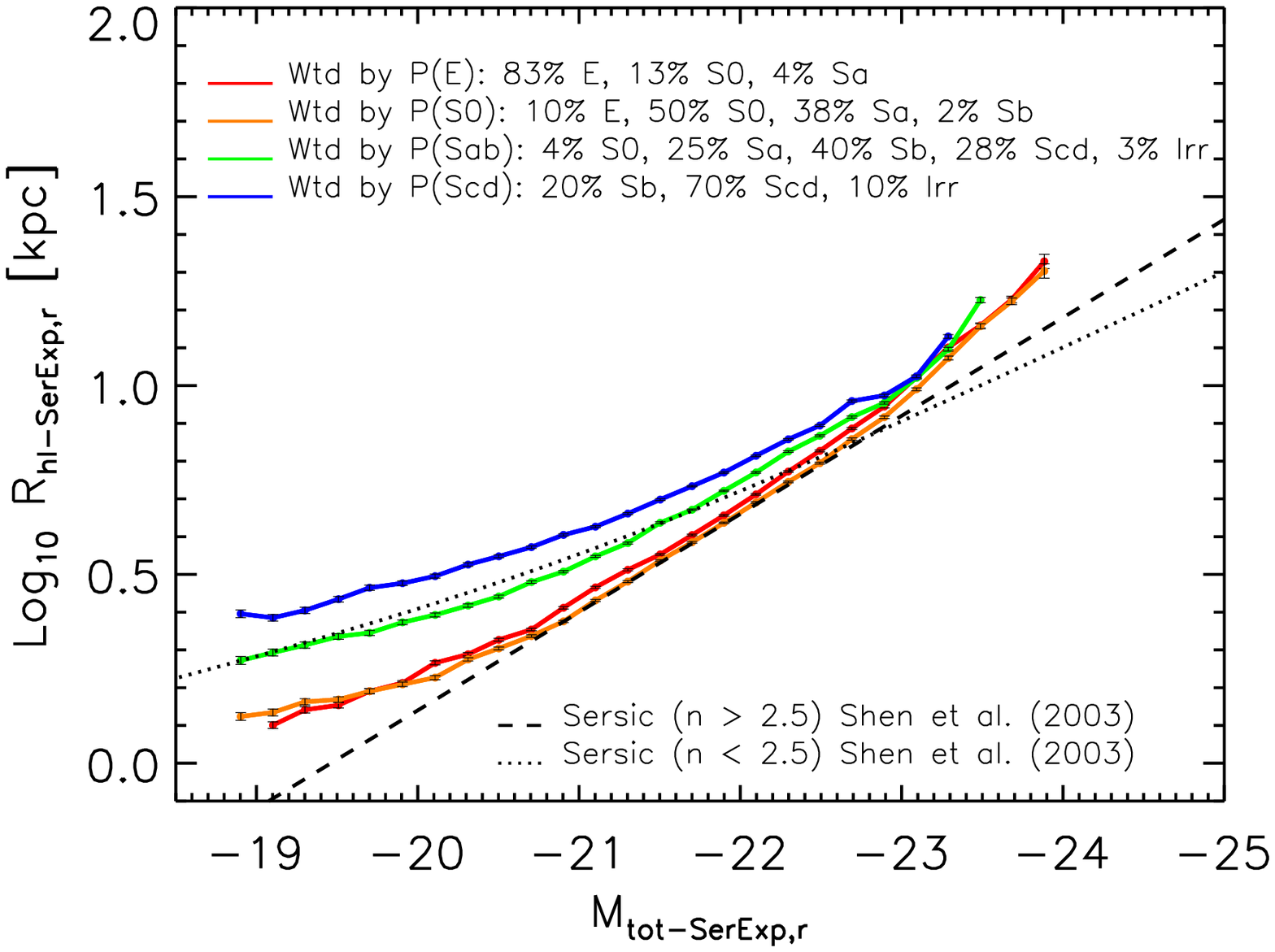}
 \includegraphics[scale = .425]{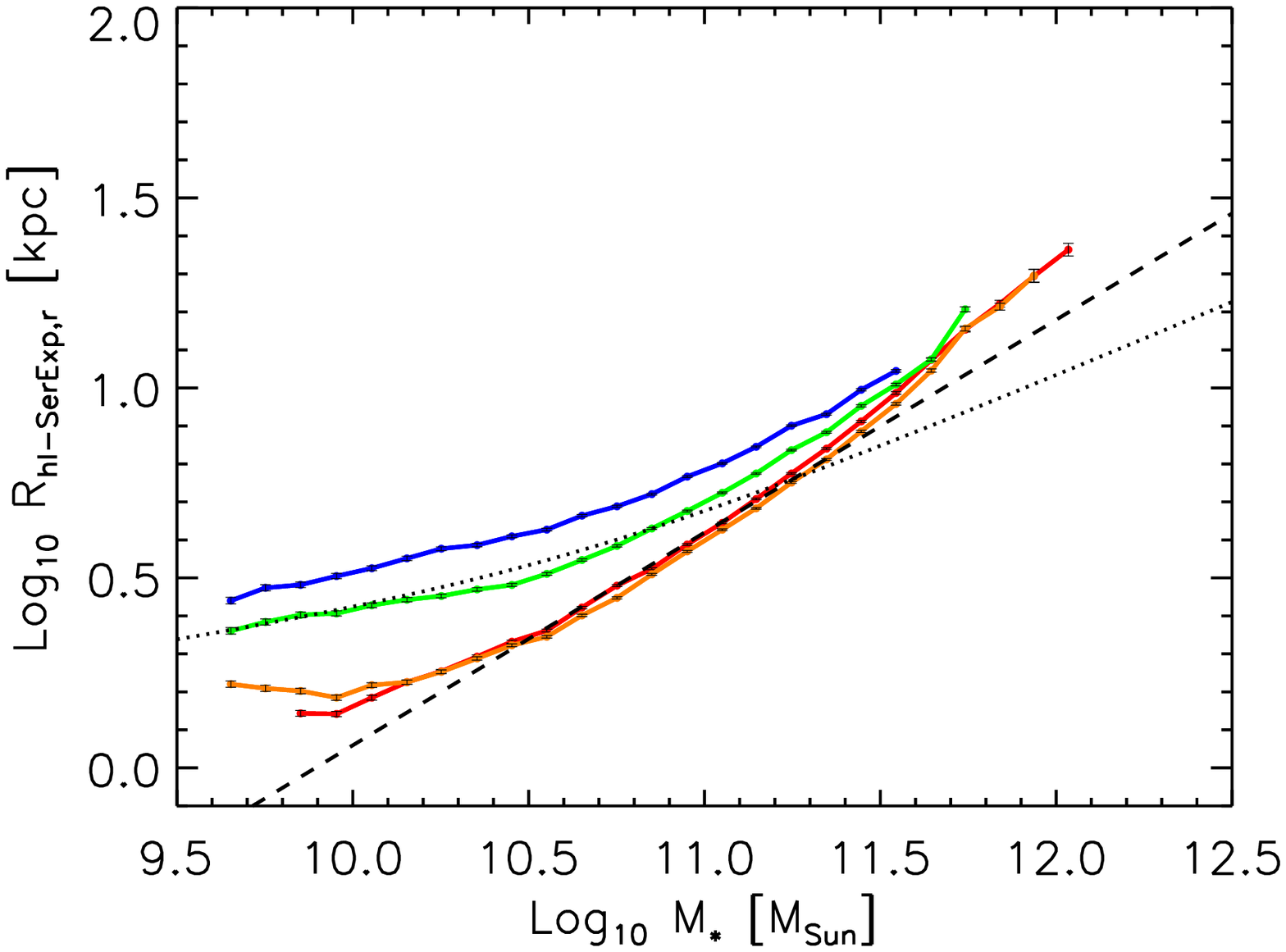}
 \caption{SerExp based size-luminosity (left) and size-$M_*$ (right) 
          relations, obtained by weighting objects by the BAC $p$(type).  
          The low $L$ or $M_*$ part of the relation for Scds has the 
          same slope as that reported by Shen et al. (2003) for their 
          $n<2.5$; and the intermediate $L$ or $M_*$ part has the same 
          slope they report for $n>2.5$.  
          The relations for S0s are very similar to those for Es, and 
          the Sab relations always lie between the E and Scd relations.
          Numbers in legend show the percentage of E, S0, Sa, Sb, 
          Sbc and Irr galaxies classified by F07 with BAC $p>0.6$.
          Using this selection we miss about $18\%$ of Es, 
          $60\%$ of S0s, $64\%$ of Sab ($37\%$ Sa and $27\%$ Sb) 
          and $56\%$ of Scd, respectively.  
          }
 \label{LRw}
\end{figure*}

Recently, Nair et al. (2011) have used just such an argument to 
claim that the $R-L$ relation has no scatter.  However, their 
argument is based on Petrosian sizes and luminosities; these are 
known to be inaccurate at large $L$, so the analysis above 
illustrates why their claim should be treated with skepticism.  
Indeed, the upper curve shows the result of subtracting (in quadrature) 
the upper dotted curve from the lower solid one, since both these 
are based on fitting to what we argued were more realistic models 
of the light profile (i.e. SerExp).  In this case, the intrinsic 
scatter is well-behaved:  although it decreases steadily with $M_r$, 
it does not go negative.  

Of course, since our estimate of the measurement error is really 
an underestimate, it is still possible that the intrinsic scatter 
is smaller than we show.  
Therefore, we turn to what we believe is a much more effective way 
of showing that there is some intrinsic scatter.  This method studies 
if the residuals from the relation correlate with other parameters, 
once correlations between the measurement errors have been accounted 
for.  If they do, then there must be some intrinsic scatter.  

Figure~\ref{LRfixS} shows the $R-L$ relation for a number of narrow bins 
in velocity dispersion $\sigma$.  At fixed $\sigma$, the $R-L$ relation 
is a power law whose slope is 0.85 for all $\sigma$ but whose 
zero-point increases as $\sigma$ decreases.  The scaling is quite well 
described by 
\begin{equation}
 \label{R|LV}
 \langle R|L,\sigma\rangle \propto L^{0.85}\sigma^{-0.73}.
\end{equation}
The dependence on $\sigma$ shows clearly that the scatter around the 
mean $R-L$ relation correlates with $\sigma$; it is not all due to 
measurement errors.  We are not the first to have made this point 
(see Bernardi et al. 2003b for an explicit discussion of this, although 
the long-studied Fundamental Plane is the result of this correlation), 
so it is surprising that Nair et al. ignored it.  The slope of 0.85 
(at fixed $\sigma$) is consistent with previous work (Bernardi et al. 2003a; 
Bernardi 2009).  While steeper than the slope of 0.64 associated with 
averaging over all $\sigma$, it is less than unity -- a fact we return 
to later, when we discuss the bulges of early-types (see 
Figure~\ref{LbRbfixS}).  


We end this subsection with the observation that the intrinsic scatter 
appears to be smallest for the most luminous objects.  Since it is 
commonly believed that mergers will affect the scatter of scaling 
relations such as this one, our overestimate of the intrisic scatter 
in the $R-L$ relation provides a new constraint on models of how the 
most massive galaxies must have formed.  E.g., Shen et al. (2003) 
argue that many minor mergers may be more consistent with the shape 
and scatter of the $R-L$ relation than are few major mergers.  
Other work has also explored constraints which come from the scatter 
(Shankar et al. 2012); it will be interesting to revisit this question 
in light of the mass-dependence we believe we see.

\begin{figure*}
 \centering
 \includegraphics[scale = .8]{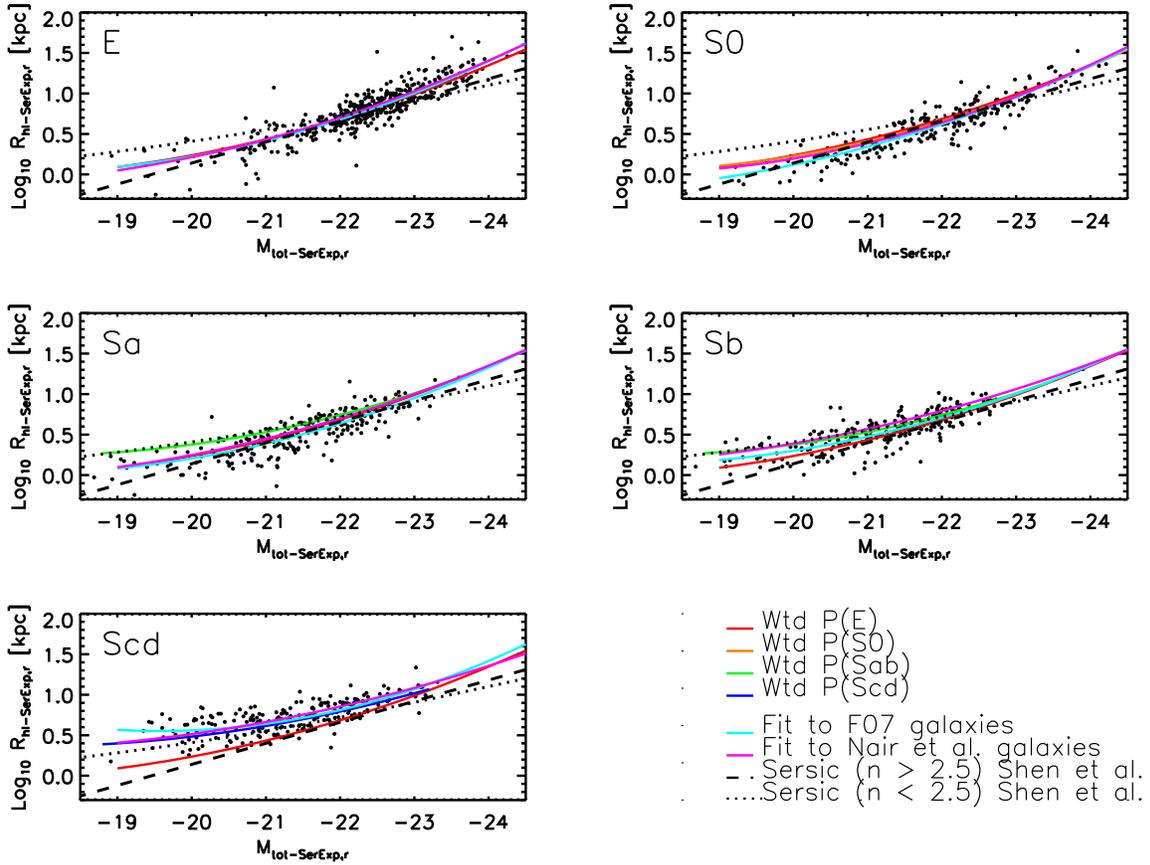}
 \caption{Comparison of the $R-L$ relation in the morphologically defined 
          samples of F07 (symbols and cyan curve) 
          and N10 (magenta), with the fits defined by 
          the BAC of Huertas-Company et al. (2011).  
          All relations are in good agreement for E and S0 galaxies; 
          for comparison, the E relation is also shown in the other panels.  
          F07 and N10 agree that Sa's define the same relation as Es and S0s, 
          whereas Sb's are offset to larger sizes at smaller $L$.  
          The BAC-based results for Sab lie further from that for Es 
          compared to those based on F07 and N10 for Sa's, but are in 
          good agreement for Sb's; however, for Scd's they lie closer 
          to the E relation than do F07 or N10. }
\label{LRf07}
\end{figure*}

\section{Dependence on Bayesian Automated Classifier morphologies}\label{lrMorph}
In the previous section we used a hard cut on the BAC probability to 
determine morphology.  Since this is not quite in the spirit of why 
such probabilities were generated in the first place, this section 
shows the result of using the BAC probabilities as weights when 
determining the $R-L$ scaling relation.  As this is one of the first 
such studies of weight-based scaling relations, and Tables~\ref{F07} 
and~\ref{N10} indicate that the relation between these weights and 
eye-ball classifications can be complex in detail, we include an 
explicit comparison of these BAC-based scaling relations with those 
based on eye-ball classifications.  

Figure~\ref{LRw} shows the size-luminosity (left) and size-$M_*$ (right) 
relations obtained by weighting objects by $p$(type) as determined by BAC.  
The results of fitting equation~(\ref{RLcurved}) to these curved 
relations are reported in Tables~\ref{LRfits} and~\ref{MsRfits}.  


Notice that galaxies weighted by $p$(Sab) define a relation which lies between that defined by $p$(Scd) on the one hand and $p$(E) and $p$(S0) on the other.  However, the Sa/Sb class is difficult to define (c.f. discussion of Tables~\ref{F07} and~\ref{N10}).  Figure~\ref{LRw} reports the fraction of E, S0, Sa, Sb, Sbc and Irr galaxies classified by F07 with the corresponding BAC $p>0.6$.

\subsection{Comparison with eye-ball classifications}
To address this more closely, Figure~\ref{LRf07} shows the $R-L$ 
relations in the F07 eye-ball classified subsamples.  The cyan curves 
show fits to these subsamples, and the magenta curves show fits based on 
the N10 (eyeball) classifications.  The two are in quite good agreement.  
To emphasize the fact that the relation is different for the different 
subsamples, the red solid curve, which is the same in each panel, shows 
the BAC-based relation for $p$(E).  
The orange, green and blue curves (in the relevant panels) show the 
BAC-based relations for $p$(S0), $p$(Sab) and $p$(Scd).  These are in 
good agreement with the F07 and N10 based relations for E and S0 
galaxies.  

Note that F07 and N10 agree that Sa's define the same relation as Es 
and S0s, whereas Sb's are offset to larger sizes at smaller $L$.  
This suggests that combining Sa's and Sb's into a single type may be 
problematic.  Indeed, the BAC-based results for Sab lie further from 
that for Es compared to those based on F07 and N10 for Sa's, but are 
in good agreement for Sb's; however, for Scd's they lie closer to the 
E relation than do F07 or N10.  These small but systematic differences 
between the BAC and eye-ball based results suggest that combining Sab's 
into a single class as is done by BAC has resulted in a weighted sum of 
the relations defined by E's and Scd's. 

Figure~\ref{LRf07} shows that the curvature in the $R-L$ relation 
is such that, for Scds, there is almost no correlation at $M_R>-20.5$.  
This flattening at low luminosities is also evident for the other 
morphological types, and is more pronounced in the $R-M_*$ relation 
shown in the right hand panel of Figure~\ref{LRw} 
(see also Figure~\ref{LRbd} below).  Indeed, Figure~\ref{LRw} shows 
that at $\log_{10}M_*/M_\odot < 10.5$, even the samples weighted by 
$p$(E) and $p$(S0) tend to have essentially no correlation between 
$R$ and $M_*$.  
This is the same mass scale at which a number of other early-type 
galaxy scaling relations change qualitatively (Bernardi et al. 2011a,b 
and our Figure~\ref{MsRserexp}). 
Since Figures~\ref{LRw} and~\ref{LRf07} indicate that it also appears 
to be significant for late-type galaxies, it is interesting to ask if 
the other, higher mass scale identified by Bernardi et al., 
$M_* = 2\times 10^{11}M_\odot$, is also significant for late-types.  

Figure~\ref{LRw} shows that, in fact, this higher mass scale seems to 
set the limit above which there are essentially no late-type galaxies.  
Figure~\ref{LRf07} tells a consistent story:  although there are 
many Es and S0s brighter than $M_r = -23$, there are no Sa, Sb or Scds 
with luminosities this large.  
Bernardi et al. suggested that this mass scale was associated with 
merger histories that were dominated by major dry mergers; since such 
mergers would destroy disks, the fact that we see no late-types above 
this mass scale is, perhaps, not surprising.

%


\subsection{Small but statistically significant difference between Ellipticals and S0s}
\begin{figure}
 \centering
 \includegraphics[scale = .4]{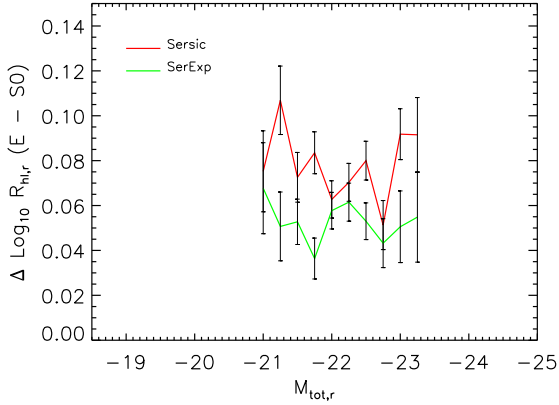}
 \caption{At fixed luminosity, Es tend to be about 0.06~dex larger 
          than S0s, although this offset depends slightly on how 
          $R$ and $L$ were determined.  }
\label{LRes0}
\end{figure}

Above, we noted that there is essentially one $R-L$ relation for E, S0 and Sa galaxies.  However, our sample is large enough to detect small but significant differences within the early-type (E and S0) sample.  A closer look at Figures~\ref{LRw} and~\ref{LRf07} indicates that S0s are slightly smaller than Es of the same luminosity.  Figure~\ref{LRes0} shows that this offset is about 0.06~dex, although it depends slightly on how $R$ and $L$ were determined.  This is particularly interesting in view of recent work at $z\sim 1$, based on the S11 reductions, which shows a similar offset of about 15\% for the SDSS sample growing to $\sim 40\%$ at $z\sim 1$ (Huertas-Company et al. 2012).  Both the sign of the trend and its evolution deserve further study, because, as we show below (see Section~\ref{smallbulges} and Figure~\ref{L_RdRb}), the sign of the trend is not what one might naively have expected.  

\section{Bulges and disks}\label{bd}
One of the virtues of our SerExp decompositions is that it allows us to study the scaling laws of disks and bulges.  Recall that the second components required to fit bulge dominated galaxies may not be disks.  In addition, although it makes sense to speak of the central component in a disk dominated galaxy, {\tt PyMorph} does not distinguish between central bulges and bars of such galaxies.  Therefore, in what follows we would like to contrast the $R-L$ and $R-M_*$ relations for early-types with those for their bulges, and these relations for late-types with those for their disks.  I.e., we focus on the change which results from adding a small second component to the dominant one, rather than on the properties of the secondary component.  See Gadotti (2008; 2009) and references therein for a discussion of the bars and pseudo-bulges in the centers of disks.  

The structural properties of bulges can be reliably estimated only if the effective radius of the bulge is larger than about 80 per cent of the PSF HWHM (Gadotti 2008; 2009).  The SDSS PSF HWHM is about 0.7 arcsec, and corresponds to about 1.3 kpc at the median redshift of $z\sim 0.1$.  On the basis of a nearer-by sample ($z\sim 0.05$), Gadotti identified that about half of the bulges are smaller than $\sim 1$~kpc which is the resolution limit of our work at $z\sim 0.1$.  This raises the question of how reliable the {\tt PyMorph} estimates are for (small) bulges, and if resolution effects will bias our results. Note that in this section we are specifically studying the bulges of bulge-dominated systems while Gadotti analyzed bulges for all galaxies with M$_* > 10^{10}$~M$_\odot$, therefore we expect our sample to be dominated by larger bulges.

\begin{figure}
 \centering
 \includegraphics[scale = .4]{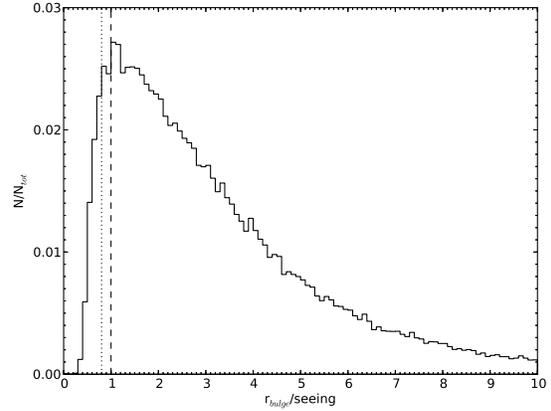}
 \caption{Distribution of the ratio between bulge effective radius and PSF HWHM for all galaxies in the sample (excluding bulge-less galaxies). The two vertical lines mark the positions where this ratio is 0.8 (dotted line) and 1 (dashed line). Only 7 per cent (11 per cent) of the bulges have effective radii below 0.8 (1) times the PSF HWHM.  Had we used the semi-major axis rather than the circularized radius, these numbers drop to 3 and 7 percent, indicating that our bulge decompositions should be reliable for the vast majority of our sample.}
 \label{rbvspsf}
\end{figure}

The reliability of the {\tt PyMorph} bulges was partially addressed in Meert et al. (2013).  But to allow a more direct comparison, Figure~\ref{rbvspsf} shows the distribution of bulge sizes in units of the PSF HWHM for galaxies in the early-type sample we study in this section, in a format which is similar to Gadotti (2009)'s Figure~7.  We find a similar fraction of galaxies have bulges below the HWHM in our sample (11\% vs 10\% for Gadotti) and about twice as many galaxies below the 80\% mark (7\% vs 3\% for Gadotti). These numbers drop to 7\% (below the HWHM) and 3\% (below the 80\% mark) in our sample if we use the semi-major axis of the bulge rather than the circularized radius. Therefore, reliable estimates of bulge parameters should be possible for the vast majority of our sample. We have similar fractions to Gadotti, despite having a deeper sample, because we are specifically studying the bulges of bulge-dominated systems, and these tend to be the more massive galaxies with larger half light radii. We will see below that the half-light radii we recover are typically larger than 1~kpc, with no particular feature to indicate problems at around 1~kpc.  Nevertheless, we reexamined the results reported below without including the bulges smaller than the HWHM and found no difference in the results.  

\begin{figure*}
 \centering
 \includegraphics[scale = .8]{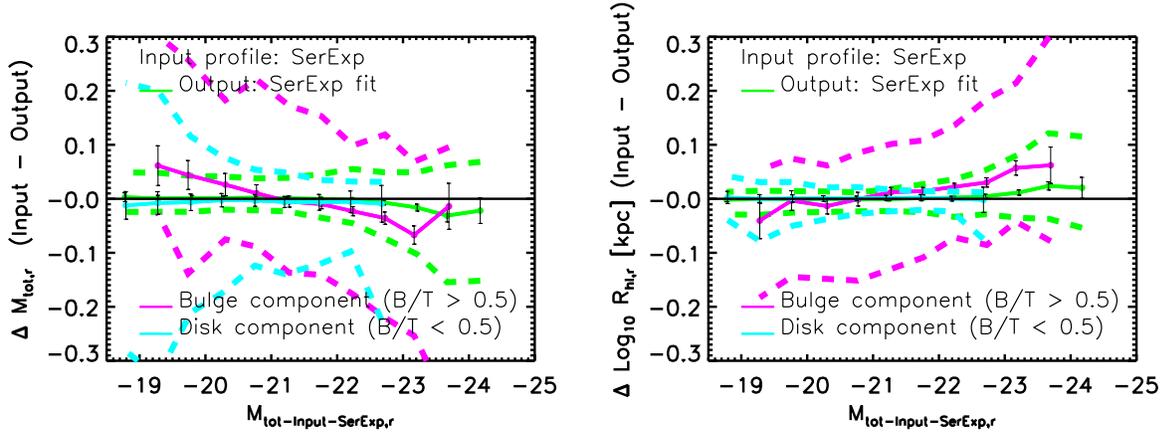}
 \caption{Biases in the estimated luminosities (left) and sizes (right) 
          of the total (green), bulge (magenta) and disk (blue) 
          components in SerExp fits to SerExp images (we restrict our 
          analyses of the bulges and disks to those objects when they 
          are the dominant component).  The parameters of the total 
          are usually unbiased, as are the disks of objects where B/T$<0.5$.  
          However, in objects where B/T$>0.5$, the estimated bulge 
          luminosities and sizes are slightly too faint and small at 
          the bright end, and a little too bright at the faint end.}
 \label{SimulationsLRBD}
\end{figure*}

Figure~\ref{SimulationsLRBD} shows a related test of {\tt PyMorph} systematics:  
SerExp fits to noisy SerExp images tend to return unbiased estimates of the disks in systems with B/T~$<0.5$, but in objects where B/T~$>0.5$, the estimated bulge luminosities and sizes are slightly too faint and small at the bright end, and a little too bright at the faint end.  However, these are rather small effects, so that the following study of the $R-L$ relations of disks and bulges is meaningful.  

The top and bottom panels of Figure~\ref{LRbd} show the relevant $R-L$ and $R-M_*$ relations.  As we cautioned before, the conversion from $L$ to $M_*$ depends on $M_*/L$, which in turn depends on stellar population modelling as well as on an assumption about how the IMF depends on galaxy mass.
But here, there is another reason to be cautious:  our $M_*$ estimates assume that $M_*/L$ for the individual components is the same as that for the total.  Since we are looking at bulges of early types and disks of late types, this assumption, while crude, should not be wildly wrong (the same would not be true for, e.g., the bulges of late types).  Nevertheless, one might imagine that, as a result, we slightly under(over)-estimate the mass in the bulge(disk) component.

\subsection{Little or no correlation for disks at low masses}
There are two striking features about late-types and their disks.  
First, although the $R_{\rm disk}-L_{\rm disk}$ and $R_{\rm disk}-M_{*{\rm disk}}$ 
relations are very curved, they run parallel to the corresponding 
$R-L$ and $R-M_*$ relations for late-types; $R_{\rm disk}$ tends to 
be 0.1~dex larger than $R_{\rm hl}$.  
That $R_{\rm disk} > R_{\rm hl}$ is not surprising, since we know that 
late-type galaxies host small bulges which will contribute to the 
light at small radii.  But that this should have produced a constant 
offset is not obvious.  We address this question shortly.  


Second, at $\log_{10}M_*/M_\odot < 10.5$, the flattening of the 
$R_{\rm disk}-M_{*{\rm disk}}$ relation with respect to the slope at 
large $M_*$ is quite pronounced: there is almost no correlation 
between $R_{\rm disk}$ and $M_{*{\rm disk}}$ at low masses.  
This flatness at the faint, low mass end is similar to that for 
Scds (see Figures~\ref{LRw} and~\ref{LRf07} and related discussion):  
these disks are far from having constant surface brightness.  
Note that these sizes are significantly larger than the $\sim 1$~kpc 
scale below which resolution effects might introduce a bias.  
Moreover, a flatter relation for disks is also seen in the lower 
panel of Figure~13 in Gadotti (2009).  

\subsection{A pure power-law for bulges}
The bulges are also interesting.  
In contrast to when the total light was used, there is almost no 
curvature in the relation for bulges which is well approximated 
by a single power-law:  
\begin{equation}
 \label{Rb|Lb}
 \langle R_{\rm bulge}|L_{\rm bulge}\rangle\propto L_{\rm bulge}^{0.85}.
\end{equation}
The amplitude of the power law is such that the relation for bulges 
is approximately the same as for the total at very large luminosities; 
as $L$ decreases, the $R-L$ relation curves away from the 
$R_{\rm bulge}-L_{\rm bulge}$ relation, towards larger sizes.  (There is 
a slight flattening of the relation at sizes smaller than about 1 kpc; 
these are the few objects for which the SDSS seeing may be becoming 
problematic.)

The power-law nature of the bulge relation suggests a picture in which 
the curvature in the early-type $R-L$ relation arises as a consequence 
of adding a second component.  However, there is an interesting puzzle:  
recall that Figure~\ref{LRfixS} shows the $R-L$ relation for a few 
narrow bins in velocity dispersion.  This relation also has almost no 
curvature and, remarkably, it runs parallel to the 
$R_{\rm bulge}-L_{\rm bulge}$ relation (the $L$ dependence of 
equation~\ref{Rb|Lb} is the same as of equation~\ref{R|LV}).  

To explore this coincidence further, Figure~\ref{LbRbfixS} shows the 
analogue of Figure~\ref{LRfixS}:  the $R_{\rm bulge}-L_{\rm bulge}$ for 
fixed bins in $\sigma$.  We find 
\begin{equation}
 \label{Rb|LbV}
 \langle R_{\rm bulge}|L_{\rm bulge},\sigma\rangle\propto L_{\rm bulge}\, \sigma^{-1}.
\end{equation}
Notice that the size is proportional to $L$.  Replacing $L_{\rm bulge}$ 
with $M_{*{\rm bulge}}$ makes no difference.  I.e., our SerExp bulges 
exhibit the scaling expected from the virial theorem, although the 
dependence on $\sigma$ is different.  


\begin{figure}
 \centering
 \includegraphics[width = \hsize]{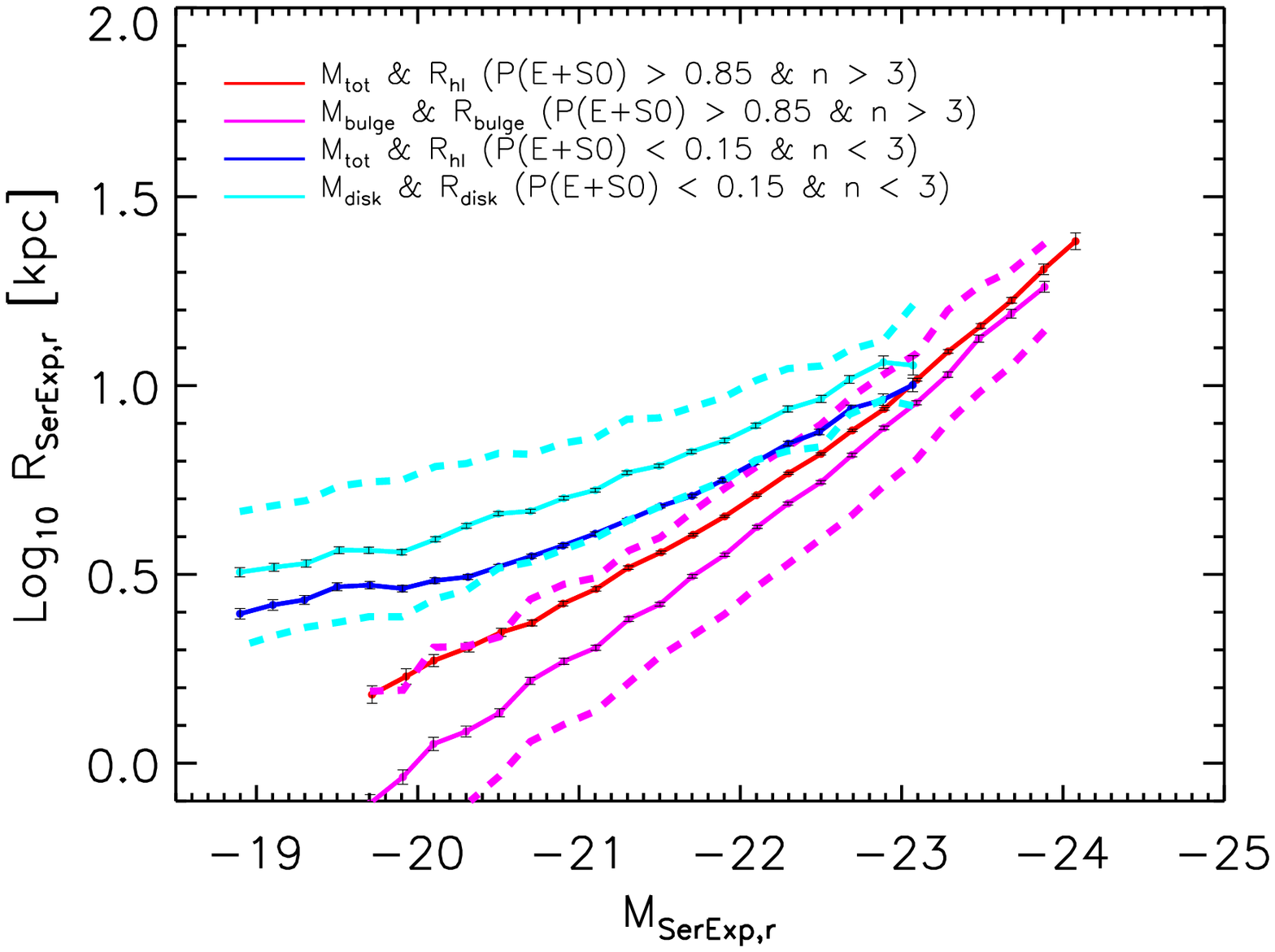}
 \includegraphics[width = \hsize]{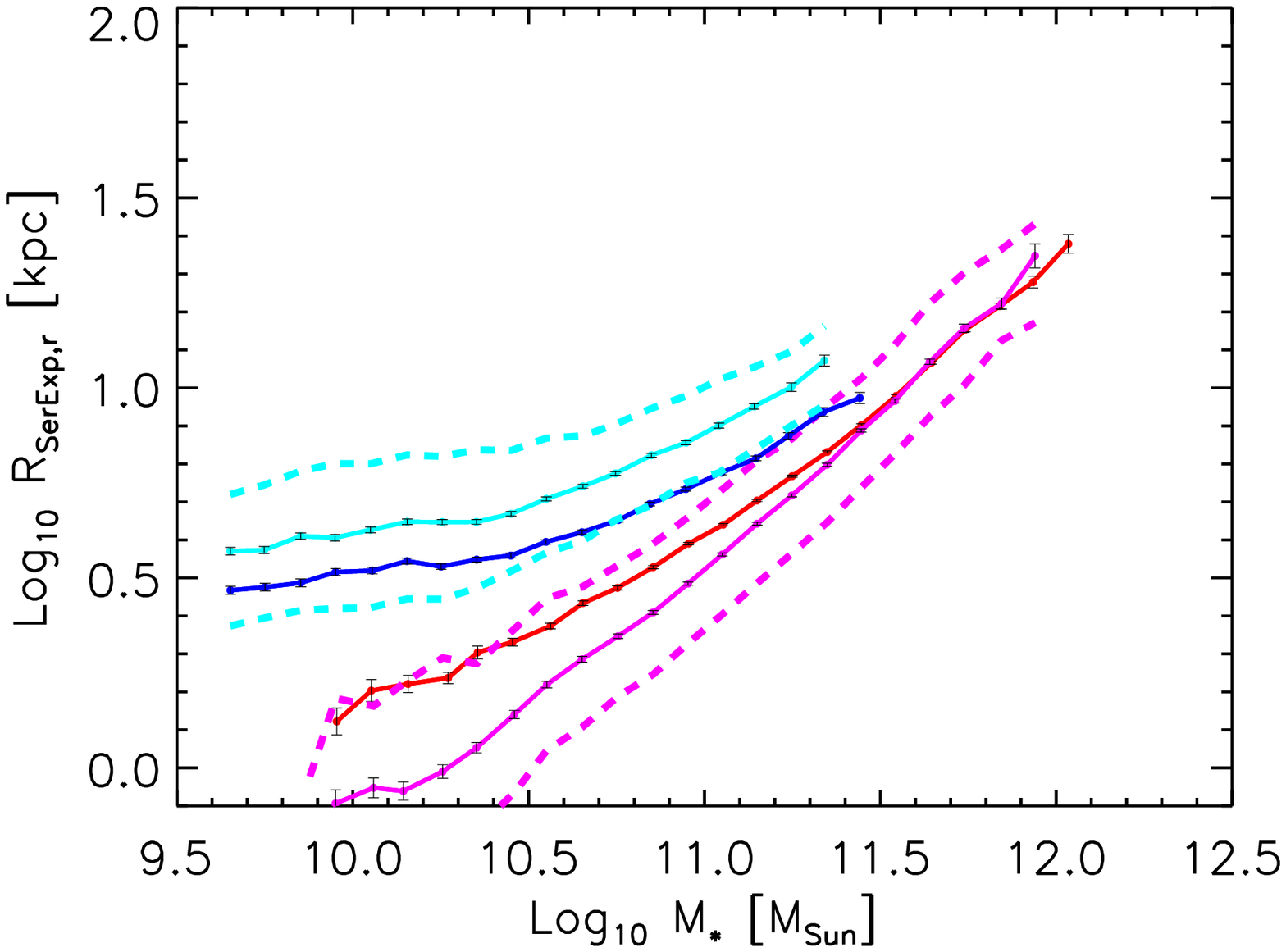}
 \caption{Similar to Figure~\ref{LRes}, but now contrasting the 
          $R-L$ (top) and $R-M_*$ (bottom) relations for early-types 
          with those for their bulges, and these relations for 
          late-types with those for their disks.  
          Note the severe flattening at $M<3\times 10^{10}M_\odot$ in 
          the relations for late-types, and the absence of late-types at 
          the mass scales on which the bulge and total $R-M_*$ relations 
          become the same:  $M>2\times 10^{11}M_\odot$.
          }
 \label{LRbd}
\end{figure}

Finally, the bottom panel of Figure~\ref{LRbd} also shows that the 
$R_{\rm bulge}-M_{*{\rm bulge}}$ relation sits on top of that for early 
types at the largest masses, suggesting that the second component which 
contributes somewhat to the light contributes little to the mass.  
It is worth noting that this happens at the same mass scale,
 $M_* = 2\times 10^{11}M_\odot$, which Bernardi et al. (2011a) noted was 
significant for early-types, and above which there appear to be no 
late-type galaxies (as is clear from this figure, as well as from 
Figures~\ref{MsRserexp}, \ref{LRw} and~\ref{LRf07}).  

In this context, it is worth noting that one gets approximately the 
same steep $R_{\rm bulge}-M_{*{\rm bulge}}$ relation if one combines the 
`ellipticals' and `classical bulges' shown in the upper panel of 
Figure~13 in Gadotti (2009).  (The pseudo-bulges which are also shown 
in that panel are associated with small B/T~$<0.3$ values, so they 
would not be included in our sample of bulge dominated systems.)  
While this is reassuring, Gadotti argues that it may better to think 
of the two populations as being physically distinct.  
This may be related to our finding of a small but systematic offset 
between the relations for Ellipticals and S0s (Figure~\ref{LRes0}).  
In addition, our Figure~\ref{LRbd} shows that the scatter around 
the mean bulge relation increases at smaller $M_*$, perhaps indicating 
that we are indeed combining two populations.  On the other hand, at 
small $M_*$ the sizes are increasingly prone to being biased by the 
seeing, so this may also be contributing to the increase in scatter.

\subsection{The smallness of bulges}\label{smallbulges}
Why are the $R-L$ relations for the disks and bulges so different 
from those for the total light?   

To address this, we selected a subset of the early-type sample with 
$-21.5 > M_r > -22.5$.  Figure~\ref{BTRshifts} shows that there is a 
strong correlation between $R_{\rm hl}/R_{\rm bulge}$ and B/T 
(at this fixed $M_r$) for the early-type sample.  If the total is 
0.55~mags brighter than the bulge (B/T$=0.6$), then the half-light 
radius of the total is about 0.35~dex larger than that of the bulge.  
A similarly tight correlation is seen for other $M_r$.  What causes 
this?




Suppose we start from the power-law $R_{\rm bulge}-L_{\rm bulge}$ relation.  
$L_{\rm bulge}$, with a given value of $B/T$, specifies a total magnitude 
$M_{\rm bulge} + 2.5 \log_{10}$(B/T).  If $n$ of the bulge is known, and 
we assume that the second component has an exponential profile, then 
the half-light radius of the combined profile is fully specified only 
if we also know $R_{\rm disk}/R_{\rm bulge}$ (see Appendix~\ref{btanalytic}).  
(Recall that, although we use the word disk, we do not mean to imply 
that the second component is necessarily a disk.)  The curves show the 
expected relations between $R_h/R_{\rm bulge}$ and B/T for a deVaucouleur 
bulge with exponential disk for $R_{\rm  disk}/R_{\rm bulge} = 2, 4$ and 6
(bottom to top).  Although they are independent of the total luminosity 
(Appendix~\ref{btanalytic} shows why), the curves shift upwards slightly 
if $n>4$ and downwards if $n<4$.  

\begin{figure}
 \vspace{-2cm}
 \centering
 \includegraphics[width = \hsize]{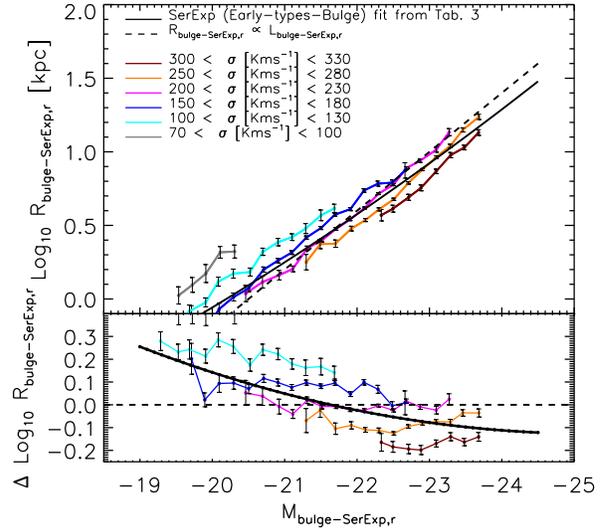}
 \caption{Same as Figure~\ref{LRfixS}, but showing  
          $R_{\rm bulge}-L_{\rm bulge}$ for a number of bins in total 
          velocity dispersion $\sigma$.  
          Replacing $L_{\rm bulge}$ with $M_*$ of the bulge yields the 
          same result:  The relation is a power law whose slope is 1 
          for all $\sigma$, but whose zero-point increases as $\sigma$ 
          decreases.}
 \label{LbRbfixS}
\end{figure}

Matching the data indicates that $R_{\rm  disk}/R_{\rm bulge}\sim 5$ at 
B/T~$<0.7$ (if $n$ is 5 rather than 4, then $R_{\rm  disk}/R_{\rm bulge}\sim 4$), 
suggesting that the correlation is caused by the fact that PyMorph uses 
second components with rather large scale lengths to account for the 
fact that a Sersic bulge is not, by itself, always a good match.  

\begin{figure}
 \centering
 \includegraphics[width = \hsize]{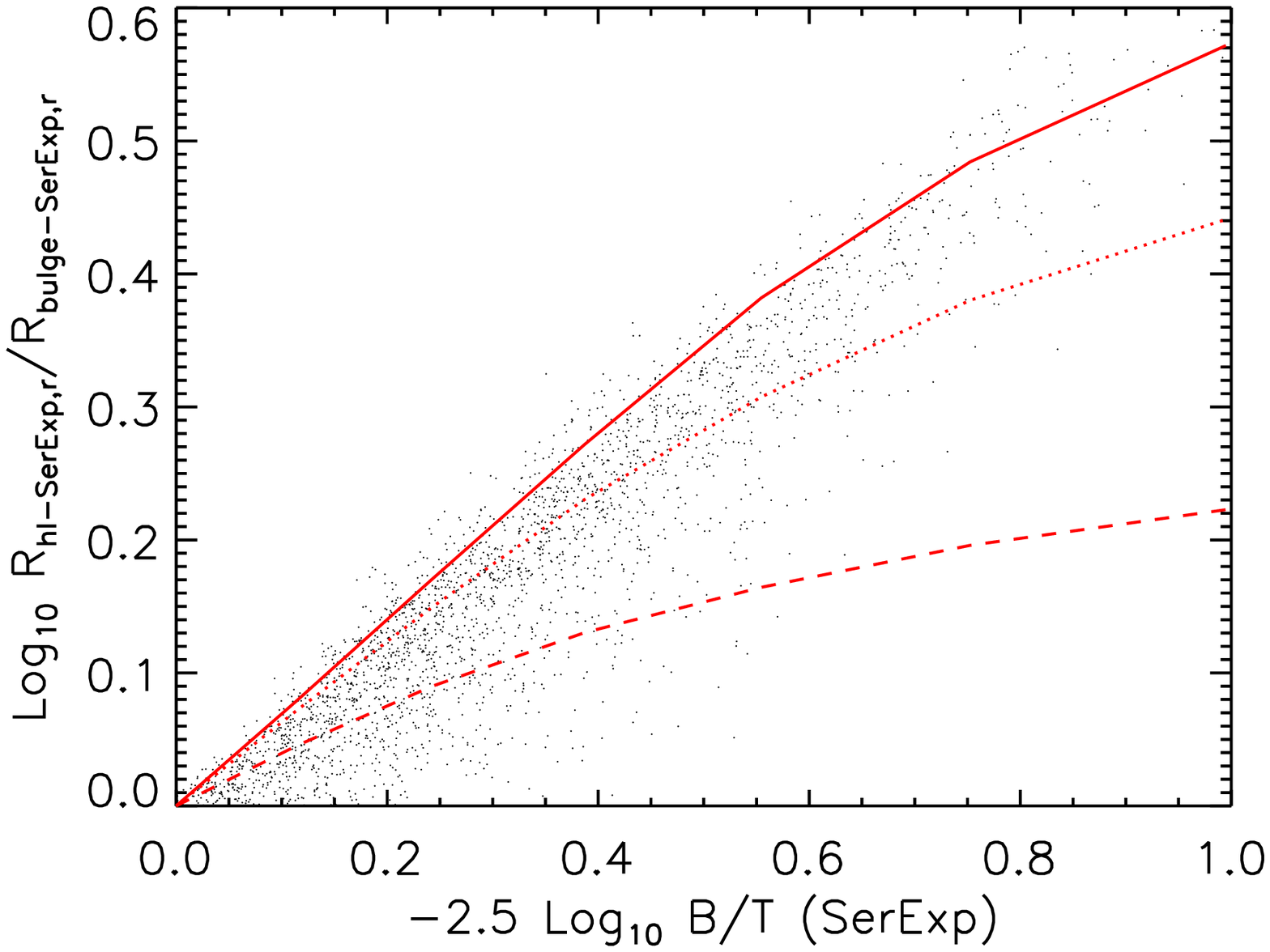}
 \includegraphics[width = \hsize]{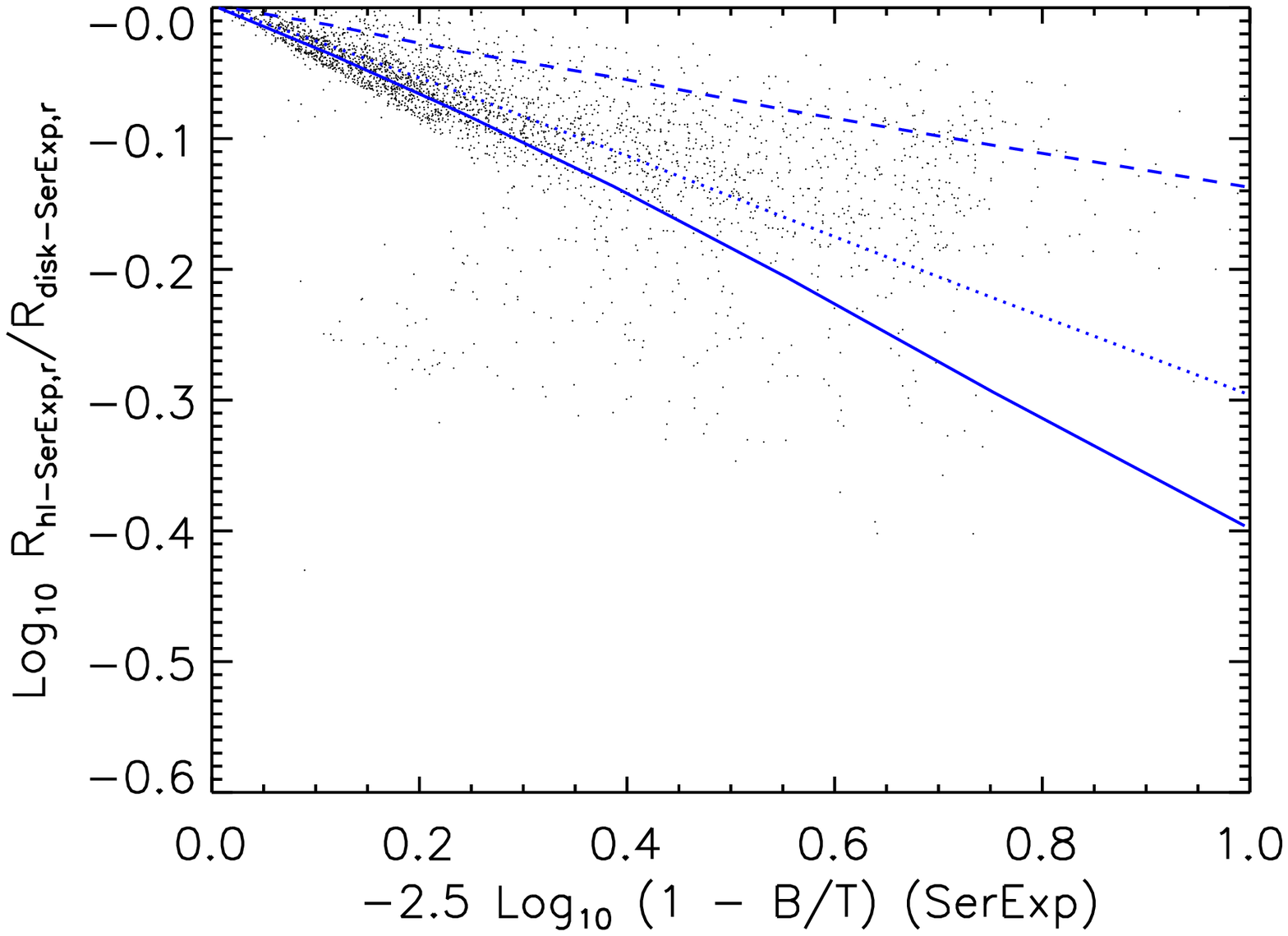}
 \caption{Correlation between $R_{\rm hl}/R_{\rm bulge}$ and B/T for 
          early-types (top panel) and between $R_{\rm hl}/R_{\rm disk}$ 
          and B/T for late-types (bottom panel).
          Although only objects with $-21.5 > M_r > -22.5$ are shown, 
          we see qualitatively similar behaviour at other luminosities.
          Dashed, dotted and solid curves show 
          the expected scaling for $n=4$ bulges with exponential disks 
          having $R_{\rm disk}/R_{\rm bulge} = 2, 4$ and 6.}
 \label{BTRshifts}
\end{figure}

\begin{figure}
 \centering
 \includegraphics[width = \hsize]{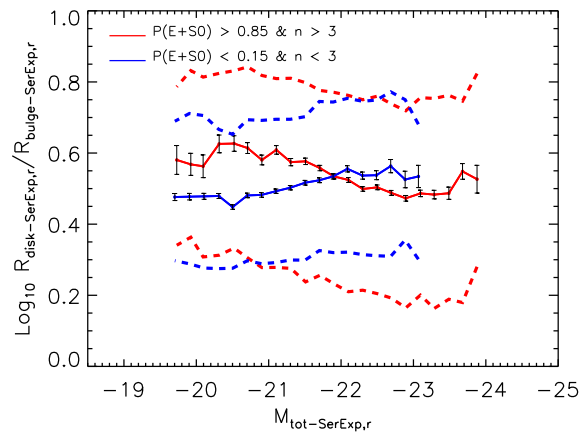}
 \caption{Dependence on total luminosity of the ratio of the size of 
          the extended component to that of the bulge for 
          early- (red, decreases slightly as luminosity increases) and 
          late-type (blue, approximately independent of luminosity) 
          galaxies.  Dashed lines show the range which encompasses 
          86\% of the data.}
 \label{L_RdRb}
\end{figure}

Whether or not these large scale lengths are physically reasonable is 
an open question, but we show in Appendix~\ref{btanalytic} that these 
tend to be objects for which the single Sersic fit returns large values 
of $n>5$; these extended second components do appear to be necessary to 
provide a good fit.  Indeed, fitting SerExp images with a single Sersic 
profile requires large values of $n$ if $0.4<$B/T$<0.7$ (bottom left 
panel of Figure~\ref{sersic}).

The bottom panel of Figure~\ref{BTRshifts} shows a similar analysis of 
the late-type sample:  
$R_{\rm hl}/R_{\rm disk}$ as a function of (1-B/T).  Most of the sample 
has B/T $<0.2$ for which $\log_{10}(R_{\rm hl}/R_{\rm disk})$ differs 
from zero by $-0.05$~dex or less.  
Although this is in the opposite direction to the shifts for 
early-types (as it should be), the resulting estimate of 
$R_{\rm  disk}/R_{\rm bulge}\sim 4$ is similar.  Of course, in this 
case, we expect $R_{\rm  disk}\gg R_{\rm bulge}$, so the value of 4 
does not require further explanation.  

We can, of course, directly measure the ratio $R_{\rm  disk}/R_{\rm bulge}$ 
for the objects in our early- and late-type samples.  
Figure~\ref{L_RdRb} shows that this ratio is indeed large, with only a 
weak dependence on $L$, and a somewhat larger scatter for early-types.  
The actual median value, $\sim 3-4$, is slightly smaller than the 
value we derived from the previous figure on the basis of the 
idealization that all galaxies were deVaucouleur bulges with 
exponential disks.  Hence, we conclude that the differences between 
the relations shown in Figure~\ref{LRbd} can be traced to the fact 
that the bulge component in a galaxy tends to be substantially smaller 
than the second component.  

Finally, recall from Figure~\ref{LRes0} that Es appear to be slightly 
larger than S0s of the same luminosity or mass.  We remade 
Figure~\ref{L_RdRb} separately for Es and S0s; although we do not 
show this, the S0s lie slightly below the Es on this plot too.  
Thus, the small but significant trend shown in Figure~\ref{LRes0} 
results from the combined facts that the second component in Es is 
a larger multiple of the bulge size than it is in S0s, and Es (being 
more bulge-dominated by definition) have larger bulges anyway.  
This strongly suggests that the second component in S0s and Es is 
not a disk, since, if anything, one expects S0s to have larger 
$R_{\rm disk}/R_{\rm bulge}$.  
While it is tempting to conclude that this second component may be 
intercluster light, as is almost certainly the case for cDs 
(Oemler 1976; Schombert et al. 1986; Gonzalez et al. 2005) --
the larger ratios at lower $L$ perhaps indicating that this is 
more difficult to mask out for low $L$ satellites in clusters --
we do not have enough confidence in either our two-component 
decompositions or our BAC classifications to discuss ten percent 
effects.  Nevertheless we do believe this is an 
interesting question particularly because Huertas-Company et al. 
report that the size difference is more dramatic at $z\sim 1$, and 
it is not obvious that the intercluster light hypothesis is even 
qualitatively consistent with this evolution.  

\subsection{Scatter}
\begin{figure}
 \centering
 \includegraphics[width = \hsize]{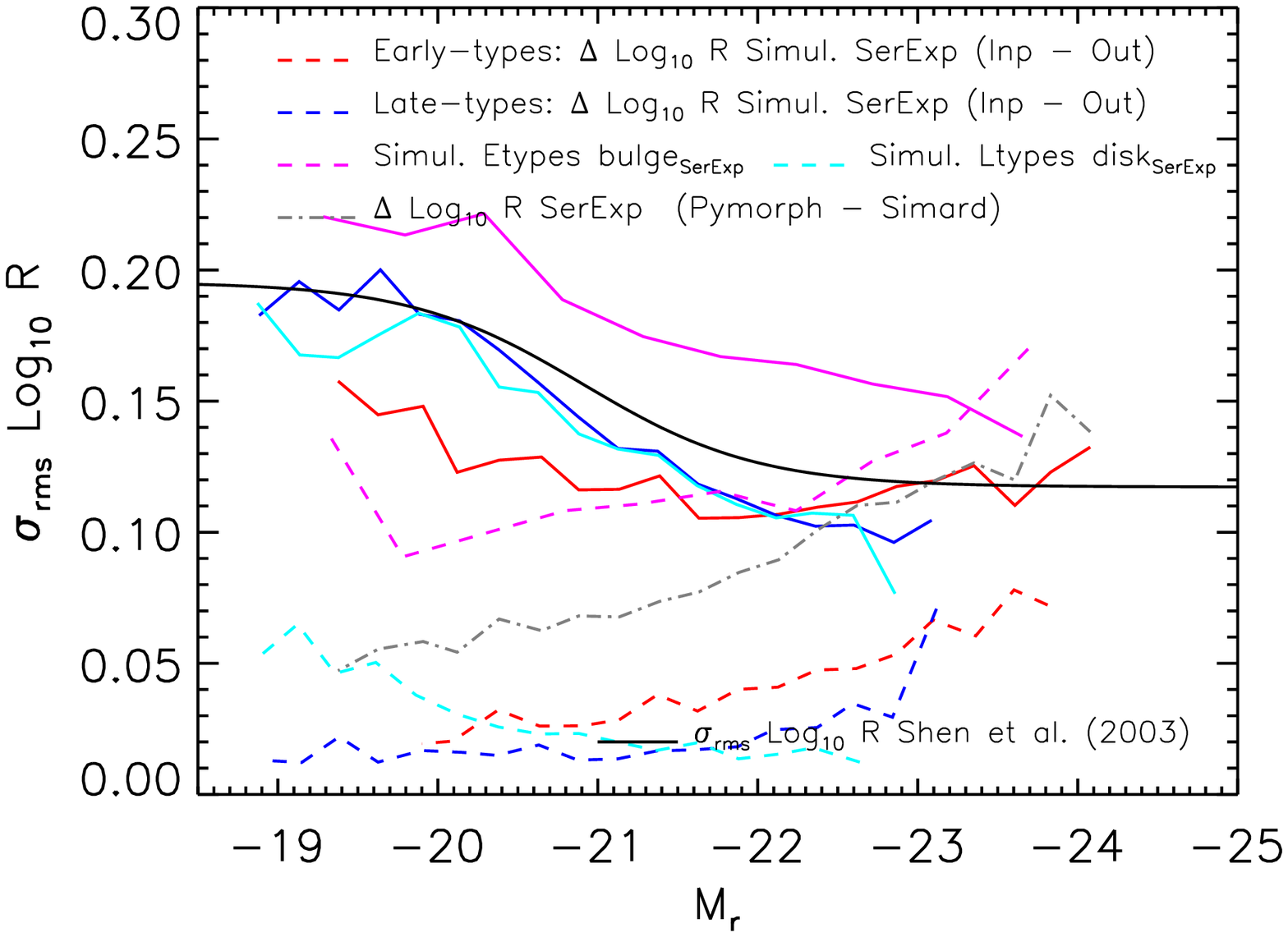}
 \includegraphics[width = \hsize]{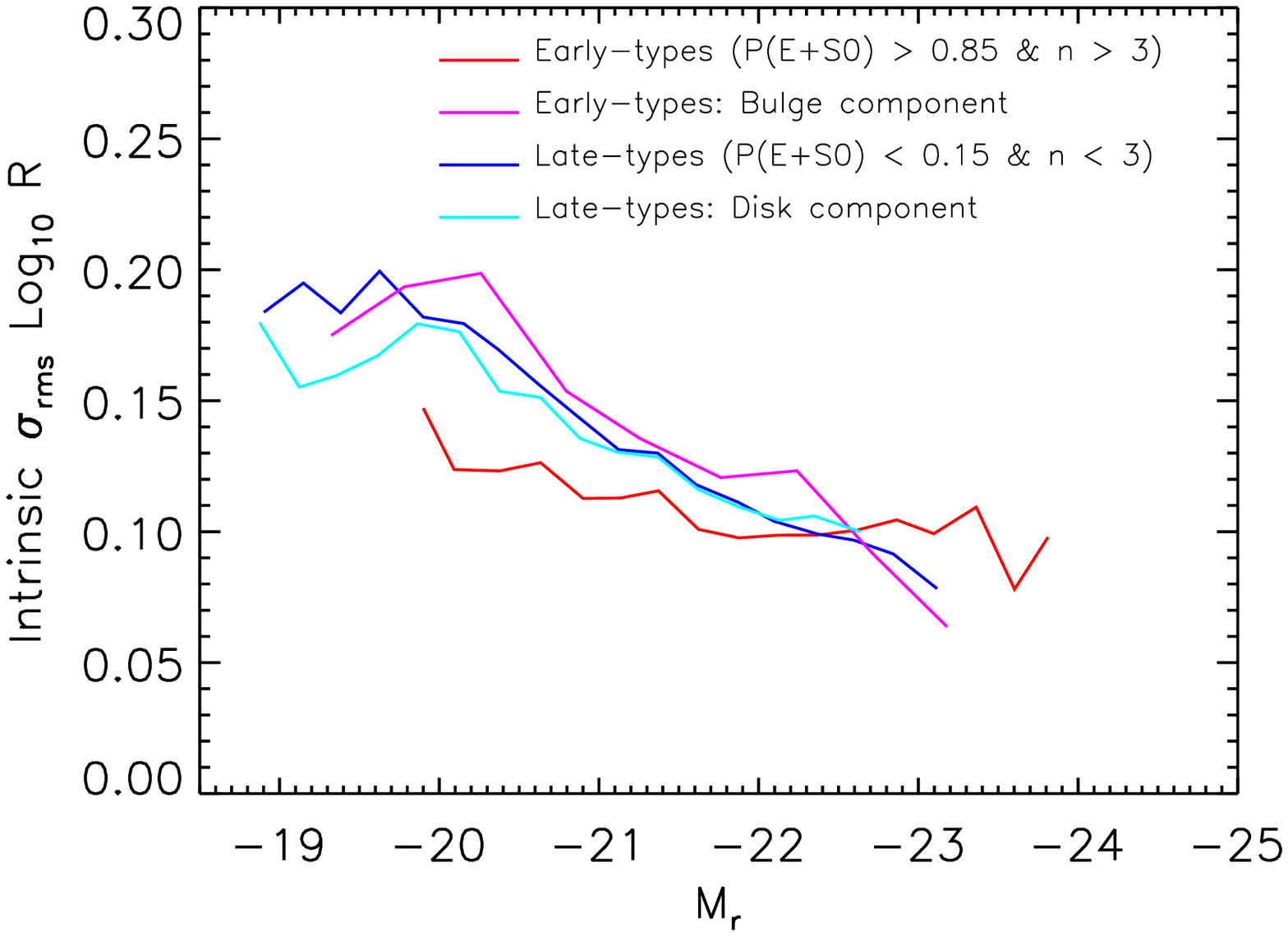}
 \caption{Top:  Observed (top) and intrinsic scatter (bottom) around 
          various $R-L$ relations as labelled (format similar to 
          Figure~\ref{LRscatter}).
          In all cases, our upper limit to the intrinsic scatter 
          decreases at large luminosities; this is particularly 
          dramatic for later-type galaxies. 
 }
 \label{intScatter}
\end{figure}

Before ending this section, Figure~\ref{intScatter} shows our 
estimate of the measured and intrinsic scatter around the mean 
$R-L$ relations defined by bulges and disks, and compares them 
with corresponding estimates for early-types and late-types.  
Notice that the measured scatter is substantially smaller around 
the early-type relation than around any of the others.
Since we argued earlier that the Shen et al. (2003) early-type 
sample is contaminated by later-types, we believe this explains the 
difference between their results and ours in Figure~\ref{LRscatter}.
Note also that the scatter around the relation for bulges is 
substantially larger than for the others.

Our estimates of the intrinsic scatter (shown in the bottom panel) 
come from subtracting, in quadrature, the measurement errors seen 
in simulations (dashed lines) from the total scatter measured in 
the data (corresponding solid lines), following the method described 
in Section~\ref{scatter}.  For this reason, we are almost certainly 
overestimating the intrinsic scatter.  Nevertheless, it is interesting 
that for late-types, disks and bulges, our estimates indicate that 
the intrinsic scatter decreases at large luminosities.
For early-types this decrease is less dramatic, with the scatter 
perhaps even levelling out at large luminosities.  We believe these 
differences, along with the power-law nature of the bulge $R-L$ relation, 
will prove to be useful for improving our understanding of how massive 
galaxies assembled their mass (e.g. Shankar \& Bernardi 2009; 
Shankar et al. 2010).

\section{Summary}
\label{disc}

We used our automated image decomposition algorithm {\tt PyMorph} to study the effects of systematics in the size-luminosity relation of galaxies in the SDSS main sample (i.e. at $z\sim 0.1$) which arise from fitting different models to the images.  We argued that when fitting to a {\tt SerExp} profile, {\tt PyMorph} returns more physically reasonable results than does the algorithm of Simard et al. (2011) (e.g. Figures~\ref{LRzn} and~\ref{LRzR} and related discussion).  

We presented a novel diagnostic of whether or not the surface brightness profiles of galaxies are better thought of as having one or two components.  The method works by fitting a number of single and two-component models to the image, and then studying the distribution in the $n$-B/T plane defined by the Sersic index $n$ associated with the single component fit and the ratio B/T of bulge to total light in the two component fit.  The way SDSS galaxies populate this plane suggests that they are not single component Sersic systems.  Rather, their distribution in $n$-B/T is more similar to that expected of two-component systems, with a Sersic + exponential model faring somewhat better than the traditional deVaucouleurs bulge with exponential disk model (Figures~\ref{sersic} and~\ref{sdss}).  I.e., in bulge dominated systems, allowing $n\ne 4$ provides a significantly improved fit.  Indeed, we even find bulges with $n>4$ in the SDSS (Figure~\ref{simLumB}).  Our conclusion that the {\tt SerExp} model is preferred is consistent with a recent analysis of the MGC, indicating that at least half of the galaxies at $z\sim 0.1$ are two component {\tt SerExp} systems (Allen et al. 2006).  


Note that we do not conclude that, for relatively noisy datasets such as the SDSS, two component models provide statistically better fits:  e.g., Meert et al. (2013) have shown that for a large fraction of our DR7 sample, $\chi^2$ per degree of freedom is not much better than it is for one component fits.  (E.g., Figure~\ref{select2} shows an example of a galaxy for which the two fits are indistinguishable, but the estimated half-light radii are very different.)  This is in agreement with Simard et al. (2011) for the SDSS, and analyses of, e.g., the MGC by Allen et al. (2006), and the GAMA dataset by Kelvin et al. (2012).  Rather, the point is that the sizes and luminosities returned by the two component SerExp model are less biased than those returned by fitting to a single Sersic or deVaucouleur profile.

For objects brighter than $L_*$, the commonly adopted procedure -- of fitting a single Sersic profile to what is really a two-component SerExp system -- leads to biases (Figures~\ref{dataM} and~\ref{dataR}).  The half-light radius is increasingly overestimated as $n$ of the fitted single component increases; it is also overestimated around B/T$_{\rm SerExp}\sim 0.6$ (Figure~\ref{dataMnpbt}).  For such objects that are likely to be early-type, but have a significant exponential component, the assumption of a single Sersic component is particularly bad.  However, the net effect on the size-luminosity relation is small, except for the most luminous tail (Figure~\ref{LR}).

On the other hand, fitting a realistic model is necessary to obtain 
sensible estimates of the intrinsic scatter around the mean $R-L$ 
relation.  Having done this, we showed that the scatter in sizes 
correlates with velocity dispersion (Figure~\ref{LRfixS}), and 
the rms scatter decreases at large luminosity (Figure~\ref{LRscatter}), 
although for early-types it may level off to a constant value of about 
0.1~dex at large luminosities.  This should provide tight constraints 
on the nature and number of mergers required to assemble the most 
massive galaxies.  

We also studied how the $R-L$ relation depends on galaxy morphology.  
Our Figure~\ref{LRw} shows one of the first uses of Bayesian 
classifier-based weights in the estimation of the $R-L$ scaling 
relation for different morphologies (e.g. Aguerri et al. 2012).  
We found that, even if we allow for finer bins in morphology, there 
seem to be only two fundamental $R-L$ relations, both of which are 
slightly but statistically significantly curved (Figures~\ref{LRes}, 
\ref{LRw} and~\ref{LRf07} and Tables~\ref{LRfits} and~\ref{MsRfits}).  

Of course, a closer inspection does reveal subtle dependences on 
morphology.  Amongst early-types, S0s tend to be about 0.06~dex 
smaller than Es of the same luminosity (Figure~\ref{LRes0}).  
This difference is smaller than the $\sim 40\%$ reported by 
Huertas-Company et al. (2012) at $z\sim 1$.  
We argued that this subtle difference argues against interpretting 
the second component in SerExp fits as a disk.  It is particularly 
interesting in view of the fact that the two types show very different 
trends as a function of age (Bernardi et al. 2010), so we expect that 
it, and its evolution, should yield interesting new constraints on 
models of how early-type galaxies assembled their stellar mass.  
Similarly, amongst late-types, faint Sbs tend to be $\sim 0.1$~dex 
smaller than Scds of the same luminosity, but these differences 
decrease as luminosity increases.  

Our two-component fits allowed us to study the $R-L$ relations for the 
bulges of early-types and disks of late types.  Although the $R-L$ relations 
for the total light in early- and late-types are curved (Figure~\ref{LRes}), 
the relation defined by the bulges in bulge dominated (typically early-type) 
galaxies is remarkably straight:
 $\langle R_{\rm bulge}|L_{\rm bulge}\rangle\propto L_{\rm bulge}^{0.85}$
(Figure~\ref{LRbd}).  The relation for disks of disk-dominated galaxies runs parallel to the $R-L$ relation for late type galaxies, being offset upwards 
by about 0.1~dex.  For disks, this curvature 
is so pronounced that, at the faint, low mass end, there is almost 
no correlation between $R$ and $L$ or $M_*$ (Figures~\ref{LRf07} 
and~\ref{LRbd}).
We argued that, both for early and late type galaxies, these differences 
arise because 
{\tt PyMorph} uses second-components for which the half light radius is 
$\sim 3-4$ times larger than that of the central bulge 
(Figures~\ref{BTRshifts}, \ref{L_RdRb} and Appendix~\ref{btanalytic}).  
It is not clear if for early-types this is physically reasonable -- but 
extended second components are clearly necessary for the {\tt SerExp} 
fits (Figures~\ref{select1} and~\ref{select2}).  

The two mass scales, $M_* \sim 3\times 10^{10}M_\odot$ and 
$M_* \sim 2\times 10^{11}M_\odot$, previously identified by 
Bernardi et al. (2011a,b), are clearly visible in our $R-L$ 
relations (e.g. Figure~\ref{MsRserexp}).  
For early-types, the former, is, among other things, the mass scale 
at which galaxies are maximally dense.  Below this scale the $R-M_*$ 
relation curves upwards with respect to the power law which best 
describes the full range of $M_*$ (Figures~\ref{LR} and~\ref{LRw}).  
Bernardi et al. suggest that this is because the disk component 
becomes more significant at these low masses.  
The larger mass scale ($M_* \sim 2\times 10^{11}M_\odot$) is where the $R-L$ 
relation of early-types curves upwards with respect to the power law 
which best describes the full range of $M_*$ (Figure~\ref{MsRserexp}).  

Whether or not these scales are associated with the onset of different 
physics is the subject of ongoing debate.  E.g. Graham \& Worley (2008) 
have pointed out that curvature in the $R-L$ relation may arise as a 
consequence of linear $n-L$ and central surface brightness-$L$ relations; 
the linearity of these would not be suggestive of changing physics along 
the relation.  Our single Sersic-based fits do indeed yield pure power 
laws for these relations, at least for Early-types above 
$M_* = 3\times 10^{10}M_\odot$.  But they are not pure power laws for 
our SerExp reductions, which we believe we have demonstrated are more 
appropriate.  

Bernardi et al. attribute the change at $M_* \sim 2\times 10^{11}M_\odot$ 
to a change in the assembly histories -- to ones in which major dry 
mergers become important.  So it is interesting that it is at this 
mass scale that the bulge and total $R-M_*$ relations become the same, 
despite being very different at smaller masses (Figure~\ref{LRbd}).  
This is particularly remarkable in light of recent work showing that 
early-types below this mass scale tend to be fast rotators 
(Cappellari et al. 2012).  It may be that our {\tt SerExp} bulge-disk 
decompositions of the images are reflecting this change in the 
kinematics.  Although running {\tt PyMorph} on the images in other SDSS 
bands would allow us to determine colors and color gradients of the 
individual components -- perhaps providing further insight -- this is 
beyond the scope of this work.  

Our analysis indicates that these same two mass scales are also 
significant for late-type galaxies.  At $M_* < 3\times 10^{10}M_\odot$, 
the $R-M_*$ relation for late-types (and their disks) flattens 
significantly (Figures~\ref{LRw}, \ref{LRf07} and~\ref{LRbd}); 
and $M_* = 2\times 10^{11}M_\odot$ marks the mass scale above which 
there are almost no late-types (Figures~\ref{LRf07} and~\ref{LRbd}).

Given the large differences between the relation for bulges 
and that for early-types at smaller masses and luminosities 
(Figure~\ref{LRbd}), it is remarkable that the slope of the $R-L$ 
relation for bulges is essentially the same as that for early-types 
within a fixed bin in velocity dispersion (Figure~\ref{LRfixS}).  
In fact, at fixed $L_{\rm bulge}$ and $\sigma$, the mean size scales as 
$R_{\rm bulge}\propto L_{\rm bulge}/\sigma$ (Figure~\ref{LbRbfixS}).  
Although the scaling proportional to $L_{\rm bulge}$ is consistent with 
the virial theorem (if $L_{\rm bulge}\propto M_{\rm bulge}$), the dependence 
on $\sigma$ is not.  Why this should be so is an open question.

Finally, we find that the scatter around the mean $R-L$ relation 
decreases as $L$ increases (and similarly for $R-M_*$), except 
for early-types, where it may flatten at 0.1~dex (Figure~\ref{intScatter}).  
We expect this to provide a useful probe of how massive galaxies 
assembled their mass (e.g. Shankar \& Bernardi 2009; 
Shankar et al. 2010).

\section*{Acknowledgments}
This work was supported in part by NASA grant ADP/NNX09AD02G
and NSF/0908242.
MB and RKS are grateful to the Meudon Observatory for its 
hospitality during June 2011 and 2012.  
FS acknowledges support from a Marie Curie grant.

\appendix

\section{Systematic effects in the Simard et al reductions}\label{noSimard}

\begin{figure*}
 \centering
 \includegraphics[width = \hsize]{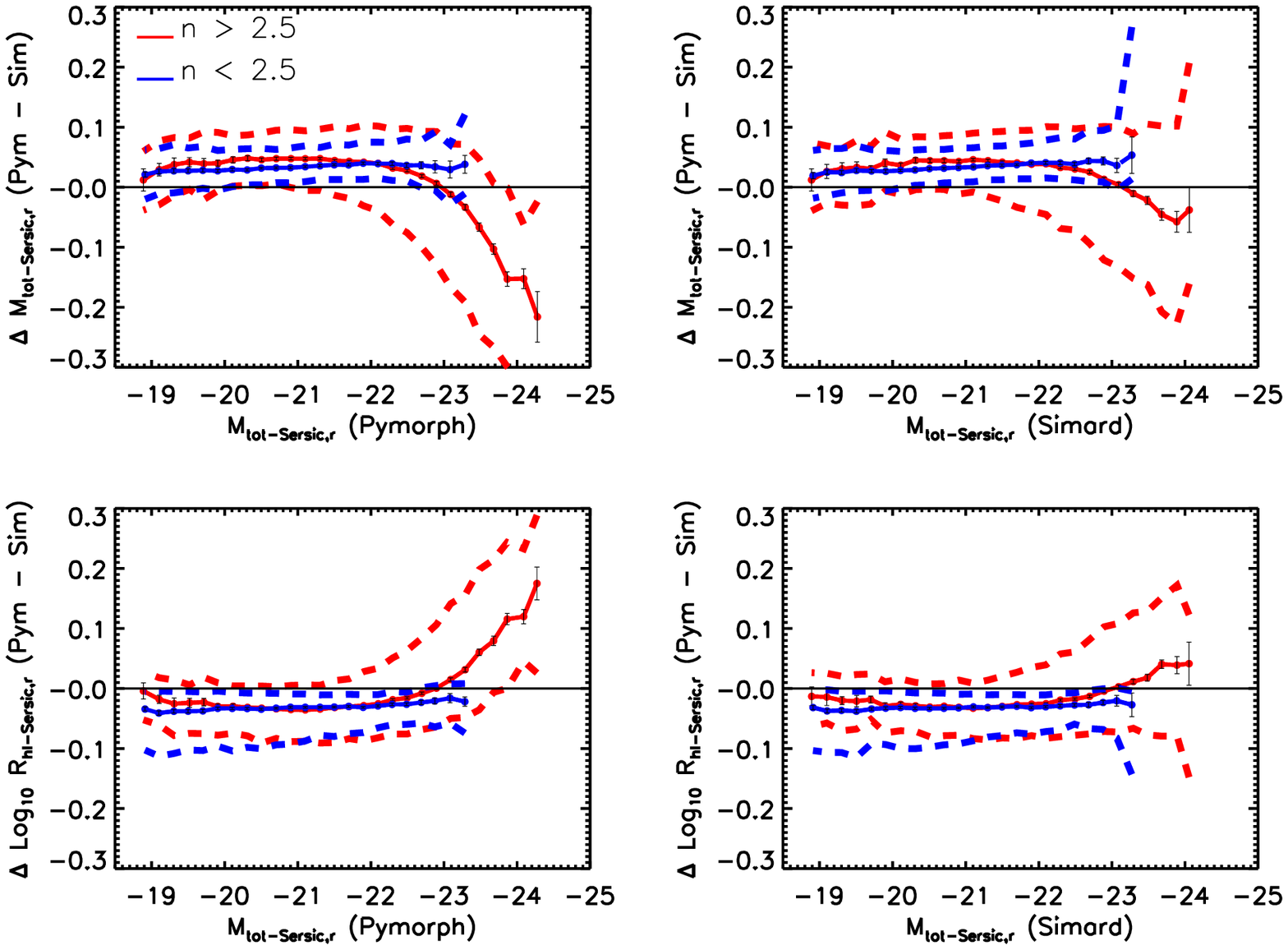}
 \caption{Differences between the S11 and PyMorph reductions tend to 
          be of order 0.04~mags fainter or 0.03~dex smaller in size, 
          and approximately the same for all Sersic indices, except 
          for $M_r<-22.5$ where {\tt PyMorph} tends to be bigger 
          and brighter if the Sersic index $n>2.5$ (i.e. if the galaxy 
          is more likely to early-type).  Solid lines and error bars 
          show the median and the error on it; dashed lines 
          show the range which encloses 86\% of the sample.}
 \label{DLRps}
\end{figure*}

The main text showed that the $R-L$ relation from single-Sersic fits 
using {\tt PyMorph} is in reasonably good agreement with that based 
on parameters from Simard et al. (2011).  However, Figure~\ref{DLRps} 
shows that, although the two algorithms return similar sizes and 
luminosities for objects with $n<2.5$ ({\tt PyMorph} is about 
0.03~dex smaller and 0.03~mags fainter), the {\tt PyMorph} sizes and 
luminosities are systematically larger at large $M_{\rm tot}$.  This 
bias for the biggest galaxies is particularly evident when shown as 
a function of $M_{\rm PyMorph}$.  

\begin{figure*}
 \centering
 \includegraphics[width = \hsize]{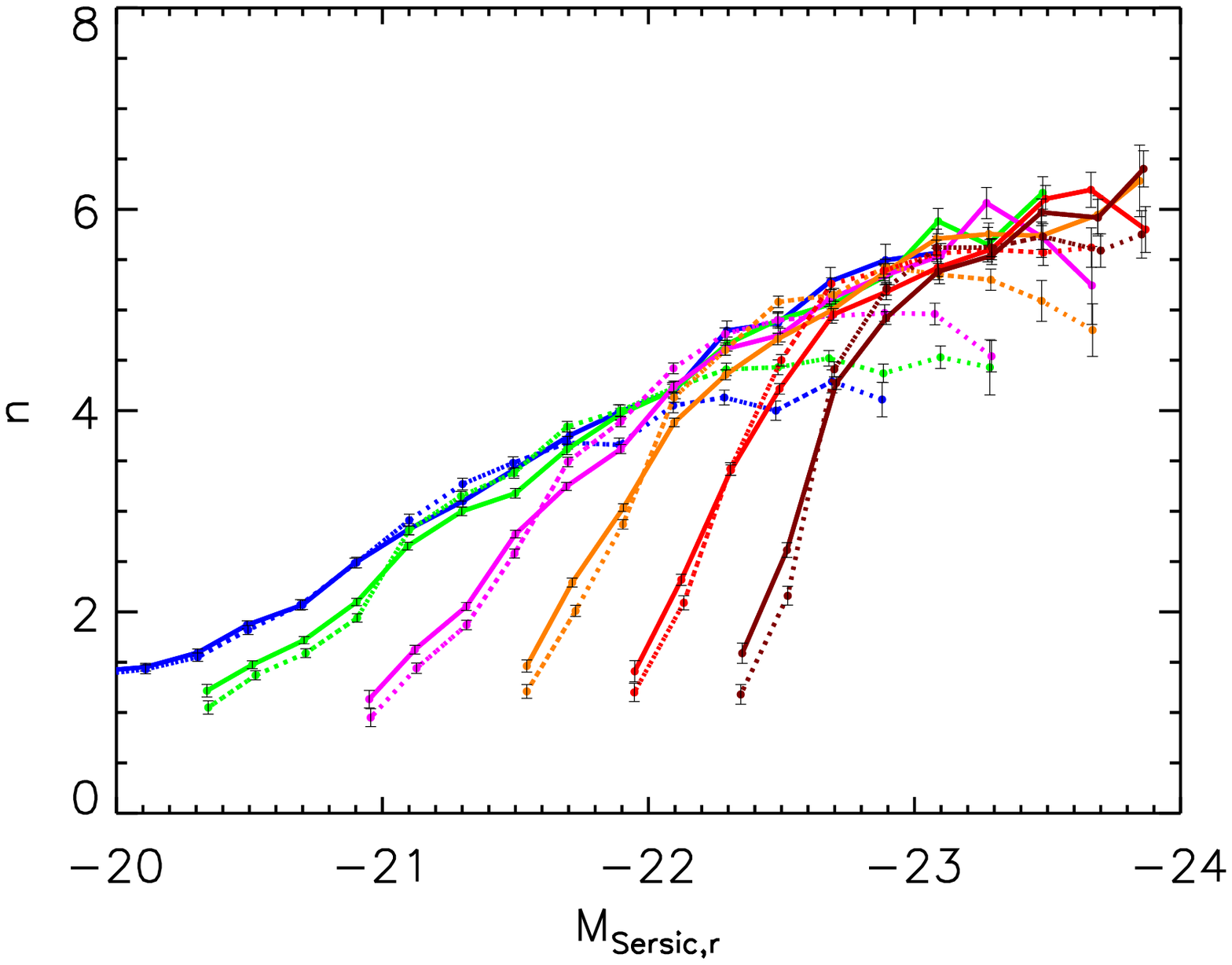}
 \caption{Our determination of the $n-L$ relation (symbols connected 
          by solid lines) shows little or no redshift dependence 
          (curves show results for adjacent redshift bins of width 0.02, 
          starting from a bin centered on $z=0.06$; 
          the sample is magnitude limited so the lower $L$ objects 
          are missing from the higher $z$ bins).  
          The sudden drop in $n$ at the faint end of each redshift 
          sample is due to the bimodal distribution in $n$ at each $L$; 
          it has nothing to do with evolution.  Except for this, 
          our determination shows little or no redshift dependence. 
          In contrast, at high luminosities, the Simard et al. 
          reductions (dotted curves) lead to systematically smaller 
          $n$ as redshift decreases.  We believe the implied evolution 
          is unphysical, so conclude that the Simard et al. reductions 
          are systematically biased.}
 \label{LRzn}
\end{figure*}

\begin{figure*}
 \centering
 \includegraphics[width = \hsize]{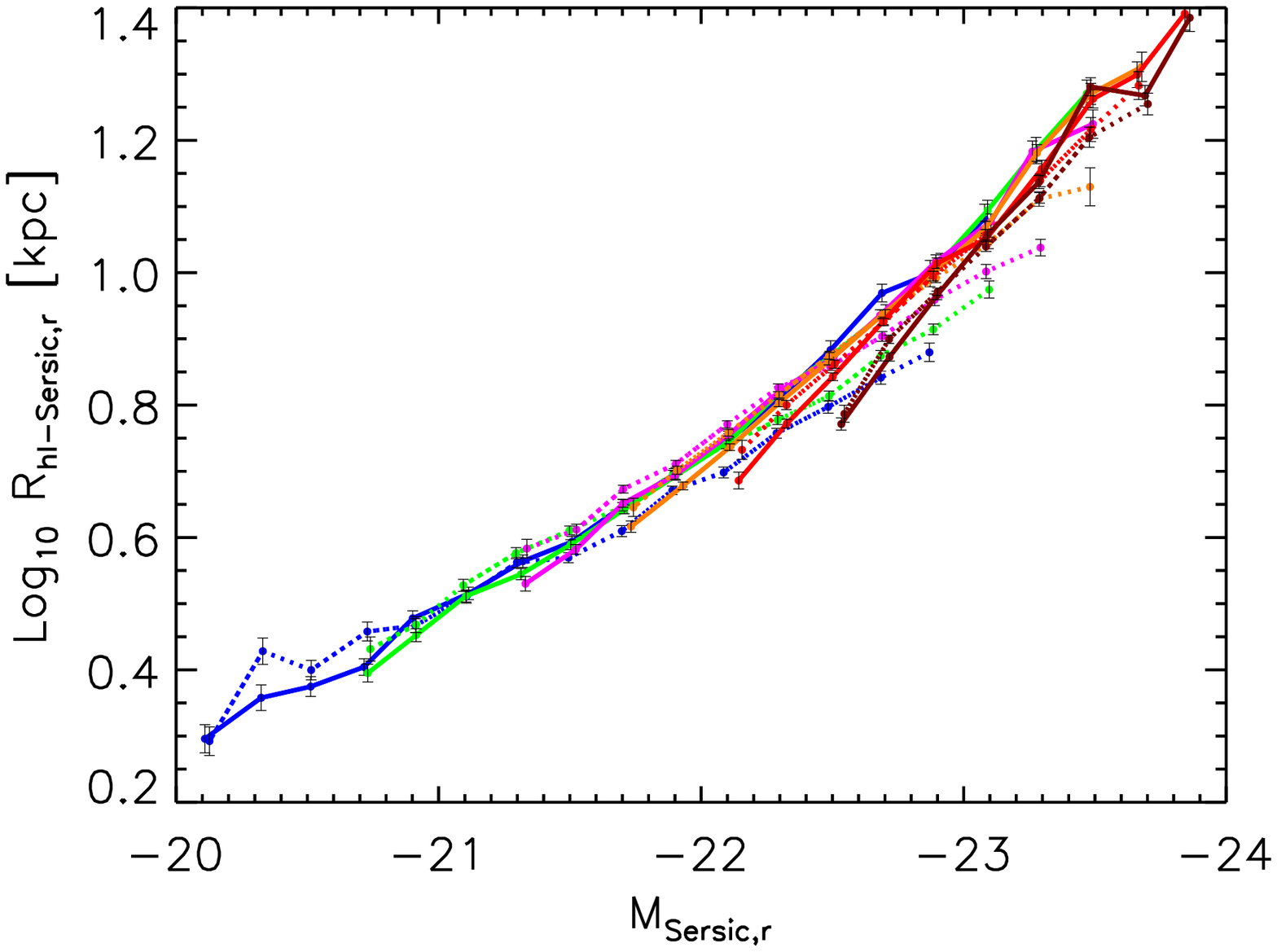}
 \caption{Similar to Figure~\ref{LRzn}, but now for the $R-L$ relation: 
          little or no redshift dependence is seen in our sample (symbols 
          connected by solid lines); 
          in contrast, at high luminosities, the Simard et al. reductions 
          (symbols connected by dashed lines) imply evolution towards 
          smaller sizes as redshift decreases.  We believe this implied 
          evolution is unphysical, so conclude that the Simard et al. 
          reductions suffer from systematic biases.  }
 \label{LRzR}
\end{figure*}

Since the $R-L$ relation of the largest galaxies is particularly 
timely, we would like to determine which reductions are more reliable.  
Figures~\ref{LRzn} and \ref{LRzR} show that the S11 reductions
indicate substantial recent evolution toward smaller $n$ and $R$ at 
fixed $L$ especially at larger $L$. We believe this evolution is unphysical, 
so conclude that the S11 reductions suffer from systematic biases.  
No such evolution is seen in the {\tt PyMorph} reductions, so we use 
them exclusively in the main text.

\section{Correlation between $R_{\rm bulge}/R_h$ or $R_{\rm disk}/R_h$ and B/T}\label{btanalytic}

\begin{figure}
 \centering
 \includegraphics[width = \hsize]{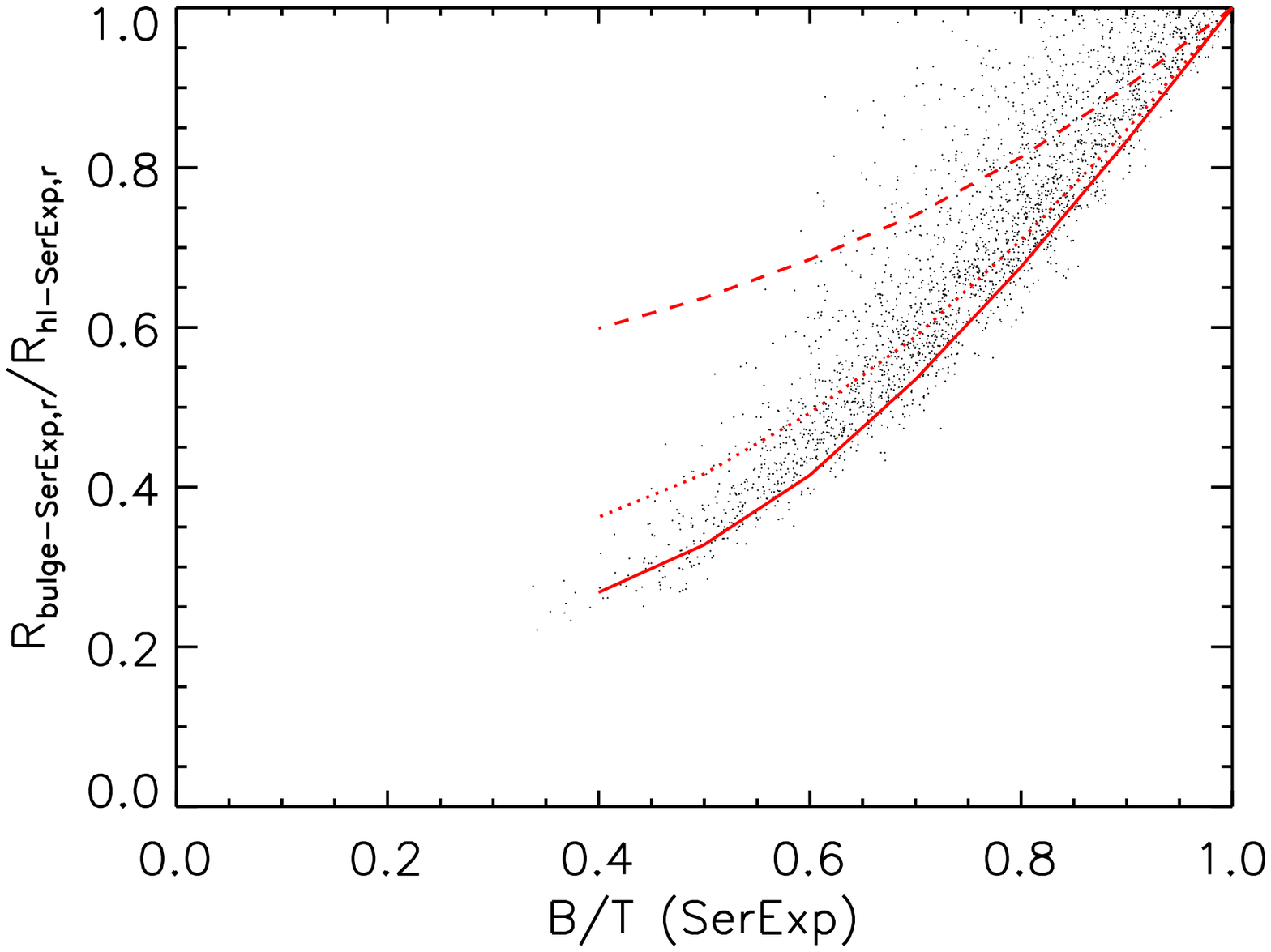}
 \includegraphics[width = \hsize]{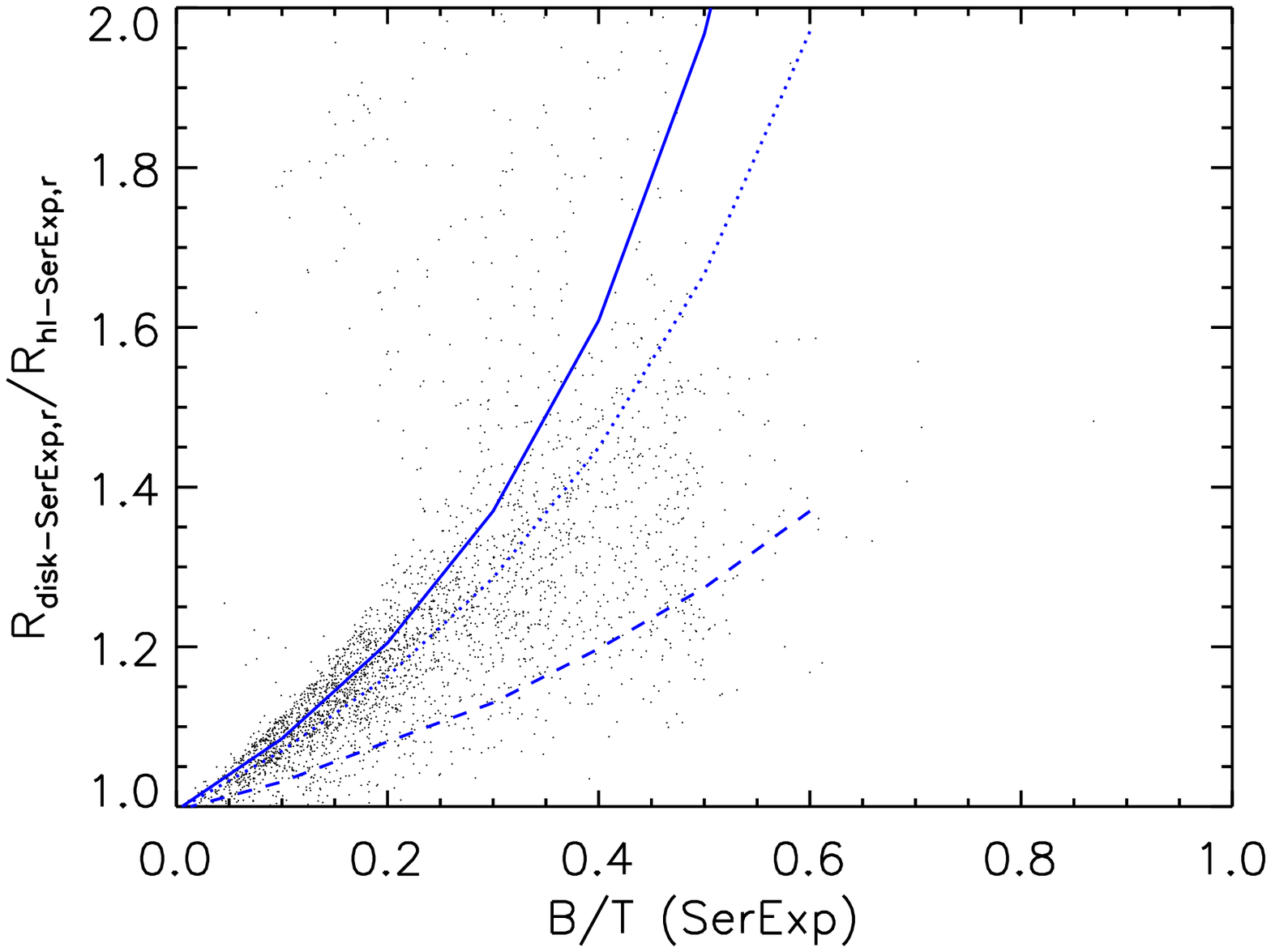}
 \caption{Correlation between $R_{\rm bulge}/R_{\rm hl}$ and B/T for 
          early-types (top panel) and between $R_{\rm disk}/R_{\rm hl}$ 
          and B/T for late-types (bottom panel).
          Although only objects with $-21.5 > M_r > -22.5$ are shown, 
          we see qualitatively similar behaviour at other luminosities.  
          Dashed, dotted and solid curves show the expected 
          correlation for $n=4$ bulges with
          $R_{\rm disk}/R_{\rm bulge} = 2, 4$ and 6.}
 \label{BTRfixmag}
\end{figure}

Figure~\ref{BTRshifts} of the main text showed a correlation between 
the bulge(disk) to total size and B/T.  This Appendix shows why it arises.  

The Sersic profile is 
\begin{equation}
 I(r) = I_n\,\exp[-(r/r_n)^{1/n}]
\end{equation} 
so the ratio of the light within $r$ to the total light in the profile is 
\begin{equation}
 \frac{L_n(<r)}{L_n} = \frac{\int_0^{r/r_n} dx\, x\,\exp(-x^{1/n})}
                           {\int_0^\infty dx\, x\,\exp(-x^{1/n})}
                     = \gamma_{2n}\left[0,(r/r_n)^{1/n}\right]
\end{equation} 
where $\gamma_{2n}$ is the incomplete Gamma function.  (For integer $n$, 
it can be written in terms of $\exp[-(r/r_n)^{1/n}]$ times a polynomial 
of degree $2n-1$ in $(r/r_n)^{1/n}$.)  

Therefore, the half-light radius $r_h$ of a SerExp profile satisfies 
\begin{equation}
 \frac{1}{2} = \frac{\rm B}{\rm T}\,
  \gamma_{2n}\left[0,\left(\frac{r_h}{r_n}\right)^{1/n}\right]
             + \left(1-\frac{\rm B}{\rm T}\right)\,
   \gamma_2\left[0,\left(\frac{r_h}{r_1}\right)^{1/n}\right].
 \label{rhBT}
\end{equation} 
For a given B/T, the right hand side is a function of $r_h/r_n$ and 
$r_h/r_1 = (r_h/r_n)(r_n/r_1)$, so it defines a different curve for 
each $r_n/r_1$, where $r_n  = r_{\rm bulge}/(1.992n-0.327)$ and 
$r_1 = r_{\rm disk}/1.67$.  Note that the curves are independent of 
luminosity $L$; therefore $L$ dependence only enters if the 
distribution of  $r_n/r_1$ and/or  B/T depend on $L$.

Figure~\ref{BTRfixmag} shows $R_{\rm bulge}/R_{\rm hl}$ as a function
of B/T for the early- (top) and late-type (bottom) samples defined  
in the main text for galaxies with $-21.5 > M_r > -22.5$; results are 
similar at other luminosities.  The curves show the predicted relations 
(equation~\ref{rhBT}) for a deVaucouleur bulge ($n=4$) with 
exponential disk.  These depend on the ratio $R_{\rm  disk}/R_{\rm bulge}$, 
for which we have chosen 2, 4 and 6.  
We argue in the main text that, despite the widespread use of the term 
`bulge-disk' decomposition for two-component fits, for bulge dominated 
galaxies, the `disk' component is almost certainly not an inclined disk; 
rather, it is an extended second component which is required to fit 
the outer parts of the profile.  But we call this second component `disk' 
anyway.

The top panel shows a very strong correlation between 
$R_{\rm bulge}/R_{\rm hl}$ and B/T (at this fixed $M_r$) for the 
early-type sample.  Clearly, if 20\% of the light is in a disk 
component, then the size is affected by at least this fraction.  
The well-known correlation between $L$ and B/T, and the fact 
that early-types span a large range of B/T, means that the 
bulge and early-type size-luminosity relations can be quite 
different.  It is perhaps surprising that the half-light radius 
of the (second) disk component is typically more than 3-5 times 
larger than that of the bulge, particularly at B/T$< 0.7$.  
We argue in the main text that these large values of $R_d/R_b$ argue 
strongly against interpretting the more extended component 
as a flat exponential disk (not always viewed edge-on).  

The bottom panel shows $R_{\rm disk}/R_{\rm hl}$ and B/T for the 
late-type sample.  Most of the sample has B/T$ < 0.2$.  
Comparison with the smooth curves indicates that 
$R_{\rm  disk}/R_{\rm bulge}\sim 5$ for most of the sample.  
In this case we do expect the disks to be substantially larger 
than the bulges, so the results are sensible.  

Figure~\ref{BTRshifts} in the main text shows this same information 
in a different format, which allows for a more direct understanding 
of the impact this correlation has on the relations shown in 
Figure~\ref{LRbd}.  And Figure~\ref{L_RdRb} in the main text shows 
that $R_{\rm disk}/R_{\rm bulge}$ is indeed $\sim 3$.

\begin{figure*}
 \centering
 \includegraphics[width = 0.8\hsize]{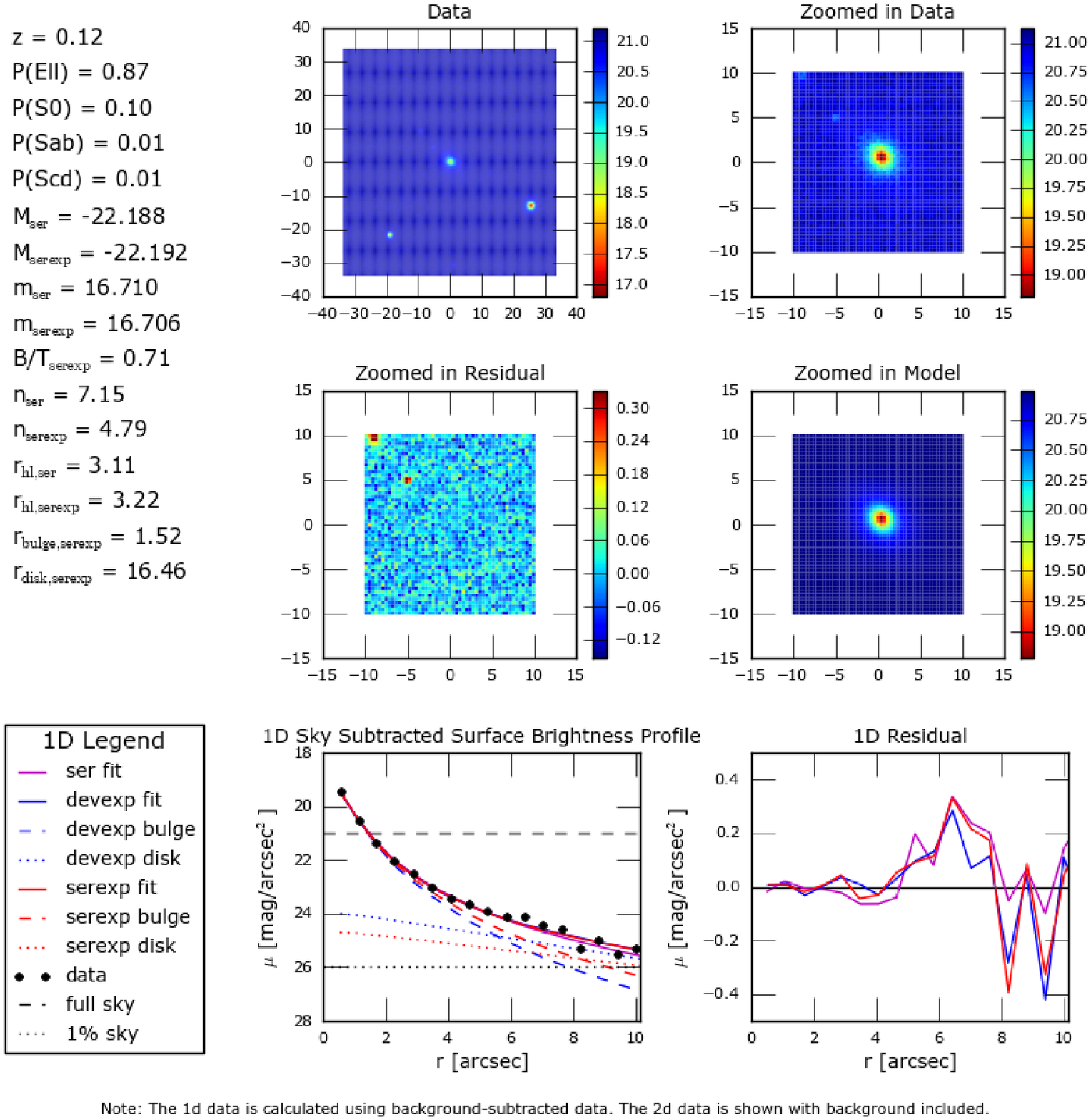}
 \caption{Example of an early-type galaxy with $M_r \sim -22$, 
 large $R_d/R_b\sim 10$ and B/T$\sim 0.7$.  Top left panel shows a 
 $\sim 20$ arcsec field centered on the object; top right panel provides 
 a closer look.  Middle right panel shows the best-fit {\tt SerExp} 
 model; middle left panel shows residuals from this fit.  
 Bottom left panel shows the one-dimensional surface brightness profile 
 (symbols), and our Sersic (solid magenta), {\tt deVExp} (solid blue) 
 and {\tt SerExp} (solid red) fits; 
 dotted and dashed curves show the corresponding disk and bulge 
 components.  Bottom right panel shows the associated residuals, 
 indicating that all three models fit similarly well.  
 Legend along the left shows the values of many quantities returned 
 by the fits, and other information, such as the BAC $p$(type), for 
 the object. }
 \label{select1}
\end{figure*}

\begin{figure*}
 \centering
 \includegraphics[width = 0.8\hsize]{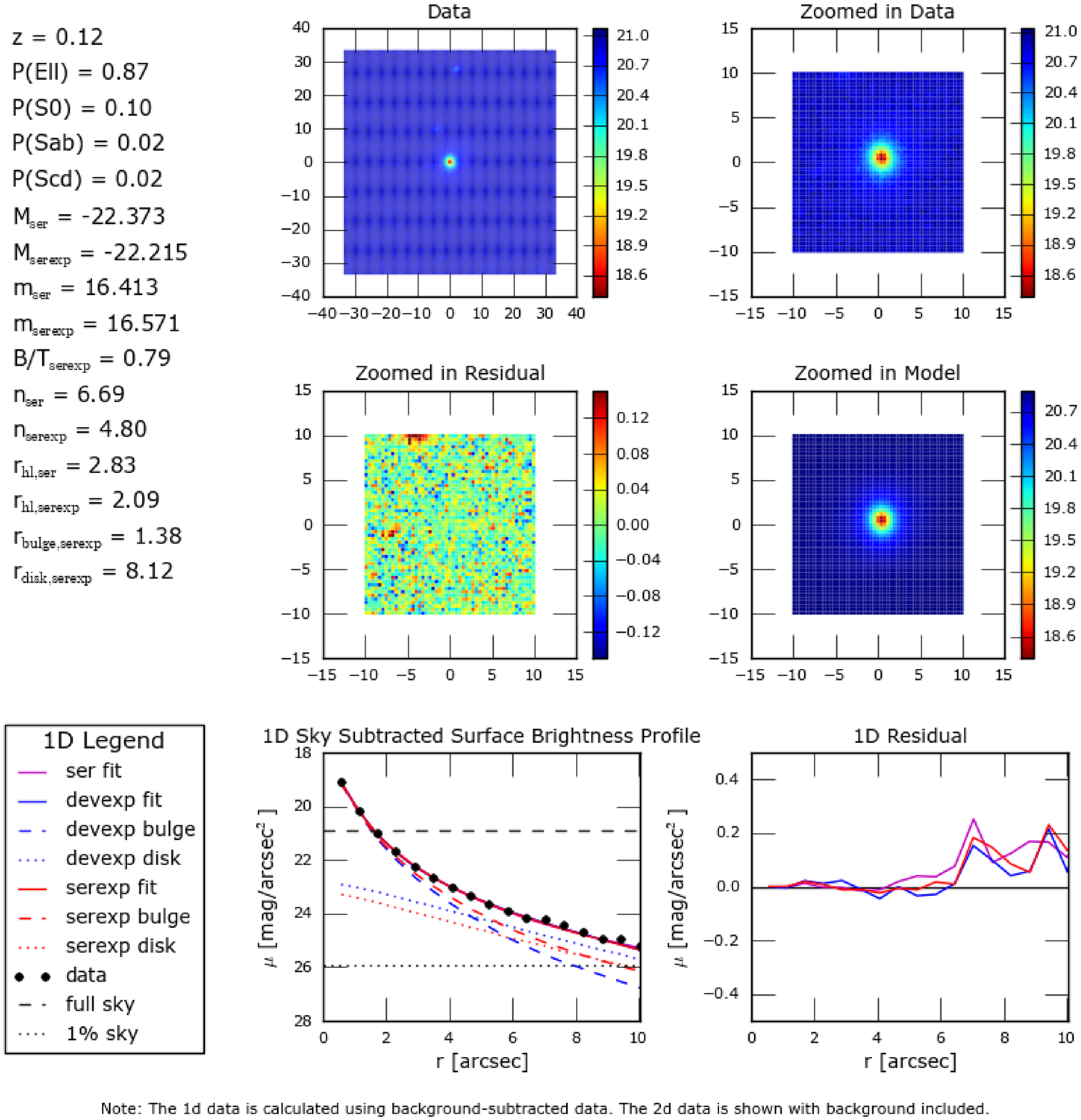}
 \caption{Same as previous figure, but for another early-type galaxy 
 selected at random from among those with the same $M_r$ and B/T range.  
 Note again that our Sersic, {\tt deVExp} and {\tt SerExp} fits all 
 describe the one-dimensional surface brightness profile rather similarly.  
 We argue in the text that this comes at the price of biased estimates 
 of the total light and half-light radius, with the {\tt SerExp} parameters 
 being the least biased.}
 \label{select2}
\end{figure*}

To address the question of large $R_d/R_b$ in our early-type sample, 
particularly at smaller B/T, Figures~\ref{select1} and~\ref{select2} 
show two examples.  The format in both cases is the same: 
The top left panel shows a $\sim 20$ arcsec field centered on the 
object, to get an idea of whether or not the object is in a crowded 
field.  The top right panel provides a closer look at the object.  
The panel just below it shows the best-fit SerExp model, and the 
middle left panel shows residuals from this fit.  The bottom left 
panel shows the one-dimensional surface brightness profile, 
and our Sersic (solid magenta), deVExp (solid blue) and SerExp (solid red) 
fits; dotted  and dashed curves show the corresponding disk and 
bulge components.  Bottom right panel shows the associated residuals.  
The legend along the left shows the values of many quantities returned 
by the fits, and other information, such as the BAC $p$(type), for the 
object.  

The object in Figure~\ref{select1}, B/T$ = 0.71$ and $R_d/R_b \sim 10 $ 
is very likely to be an elliptical:  $p$(E)$=0.87$.  
The Sersic and SerExp fits return almost the same magnitudes 
($M_r\sim -22.2$) and total half light radii ($\sim 3.15''$).  
However, $n=7.15$ for the single Sersic fit, but $n=4.79$ for the 
SerExp bulge.  For the SerExp, as for the deVExp fits, the second 
component is clearly necessary.  The $\chi^2_{dof}$ values for these 
fits are similar.  

In Figure~\ref{select2}, the Sersic magnitudes and half light radii 
are slightly larger, but otherwise the qualitative trends are the 
same:  the single Sersic fit requires large $n$, and the second 
component in the SerExp fit clearly requires large $R_d/R_b\sim 6$.  

\clearpage

\begin{thebibliography}{99}

\bibitem[]{} Abazajian, et al. 2009, ApJS, 182, 543

\bibitem[]{} Aguerri, J. A. L., Huertas-Company, M., S{\'a}nchez Almeida, J. \& Mu{\~n}oz-Tu{\~n}{\'o}n, C. 2012, A\&A, 540, 136 

\bibitem[{Allen et al.}{2006}]{mgc06} Allen, P. D., Driver, S. P., Graham, A. W., Cameron, E., Liske, J. \& de Propris, R. 2006, MNRAS, 371, 2

\bibitem[]{} Bernardi, M., et al. 2003a, AJ, 125, 1849

\bibitem[]{} Bernardi, M., et al. 2003b, AJ, 125, 1866

\bibitem[Bernardi et al. 2007]{Bernardi07}
 Bernardi M., Hyde J.~B., Sheth R.~K., Miller C.~J., Nichol R.~C.\  2007, AJ, 133, 1741 

\bibitem[]{} Bernardi, M. 2009, MNRAS, 395, 1491

\bibitem[{Bernardi et al. 2010}]{alltypes}
Bernardi M., Shankar, F., Hyde, J. B., Mei, S., Marulli, F. \& Sheth, R. K. 2010, MNRAS, 404, 2087

\bibitem[]{} Bernardi, M., Roche, N., Shankar, F. \& Sheth, R. K. 2011a, MNRAS, 412, L6

\bibitem[]{} Bernardi, M., Roche, N., Shankar, F. \& Sheth, R. K. 2011b, MNRAS, 412, 684

\bibitem[]{} Binggeli, B., Jerjen, H. 1998, A\&A, 333, 17

\bibitem[]{} Binggeli, B., Sandage, A., Tarenghi, M. 1984, AJ, 89, 64

\bibitem[]{} Blanton, M. R. et al. 2003, ApJ, 594, 186

\bibitem[]{} Bruce, V. A. et al. 2012, MNRAS, submitted (arXiv1206.4322)

\bibitem[]{} Cappellari, M. et al. 2012, MNRAS, submitted (arXiv:1208.3523)

\bibitem[]{} Capaccioli, M., Caon, N. 1991, MNRAS, 248, 523

\bibitem[]{} Ciotti L., Bertin G. 1999, A\&A, 352, 447

\bibitem[Cimatti et al. 2008]{Cimatti08}
 Cimatti, A., et al.\ 2008, A\&A, 482, 21

\bibitem[deVaucouleurs 1948]{deV48}
 deVaucouleurs, G. 1948, Annales d'Astrophysique, 11, 247

\bibitem[]{atlas11} Emsellem, E. et al. 2011, MNRAS, 414, 888

\bibitem[Fukugita et al.(2007)]{F07}
 Fukugita M., et al., 2007, AJ, 134, 579

\bibitem[Gadotti(2008)]{g08} 
 Gadotti D. A., 2008, MNRAS, 384, 420

\bibitem[Gadotti(2009)]{g09}
 Gadotti D. A., 2009, MNRAS, 393, 1531

\bibitem[]{} Gonzalez, A. H., Zabludoff, A. I. \& Zaritsky, D. 2005, ApJ, 618, 195

\bibitem[]{} Graham, A., 2013, in Planets, Stars and Stellar Systems Vol. 6, by Oswalt, Terry D.; Keel, William C., ISBN 978-94-007-5608-3. Springer Science+Business Media Dordrecht, p. 91 (arXiv:1108.0997)

\bibitem[]{} Graham, A.W., Worley, C.C. 2008, MNRAS, 388, 1708

\bibitem[]{} Graham, A.W., Erwin, P., Trujillo, I., Asensio Ramos, A., Worley, C.C. 2003, AJ, 125, 2951

\bibitem[]{} Hyde, J. B. \& Bernardi, M. 2009, MNRAS, 394, 1978

\bibitem[]{marc+11} Huertas-Company, M., Aguerri, J. A. L, Bernardi, M., Mei, S. \& S{\'a}nchez Almeida, J. 2011, A\&A, 525, 157

\bibitem[]{marc+12} Huertas-Company M., Mei S., Shankar F., Delaye L., Raichoor A., Covone G., Finoguenov A., Kneib J.-P., Le F\'evre O., Povic M., 2012, MNRAS, in press (arXiv:1207.5793)

\bibitem[]{} Johnston, E. J., Arag{\'o}n-Salamanca, A., Merrifield, M. R. \& Bedregal, A. G. 2012, MNRAS, in press (arXiv:1202.6064)

\bibitem[]{} Meert, A., Vikram, V. \& Bernardi, M. 2013, MNRAS, 433, 1344

\bibitem[]{} Nair P., Abraham R. G., 2010, ApJS, 186, 427

\bibitem[]{} Nair P., van den Bergh, S. \& Abraham, R. G. 2011, ApJL, 734, 1 

\bibitem[]{} Oemler A., Jr., 1976, ApJ, 209, 693

\bibitem[]{} Saglia, R. P. et al. 2010, A\&A, 524, 6

\bibitem[]{} S\'ersic, J. L. 1968, Atlas de Galaxias Australes, Observatorio Astron\'omico de C\'ordoba

\bibitem[]{} Schombert, J.M. 1986, ApJS, 60, 603

\bibitem[]{} Shankar F., Bernardi M., 2009, MNRAS, 396, L76

\bibitem[]{} Shankar F., Marulli F., Bernardi M., Dai X., Hyde J. B., Sheth R. K., 2010, MNRAS, 403, 117

\bibitem[]{} Shankar, F., Marulli, F., Bernardi, M., Mei, S., Meert, A. \& Vikram, V. 2012, MNRAS, in press (arXiv:1105.6043)

\bibitem[]{} Shen S., et al., 2003, MNRAS, 343, 978

\bibitem[]{} Sheth R. K., Bernardi M.,  2012, MNRAS, 422, 1825

\bibitem[]{} Simard, L., Mendel, J. T., Patton, D. R., Ellison, S. L. \& McConnachie, A. W. 2011, ApJS, 196, 11

\bibitem[]{} Stoughton C., et al., 2002, AJ, 123, 485

\bibitem[]{} Trujillo I., et al., 2006, MNRAS, 373, 36

\bibitem[van Dokkum et al. 2008]{vDokkum08}
 van Dokkum, P. G. et al. 2008, ApJL, 677, 5

\bibitem[]{} Vikram V., Wadadekar Y., Kembhavi A. K., Vijayagovindan G. V., 2010, MNRAS, 409, 1379

\end{thebibliography}
\end{document}